\documentclass[12pt,a4paper,oneside,reqno]{article}
\usepackage{adjustbox}
\usepackage{amsmath}
\usepackage[skip=2pt]{caption}  
\usepackage{amssymb}
\usepackage{float}
\usepackage[T1]{fontenc}
\usepackage{hhline}
\usepackage[utf8]{inputenc}
\usepackage{mathtools}
\usepackage{multicol}
\usepackage{multirow}
\usepackage{wasysym}
\usepackage{graphicx}
\usepackage{caption}
\usepackage{subcaption}
\usepackage{amsmath,amssymb} 
\usepackage{siunitx}
\usepackage[margin=1in]{geometry} 
\usepackage{enumerate}
\usepackage{hyperref}
\usepackage{graphicx}

\newcommand\blfootnote[1]{%
  \begingroup
  \renewcommand\thefootnote{}\footnote{#1}%
  \addtocounter{footnote}{-1}%
  \endgroup
}
\usepackage{amsmath,cases}

\usepackage{cite}
\usepackage{hyperref}
\hypersetup{
colorlinks=true,
linkcolor=blue,
filecolor=magneta,
urlcolor=cyan,
}
\begin{document}
 \nocite{*} 


\title{Analysis of a parametrically excited 2DOF oscillator with nonlinear restoring magnetic force and rotating rectangular beam}


\maketitle
\author
\begin{center}
{Muhammad Junaid-U-Rehman$^{1}$}, {Grzegorz Kudra}$^{1,*}$, Krystian Polczy\'nski$^1$, {Kevin Dekemele}$^{2}$, {Jan Awrejcewicz}$^{1}$
\blfootnote{
{E-mail address:} (M. J. Rehman) muhammad-junaid.u-rehman@dokt.p.lodz.pl, (G. Kudra${}^*$) grzegorz.kudra@p.lodz.pl, (K. Polczynski) krystian.polczynski@p.lodz.pl, (K. Dekemele) kevin.dekemele@ugent.be, (J. Awrejcewicz) jan.awrejcewicz@p.lodz.pl}
\end{center}
\begin{center}
\emph{$^1$Department of Automation, Biomechanics, and Mechatronics, Lodz University of Technology, $1/15$ Stefanowski st., $90-537$ \L\'od\'z, Poland}\\
\emph{$^2$Department of Electromechanical Systems and Metal Engineering,
Ghent University, Technologiepark 46, Ghent, Belgium}
\end{center}

\begin{abstract}
\noindent This study investigates a detailed analytical and numerical investigation of a nonlinear two-degree-of-freedom (2DOF) mechanical oscillator subjected to parametric excitation, magnetic stiffness nonlinearities, and dry friction. The considered system consists of two coupled oscillators, both of which are connected to a rotating rectangular beam that induces a time-periodic stiffness variation. The Complex Averaging (CxA) method is employed to derive approximate analytical solutions, which are thoroughly validated through time-domain simulations and bifurcation analyses. The dynamic analysis reveals a rich spectrum of nonlinear behaviors, including periodic, quasi-periodic, and chaotic responses. Detailed bifurcation diagrams, Lyapunov exponent analysis, and Poincaré maps demonstrate the influence of nonlinear stiffness degree, mass symmetry, and frictional effects on system stability and response amplitude. The obtained results give a significant understanding of the dynamic behavior of coupled nonlinear systems and establish a conceptual framework for the development of complex vibration abatement strategies, energy harvesting devices, and advanced mechanical systems.
\end{abstract}

\section{Introduction}
The significance of differential equations (DEs) in science and engineering is paramount. They function as effective methods for modeling, evaluating, and estimating the behavior of dynamic systems across different areas of study. DEs provide a mathematical framework for analyzing complex events, encouraging scientists and engineers to tackle significant problems, make informed decisions, and foster innovation and advancement in their respective fields. DEs are essential in engineering for the analysis and design of systems, including electrical circuits, mechanical structures, fluid flow systems, and control systems. DEs are utilized in mechanical engineering (ME) to analyze the dynamics of mechanical systems, such as vibrations, heat transfer, and fluid movement, allowing engineers to enhance designs for efficiency, safety, and performance. In electrical engineering (EE), DEs model the behavior of circuits comprising resistors, capacitors, and inductors, enabling engineers to predict voltage and current distributions and design circuits with specific features \cite{p2,p03,p3,p003,p0003}.

The study of mechanical oscillators, particularly those with multiple degrees of freedom (DOF), has been a fundamental area of research in the field of applied mechanics and engineering dynamics. Systems exhibiting parametric excitation and dry friction \cite{p0101} forces are particularly challenging due to their complex, nonlinear behaviors, which include bifurcations and chaotic responses. Parametrically excited systems based on Mathieu-type equations have been studied extensively in the literature \cite{p0102,p0103,p0104}, often without considering dry friction effects. These systems serve as classical examples of parametric excitation, with their dynamic behavior thoroughly investigated in numerous scholarly works \cite{p0105,p0106}. Understanding these dynamics is crucial for designing and optimizing mechanical systems used in various industries, including automotive, aerospace, and civil engineering. Examples of parametric oscillators in mechanical systems include pendulums with adjustable suspension points \cite{p10}. Such oscillations are observable in fundamental electrical systems \cite{p11}, and we can also see the parametric excitation in molecular physics \cite{p12}. The stabilizing behavior of the inverted pendulum, as investigated by P. Kapitza \cite{p14}, may resemble their behavior. A.H. Nayfeh discussed the 1DOF and 2DOF parametric oscillator by employing the multiple scales scheme and without dry friction in \cite{AH1,AH2,AH3}. J. Awrejcewicz et al. discussed the nonlinear dynamics of the 2DOF spring pendulum encompassing resonant and non-resonant oscillations. The resonance curves have been developed, and a stability study has been conducted. The kinematically excited spring pendulum is examined using the three time scales of the method of multiple scales, and both parametric and primary resonances, occurring either concurrently or independently, are explored in \cite{p7}.

Friction is a tangential reaction force that arises between two surfaces in contact and plays an important role in power transmission across a wide range of mechanical systems. It is present in nearly all mechanical engineering applications, including bearings, gear systems, hydraulic and pneumatic actuators, robotic joints, valves, disc brakes, and wheels. The phenomenon of friction stems from various underlying mechanisms, such as elastic and plastic deformation, fluid interactions, wave propagation, and material properties \cite{n01a}. Due to its inherently nonlinear nature, friction can lead to complex behaviors such as steady-state errors, limit cycles, and degraded system performance. A more comprehensive elucidation of the friction phenomenon at low velocities, particularly during the transition through zero velocity, is essential. Karnopp presented a model to resolve issues related to zero velocity detection and to prevent transitions between distinct state equations for sticking and sliding in \cite{n06}. A large number of scientists and researchers examined the friction models. Filippov \cite{dr01} examined the discontinuous dynamical characteristics of a Coulomb friction oscillator and introduced a DEs theory featuring discontinuous right-hand sides. The differential inclusion was introduced using set-valued analysis to describe sliding motion over the discontinuous border. Common instances include various combinations of Coulomb friction, viscous friction, and the Stribeck effect \cite{n02a}. Haessig and Friedland introduced a bristle model \cite{n03}, while Coulomb, hyperbolic tangent, and LuGre were explained in \cite{n04}. Shaw \cite{dr02} investigated the stability of periodic responses in friction-driven oscillators using Poincaré map techniques. In a related study, Awrejcewicz and Delfs \cite{dr03} analyzed the equilibrium stability of a two-degree-of-freedom system influenced by dry friction. Feeny and Moon \cite{dr04} explored the chaotic behavior of a dry-friction oscillator through both experimental observations and numerical simulations. Additionally, Oestreich et al. \cite{dr05} studied the bifurcation characteristics and stability of a non-smooth oscillator subject to frictional effects. A classical model addressing certain dynamic friction phenomena was proposed by Armstrong in \cite{n05}.

Damping has played a significant role in friction-induced vibrations. Hoffmann and Gaul \cite{dm01} examined a mass-on-belt system via the lens of the mode-coupling mechanism, revealing that non-proportional damping may unexpectedly destabilize the system despite expectations to the contrary. Fritz and colleagues \cite{dm04} employed an FEM of a whole braking corner, followed by a stability analysis, to investigate how damping influences the patterns of brake squeal coalescence. Sinou and Jezequel \cite{dm02} made analogous observations, indicating that an optimal structural damping ratio and pulsation ratio can reduce the unstable zone caused by mode coupling. An experimental study on the correlation between modal damping distribution and the tendency to produce squeal in a beam-on-disk configuration was conducted by Massi and Giannini \cite{dm03}. A mass-spring system that can move in three dimensions and includes friction, and they analyzed how different system parameters affect stability, especially looking at how damping affects instabilities when modes couple under flat and straight friction conditions, was discussed by Charroyer and colleagues \cite{dm05}. In addition to damping, nonlinear contact stiffness has a significant influence on the identification of unstable zones. Li et al. \cite{dm06} demonstrated that increasing the nonlinear stiffness generally leads to system destabilization; however, stability can be regained within certain higher stiffness ranges.

Typically, vibrations are considered unpleasant. Their impact on mechanical systems is detrimental and can lead to expensive malfunctions. In the last decades, there has been a growing focus on the movement of the nonlinear 2DOF oscillation system \cite{n0,n01,p6}. The behavior of interconnected vibrating systems, such as elastic beams supported by two springs and the vibration of a milling machine, can be accurately represented using these 2DOF systems \cite{n02}. Their equations of motion are comprised of two second-order differential equations that include cubic nonlinearities. The literature is described in standard vibration textbooks \cite{n1} and can also be found in References \cite{n2,n3}. M. Krack and J. Gross used the harmonic balance method for investigating nonlinear vibrations in mechanical systems and for identifying, analyzing, and controlling complex dynamical behaviors in engineering applications \cite{pp7}. Furthermore, J. Awrejcewicz and collaborators explored system stability, bifurcations, and Lyapunov exponents using Floquet theory applied to damped Mathieu-type systems \cite{lyap1,lyap2,lyap3}.

Considering the previous background, the goal of this research is to examine the different dynamics of a 2DOF parametric oscillator—a system subject to periodic variations in its parameters, resulting in rich dynamic behaviors. This study builds on previous works that have explored the dynamics of 1DOF systems \cite{p4,p5} but expands the analysis to 2DOF systems, which are more representative of real-world engineering applications. The main objective of this study is to explore and discuss the different dynamics by using the CxA method alongside numerical simulations to capture the intricate dynamics of the system. Utilizing a CxA method, we derive a series of modulation equations that are readily interpretable in physical terms and provide insight into the amplitude-frequency response \cite{n3a,n3aa}. We have computed the numerical results of the 2DOF parametric system, including time plots, phase plots, bifurcation diagrams for a deeper look at different dynamics, and Lyapunov exponents, which confirm the behavior of the considered system. By comparing analytical and numerical results, the research aims to bridge the gap between theoretical models and practical applications. The analysis will cover the influence of nonlinear stiffness and dry friction on the system's response, providing insights into how these factors contribute to the overall stability and performance of mechanical systems.

The research is structured as follows: Section \ref{sec2} introduces the dynamical system, and it is divided into four further subsections: The physical model is described in detail in subsection \ref{subsec2.1}; the mathematical model of the system and dimensional and nondimensional equations of motion are derived in subsections \ref{subsec2.2} and \ref{subsec2.3}; and dimensional and nondimensional system parameters are examined in subsection \ref{subsec2.4}. In Section \ref{sec3}, analytical solutions are computed by using the CxA method. Section \ref{sec4} explores the amplitude frequency response by the CxA method, a comparison analysis of analytical and numerical results, bifurcation analysis, and time, phase, and Poincaré plots for the considered parametric system, and it is divided into two subsections; firstly, we investigated the 2DOF system with the different masses in subsection \ref{different_masses}, and then we discussed the 2DOF system with the same masses in subsection \ref{same_masses}. At last, Section \ref{sec5} summarizes the results and conclusions of this investigation.
\section{Dynamical System}\label{sec2}
\subsection{Physical model}\label{subsec2.1}
\begin{figure}[H]
  \centering
  \includegraphics[scale=0.18]{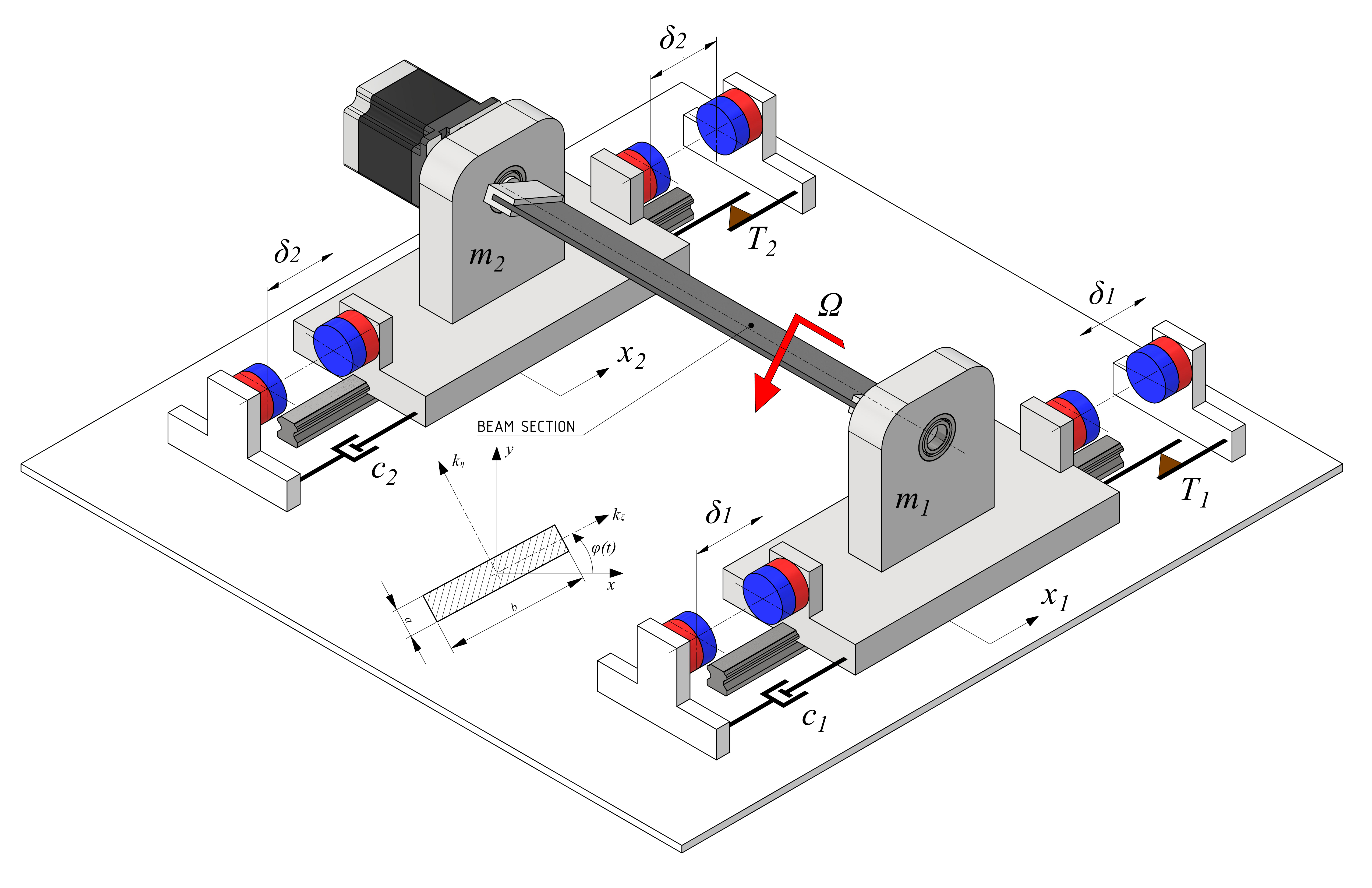} 
  \caption{Experimental setup of 2DOF mechanical parametric oscillator.}
  \label{fig1}
\end{figure}

The physical concept of the investigated system is presented in Fig. \ref{fig1}. This is a physical model of a possible experimental setup that is an extension of the 1DOF system experimentally studied in the works \cite{p4,p5} to a 2DOF system.

The system consists of two bodies with masses $m_1$ and $m_2$ moving parallel on rolling guides, whose displacement relative to the equilibrium position is defined by coordinates $x_1$ and $x_2$, respectively. The motion resistances, the main source of which are the rolling guides, are modeled as the sum of the viscous damping (with coefficients $c_1$ and $c_2$) and a component mathematically equivalent to the Coulomb dry friction (parameters $T_1$ and $T_2$). Each body features a strongly nonlinear spring realized by appropriate pairs of repulsive magnets, mounted so that in the equilibrium position the distance between them is $\delta_1$ and $\delta_2$, respectively. Both bodies are coupled by an additional linear spring implemented utilizing a beam connecting them. This beam has a rectangular cross-section with dimensions $a$ and $b$, so the spring has different stiffnesses in the two principal directions, equal to $k_\xi$ and $k_\eta$, respectively. One of the bodies is equipped with a stepper motor that controls the angular position of the beam $\varphi(t)$. Presented studies assume a constant angular velocity of the beam, $\Omega$, and angle $\varphi(t) = \Omega t$. 

\subsection{Dimensional equations of motion}\label{subsec2.2}
The physical model presented in the previous section is described by the following governing equations of motion:
\begin{equation}\label{equ1}
\begin{aligned}
m_1\ddot{x}_1+c_1\dot{x}_1+T_1 f_0({\dot{x}_1})+K(t)(x_1-x_2)+F_{s1}(x_1)=0,\\
m_2\ddot{x}_2+c_2\dot{x}_2+T_2 f_0({\dot{x}_2})+K(t)(x_2-x_1)+F_{s2}(x)=0.
\end{aligned}
\end{equation}
In the above equations, the terms $c_i \dot{x}_i+T_i f_0({\dot{x}_i});~i=1,2$, represent the resistance force in the rolling guide (for details, see \cite{p4}). The following set-valued function $f_0(\upsilon)$ is used to describe the Coulomb friction law
\begin{subequations}\label{dry1}
\begin{numcases}
   {f_0(\upsilon)} 
        =1, & \text{if } ~~ $\upsilon > 0$ \label{dry_a}, \\
       \in [-1,1], & \text{if }~~~ $\upsilon=0$ \label{dry_b},\\
=-1, & \text{if }~~~ $\upsilon < 0$ \label{dry_c}.
   \end{numcases}
\end{subequations}
The function $K(t)=\frac{k_{\xi} + k_{\eta}}{2} + \frac{k_{\xi} - k_{\eta}}{2}\cos\left(2\Omega t\right)$ represents the stiffness that varies over time due to the beam that connects the two oscillators and rotates at angular velocity $\Omega$. 
The forces $F_{si}(x_i);~i=1,2,$ are the results of the magnetic springs, which originate from the existence of four sets of repulsive magnets.

In this work, we use the following model of forces generated by magnetic springs
\begin{equation}
    \label{3.2a}
    F_{si}(x_i)=k_{i1}x_i+k_{i3}x_i^3+k_{i5}x_i^5;~i=1,2.
\end{equation}
 In previous works \cite{p4,p5} we used a more complicated model of forces generated by magnetic springs
\begin{equation}
    \label{3.2b}
    F_{si}(x_i)=F_{\rm MOi}\Big[\frac{1}{\{1 +d_i(\delta_i - x_i)\}^4} - \frac{1}{\{1 +d_i(\delta_i + x_i)\}^4}\Big];~i=1,2,
\end{equation}
where $F_{\rm MOi}$, $d_i$ and $\delta_i$  are some parameters. In Sec. \ref{parameters} it is shown that the model (\ref{3.2a}) can replace the model (\ref{3.2b}).

\subsection{Non-dimensional equations of motion}\label{subsec2.3}
Let us introduce dimensionless time $\tau = \omega_{\rm n} t$, where $\omega_n$ is the frequency of small vibrations of the first body of mass $m_1$, with the second body immobilized,  for average stiffness:
\begin{equation}
    \label{3.2aa}
    \omega_{\rm n} = \sqrt{\frac{\frac{k_\xi+k_\eta}{2}+k_{11}}{m_1}}.
\end{equation}
The subsequent connections can be established between the derivatives of actual and dimensionless time:$$\frac{d(\dots)}{dt}=\frac{d(\dots)}{d\tau}\frac{d(\tau)}{dt}=\omega_{\rm n} \frac{d(...)}{d\tau}.$$
Furthermore, we establish the non-dimensional displacements of the system as $u_i = \frac{x_i}{\delta_1}$; $i=1,2$. Thus, we may express the following relations: $\dot{x}_i = \omega_n \delta_1 {u}^{\prime}_i$, $\ddot{x}_i = \omega_n^2 \delta_1 {u}^{\prime\prime}_i$; $i=1,2$, and $\Omega=\omega_n \omega$, where it is used the notation $(...)^\prime=\frac{d(...)}{d\tau}$ and $\omega = \varphi^\prime$ denotes the non-dimensional angular frequency of parametric forcing.

By applying the previously defined relationships, the original governing equation of motion (\ref{equ1}) can be transformed into its dimensionless form:
\begin{subequations}\label{ndeqs}
\begin{align}
&{u}^{\prime\prime}_1+2\zeta_1 {u}^{\prime}_1+\sigma_1 f_0({u}^{\prime}_1)+\big(p+q \cos(2\omega \tau)\big)(u_1-u_2)+(1-p) u_1+\kappa_{13} u_1^3
+\kappa_{15} u_1^5=0,\label{2a}
\\
&{u}^{\prime\prime}_2+2\zeta_2 {u}^{\prime}_2+\sigma_2 f_0({u}^{\prime}_2)+\mu\big(p+q \cos(2\omega \tau)\big)(u_2-u_1)+\kappa_{21} u_2+\kappa_{23} u^3_2+\kappa_{25} u^5_2=0,\label{2b}
\end{align}\label{eq2}
\end{subequations}
where we defined and implemented the subsequent parameters (note that $p+\kappa_{11}=1$):
\begin{equation}
\begin{split} \label{pr0}
&\zeta_i=\frac{c_i}{2m_i\omega_n},~\sigma_i=\frac{T_i}{m_i \omega_n^2 \delta_1},~p=\frac{k_\xi+k_\eta}{2m_1 \omega_n^2},~q=\frac{k_\xi-k_\eta}{2m_1 \omega_n^2}, \kappa_{ij}=\frac{\delta_1^{j-1}}{m_i\omega_n^2}k_{ij} ,~\mu=\frac{m_1}{m_2}.
\end{split}
\end{equation}

During numerical simulations using dimensionless equations (\ref{ndeqs}), the non-smooth dry friction terms $f_0({u}^{\prime}_i)$ are replaced by their smooth counterparts $f_{0r}(\alpha {u}^{\prime}_i)$ ($i=1,2$), where the corresponding smooth function is defined as
\begin{equation}
    \label{smoothedDryFriction}
    f_{0r}(\upsilon)=\frac{\upsilon}{\sqrt{\upsilon^2+1}},
\end{equation}
and where it is assumed $\alpha=10^3$.
\subsection{{The system parameters}}\label{subsec2.4}
\label{parameters}
The parameters of the investigated dynamical system for the exemplary analyses performed in this work are based on the parameters measured and identified on the real experimental stand of the same dynamical system but having 1DOF, i.e., with a fixed cart of mass $m_2$ \cite{p4,p5}. Additionally, it is assumed that the motion resistances in the linear rolling bearings and the magnetic springs are the same along both carts with masses $m_1$ and $m_2$.

\begin{figure}[H]
    \centering
    \begin{subfigure}{0.49\textwidth}
        \includegraphics[width=\linewidth]{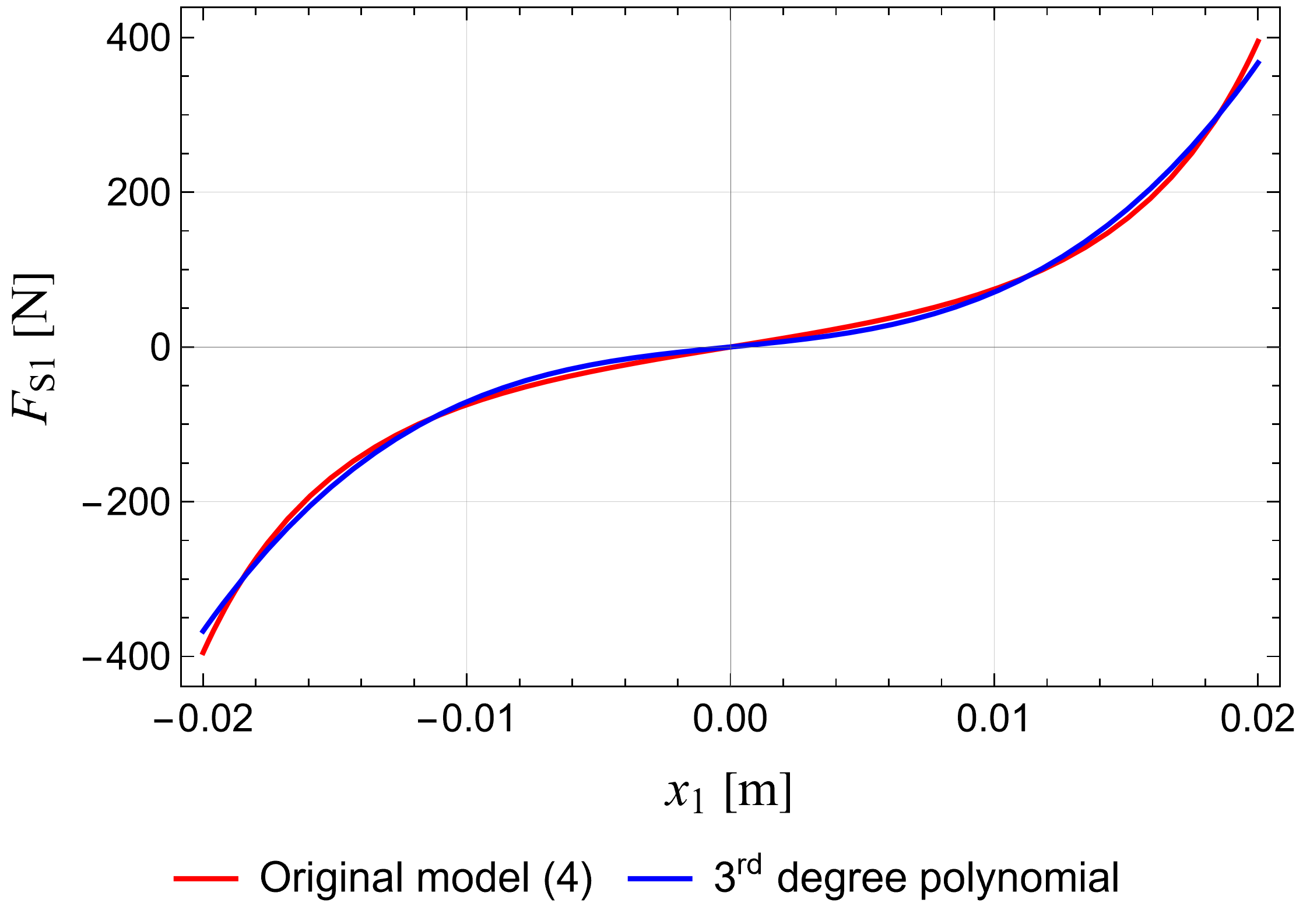}
        \caption{}
        \label{mag1a}
    \end{subfigure}
    \begin{subfigure}{0.49\textwidth}
        \includegraphics[width=\linewidth]{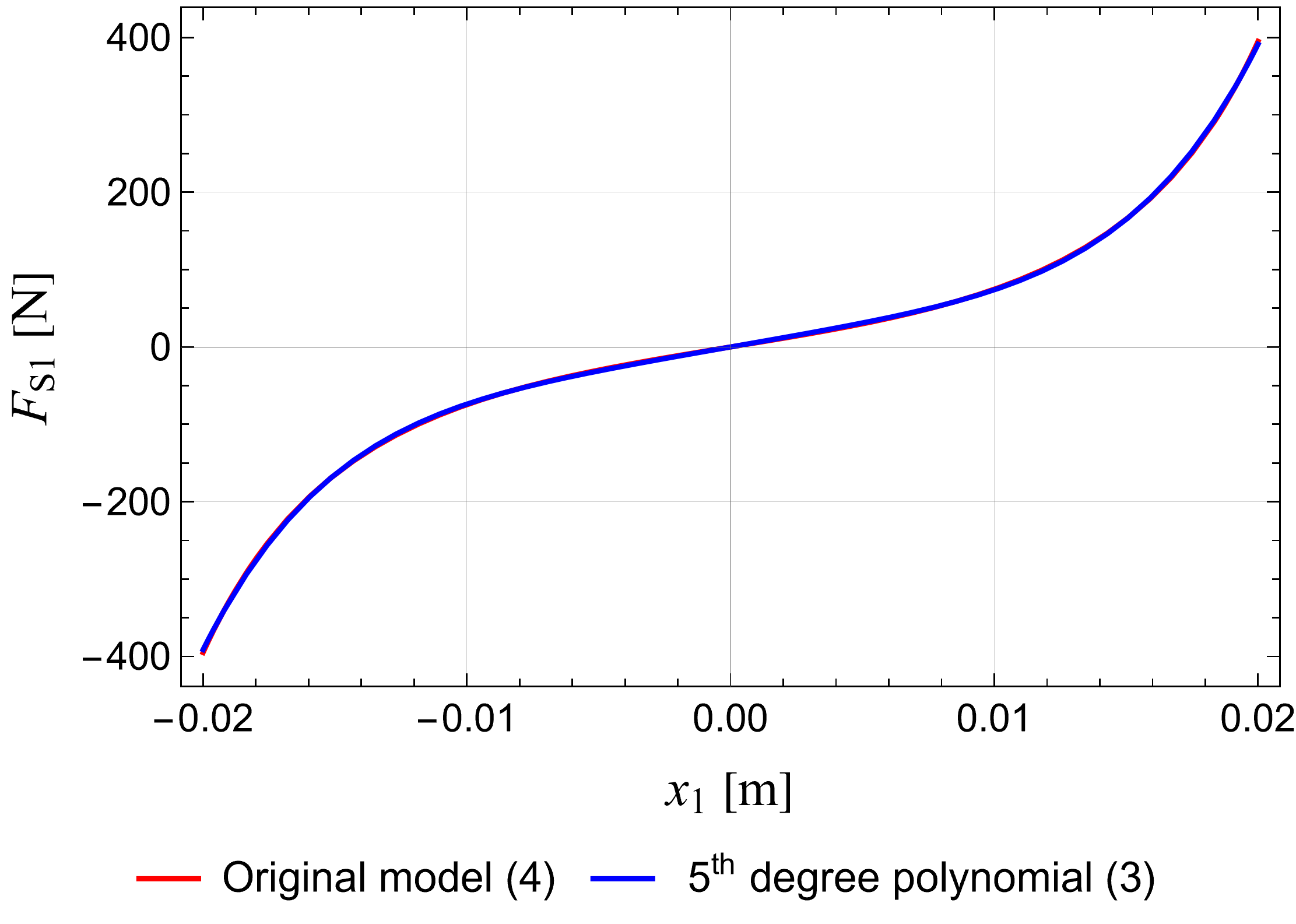}
        \caption{}
        \label{mag1b}
    \end{subfigure}
\caption{Characteristics of the magnetic force: original model (\ref{3.2b}) and its approximations by \(\text{3}^{\text{rd}}\) degree (\ref{mag1a}) and \(\text{5}^{\text{th}}\) (\ref{mag1b}) degree polynomial models.}
    \label{mag1}
\end{figure}

Based on exemplary parameters of magnetic spring identified on a real experimental stand: $F_{M01}=400.96$ N, $d_1=46.628$ $\text{m}^{-4}$, $\delta_1=0.02$ m, it can be proved that model (\ref{3.2a}) is sufficient to replace model (\ref{3.2b})—see Fig. \ref{mag1}, where two versions of the approximation of the original model (\ref{3.2b}) are shown: using a \(\text{3}^{\text{rd}}\) degree polynomial (\ref{mag1a}) and a \(\text{5}^{\text{th}}\) degree polynomial (\ref{mag1b}), i.e., using model (\ref{3.2a}). The fitting was performed by minimizing the mean square of the difference between the corresponding characteristics in the interval $\langle-\delta_1,\delta_1 \rangle$.
It was found that the \(5^{\text{th}}\)-order polynomial approximation captures the behavior of the original stiffness function with high accuracy over the full range of interest. In contrast, while the \(3^{\text{rd}}\)-order polynomial provides a reasonable approximation in terms of average fit, it exhibits significant discrepancies in the slope, particularly near \(x_1 = 0\). This deviation alters the extracted linear stiffness coefficient \(k_{11}\) and consequently affects all related dimensionless parameters. Although the underlying physical system remains unchanged, such a change in representation introduces inconsistencies in the nondimensional model. Therefore, to maintain consistency in parameter scaling and preserve the fidelity of the model, only the \(5^{\text{th}}\)-order polynomial representation is used in the following analysis.


\noindent In the remainder of the work, we will investigate two variants of the dynamic system: with different masses $m_1 =4.55$ kg, $m_2 =8.72$ kg (A) and with equal masses $m_1 = m_2=8.72$ kg (B). The corresponding real parameters of different variants of the system are gathered in Table \ref{tab1}, while the parameters of the corresponding nondimensional system can be found in Table \ref{tab2}. Such a choice of variants of the mass values of both oscillators is justified by the possibility of carrying out an appropriate experiment, because the value of $m_2=8.72$ kg corresponds to the actual mass of the second cart on the experimental stand (it contains a stepper motor driving the beam connecting both oscillators). On the other hand, it is possible to add mass to the first trolley and make it equal to the mass of the second trolley. Table \ref{tab1} also presents the relations for some parameters that link them with other dimensionless parameters of the system, which are valid for the special case under consideration, in which the dimensional parameters of the motion resistances in the bearings and the magnetic springs are the same for both carts.

\begin{table}[H]
\centering
 \caption{Parameters of dimensional model.}
\begin{adjustbox}{max width=\textwidth}
\resizebox{0.75\textwidth}{!}{ 
\begin{tabular}{p{2.37cm}p{1.63cm}p{2.4cm}p{2.4cm}}
\hline
\multicolumn{1}{|p{2.37cm}|}{\multirow{2}{*}{\parbox{2.74cm}{\centering parameter}}} & 
\multicolumn{1}{|p{1.63cm}|}{\multirow{2}{*}{\parbox{1.63cm}{\centering unit}}} & 
\multicolumn{2}{|p{4.8cm}|}{\centering model version/parameters’ values} \\
\hhline{~~--}
\multicolumn{1}{|p{2.37cm}|}{} & 
\multicolumn{1}{|p{1.63cm}|}{} & 
\multicolumn{1}{|p{2.4cm}|}{\centering A} & 
\multicolumn{1}{|p{2.4cm}|}{\centering B} \\
\hline
\multicolumn{1}{|p{2.37cm}|}{\centering\( m_{1} \)} & 
\multicolumn{1}{|p{1.63cm}|}{\centering\( \text{kg} \)} & 
\multicolumn{1}{|p{2.4cm}|}{\centering 4.55} & 
\multicolumn{1}{|p{2.4cm}|}{\centering 8.72} \\
\hline
\multicolumn{1}{|p{2.37cm}|}{\centering\( m_{2} \)} & 
\multicolumn{1}{|p{1.63cm}|}{\centering\( \text{kg} \)} & 
\multicolumn{1}{|p{2.4cm}|}{\centering 8.72} & 
\multicolumn{1}{|p{2.4cm}|}{\centering 8.72} \\
\hline
\multicolumn{1}{|p{2.37cm}|}{\centering\( k_{\xi} \)} & 
\multicolumn{1}{|p{1.63cm}|}{\centering\(\text{N/m} \)} & 
\multicolumn{2}{|p{4.80cm}|}{\centering \(4336\) } \\
\hline
\multicolumn{1}{|p{2.37cm}|}{\centering\( k_{\eta} \)} & 
\multicolumn{1}{|p{1.63cm}|}{\centering\(\text{N/m} \)} & 
\multicolumn{2}{|p{4.80cm}|}{\centering \(271\)} \\
\hline
\multicolumn{1}{|p{2.37cm}|}{\centering\( \delta_{1} = \delta_{2} \)} & 
\multicolumn{1}{|p{1.63cm}|}{\centering\( \text{m} \)} & 
\multicolumn{2}{|p{4.80cm}|}{\centering \( 0.02\)} \\
\hline
\multicolumn{1}{|p{2.37cm}|}{\centering\( F_{M01}=F_{M02} \)} & 
\multicolumn{1}{|p{1.63cm}|}{\centering\( \text{N} \)} & 
\multicolumn{2}{|p{4.80cm}|}{\centering\( 401.0\)} \\
\hline
\multicolumn{1}{|p{2.37cm}|}{\centering\( d_{1} = d_{2} \)} &
\multicolumn{1}{|p{1.63cm}|}{\centering\( \text{m}^{-4} \)} & 
\multicolumn{2}{|p{4.80cm}|}{\centering \(46.63\) } \\
\hline
\multicolumn{1}{|p{2.37cm}|}{\centering\( k_{11} = k_{21} \)} & 
\multicolumn{1}{|p{1.63cm}|}{\centering\(\text{N/m} \)} & 
\multicolumn{2}{|p{4.80cm}|}{\centering \(6.0 \cdot 10^3\)} \\
\hline
\multicolumn{1}{|p{2.37cm}|}{\centering\( k_{13} = k_{23} \)} & 
\multicolumn{1}{|p{1.63cm}|}{\centering\( \text{N/m}^3 \)} & 
\multicolumn{2}{|p{4.80cm}|}{\centering\( 7.12 \cdot 10^6\) } \\
\hline
\multicolumn{1}{|p{2.37cm}|}{\centering\( k_{15} = k_{25} \)} & 
\multicolumn{1}{|p{1.63cm}|}{\centering\( \text{N/m}^5 \)} & 
\multicolumn{2}{|p{4.80cm}|}{\centering \(6.70 \cdot 10^{10}\)} \\
\hline
\multicolumn{1}{|p{2.37cm}|}{\centering\( c_{1} = c_{2} \)} & 
\multicolumn{1}{|p{1.63cm}|}{\centering\( \text{N} \cdot \text{s/m} \)} & 
\multicolumn{2}{|p{4.80cm}|}{\centering \(16.48\)} \\
\hline
\multicolumn{1}{|p{2.37cm}|}{\centering\( T_{1} = T_{2} \)} & 
\multicolumn{1}{|p{1.63cm}|}{\centering\( \text{N} \)} & 
\multicolumn{2}{|p{4.80cm}|}{\centering \(1.512\) } \\
\hline
\end{tabular}
}
\end{adjustbox}
\label{tab1}
\end{table}

\begin{table}[H]
\centering
 \caption{Parameters of non-dimensional model.}
\begin{adjustbox}{max width=\textwidth}
\resizebox{0.60\textwidth}{!}{ 
\begin{tabular}{p{2.4cm}p{2.4cm}p{2.4cm}}
\hline
\multicolumn{1}{|p{2.4cm}|}{\multirow{2}{*}{\parbox{2.14cm}{\centering parameter}}} & 
\multicolumn{2}{|p{4.8cm}|}{\centering model version/parameters’ values} \\
\hhline{~--}
\multicolumn{1}{|p{2.4cm}|}{} & 
\multicolumn{1}{|p{2.4cm}|}{\centering A} & 
\multicolumn{1}{|p{2.4cm}|}{\centering B} \\
\hline
\multicolumn{1}{|p{2.4cm}|}{\centering\( \mu \)} & 
\multicolumn{1}{|p{2.40cm}|}{\centering\( 0.5219 \)} & 
\multicolumn{1}{|p{2.40cm}|}{\centering\( 1 \)} \\
\hline
\multicolumn{1}{|p{2.4cm}|}{\centering\( p \)} & 
\multicolumn{2}{|p{4.8cm}|}{\centering\( 0.2775 \)} \\
\hline
\multicolumn{1}{|p{2.4cm}|}{\centering\( q \)} & 
\multicolumn{2}{|p{4.8cm}|}{\centering\( 0.2449 \)} \\
\hline
\multicolumn{1}{|p{2.4cm}|}{\centering\( \kappa_{13} \)} & 
\multicolumn{2}{|p{4.80cm}|}{\centering\( 0.343 \)} \\
\hline
\multicolumn{1}{|p{2.4cm}|}{\centering\( \kappa_{15} \)} & 
\multicolumn{2}{|p{4.80cm}|}{\centering\( 1.291 \)} \\
\hline
\multicolumn{1}{|p{2.4cm}|}{\centering\( \kappa_{21} \)} &  
\multicolumn{2}{|p{4.80cm}|}{\centering\( \mu(1-p) \)} \\
\hline
\multicolumn{1}{|p{2.4cm}|}{\centering\( \kappa_{23} \)} &  
\multicolumn{2}{|p{4.80cm}|}{\centering\( \mu~\kappa_{13} \)} \\
\hline
\multicolumn{1}{|p{2.4cm}|}{\centering\( \kappa_{25} \)} & 
\multicolumn{2}{|p{4.80cm}|}{\centering\( \mu~\kappa_{15} \)} \\
\hline
\multicolumn{1}{|p{2.4cm}|}{\centering\( \zeta_{1} \)} &  
\multicolumn{2}{|p{4.80cm}|}{\centering\( 0.03062~\mu^{-1/2} \)} \\
\hline
\multicolumn{1}{|p{2.4cm}|}{\centering\( \zeta_{2} \)} & 
\multicolumn{2}{|p{4.80cm}|}{\centering\(\mu~\zeta_1 \)} \\
\hline
\multicolumn{1}{|p{2.4cm}|}{\centering\( \sigma_{1} \)} & 
\multicolumn{2}{|p{4.80cm}|}{\centering\( 0.009106 \)} \\
\hline
\multicolumn{1}{|p{2.4cm}|}{\centering\( \sigma_{2} \)} & 
\multicolumn{2}{|p{4.80cm}|}{\centering\( \mu~\sigma_1 \)} \\
\hline
\end{tabular}}
\end{adjustbox}
 \label{tab2}
\end{table}

\section{Analytical solutions}\label{sec3}
In this section, our objective is to perform an analysis of the system (\ref{eq2}) using a perturbation method. We utilize the CxA method as recommended in reference \cite{cca1}. Therefore, we utilize the methodology above to provide analytical solutions that closely correspond to the numerical solutions.

\subsection{Approximate analytical solutions by CxA method}
To obtain the amplitude–frequency response, the Complexification–Averaging (CxA) method \cite{cca2,cca3} is utilized, which has proven effective in analyzing strongly nonlinear mechanical systems. This approach involves decomposing the system's motion into fast oscillatory components and slowly varying amplitude modulations through the introduction of complex variables. Subsequently, the amplitude modulation dynamics are averaged over one period of the base excitation frequency. Recent developments have extended this method to support averaging over multiple excitation frequencies \cite{cca4}. The first step in the CxA procedure is to define the time-dependent complex variables representing the modulated amplitudes, $A_1\in \mathbb{C}$ and $A_2 \in \mathbb{C}$:
\begin{align}\label{ca1}
2 A_1(\tau) e^{i \omega \tau}=u_1(\tau)-i \frac{u_1^\prime(\tau)}{\omega},\hspace{2cm}
2 A_2(\tau) e^{i \omega \tau}=u_2(\tau)-i \frac{u_2^\prime(\tau)}{\omega},
\end{align}
where $i=\sqrt{-1}$ is the complex number. The variables $u_1$, $u_2$, and the derivatives of these variables as functions of $A_1$ and $A_2$ are derived by adding and subtracting Eq. (\ref{ca1}) and their complex conjugates:
\begin{subequations}
\begin{align}
&u_1=A_1 e^{i \omega \tau}+\bar{A}_1 e^{-i \omega \tau},\hspace{2.2cm}u_2=A_2 e^{i \omega \tau}+\bar{A}_2 e^{-i \omega \tau},\label{ca2a} \\
&u_1^\prime=i\omega\big(A_1 e^{i \omega \tau}-\bar{A}_1 e^{-i \omega \tau}\big),\hspace{1.5cm}u_2^\prime=i\omega\big(A_2 e^{i \omega \tau}-\bar{A}_2 e^{-i \omega \tau}\big),\label{ca2b}
\end{align}\label{ca2}
\end{subequations}
\noindent where $\bar{A}$ denotes the conjugate of a complex-valued function $A$. Taking the time derivative of Eq. (\ref{ca1}) to get the acceleration in terms of new variables $A_1$ and $A_2$ below:
\begin{align}\label{ca3}
2 A^\prime_1 e^{i \omega \tau}+2i\omega A_1 e^{i \omega \tau}=u_1^\prime-i \frac{u_1^{\prime\prime}}{\omega},\hspace{2cm}
2 A^\prime_2 e^{i \omega \tau}+2i\omega A_2 e^{i \omega \tau}=u_2^\prime-i \frac{u_2^{\prime\prime}}{\omega}.
\end{align}
Using Eq. (\ref{ca2b}) into Eq. (\ref{ca3}), which yields
\begin{align}\label{ca4}
u_1^{\prime\prime}=2i\omega A^\prime_1 e^{i \omega \tau}-\omega^2 u_1,\hspace{2cm}
u_2^{\prime\prime}=2i\omega A^\prime_2 e^{i \omega \tau}-\omega^2 u_2.
\end{align}
Using Eqs. (\ref{ca2}), \& (\ref{ca4}) in Eq. (\ref{eq2}), multiplying by $e^{-i \omega \tau}$ and then averaging with the forcing frequency $\omega$ results in
\begin{subequations}
\begin{align}
&
2 \zeta_1 \omega A_1 + 2 i \omega A^\prime_1 
+ A_1 - p A_2+ \frac{q}{2}(\bar{A}_1-\bar{A}_2)
+ 3 \kappa_{13} A^2_1\bar{A}_1+ 10 \kappa_{15} A^3_1 \bar{A}_1^2 
- \omega^2 A_1\nonumber
\\& + \sigma_1 f_0\big(i\omega A_1 e^{i \omega \tau}-i\omega\bar{A}_1 e^{-i \omega \tau}\big) = 0,\label{ca5a} \\
&2i\zeta_2 \omega A_2 + 2 i \omega A^\prime_2 
+ \kappa_{21}A_2 - \mu p A_1+\mu p A_2+ \frac{\mu q}{2}(\bar{A}_2-\bar{A}_1)+ 3 \kappa_{23} A^2_2\bar{A}_2+ 10 \kappa_{25} A^3_2 \bar{A}_2^2\nonumber\\& - \omega^2 A_2 + \sigma_2 f_0\big(i\omega A_2 e^{i \omega \tau}-i\omega\bar{A}_2 e^{-i \omega \tau}\big) = 0.\label{ca5b}
\end{align}\label{ca5}
\end{subequations}
\noindent We can expand the dry friction function in a Fourier series to capture all the resonant aspects of the following form
\begin{align}\label{eq177a}
& f_0(u^\prime_1)=f_0\big(i\omega A_1 e^{i \omega \tau}-i\omega\bar{A}_1 e^{-i \omega \tau}\big)=\sum_{-\infty}^{+\infty}r_{m}e^{im\omega\tau},
\end{align}
and
\begin{align}\label{eq177aa}
& f_0(u^\prime_2)=f_0\big(i\omega A_2 e^{i \omega \tau}-i\omega\bar{A}_2 e^{-i \omega \tau}\big)=\sum_{-\infty}^{+\infty}s_{n}e^{in\omega\tau}.
\end{align}
The Fourier coefficients \( r_{m}\) and \( {s}_n \) in Eqs. (\ref{eq177a}) and (\ref{eq177aa}) respectively, can be expressed as follows:
\begin{align}\label{eq1777a}
&r_m=\frac{1}{2\pi}\int_{0}^{2\pi} f_0\big(i\omega A_1 e^{i \omega \tau}-i\omega\bar{A}_1 e^{-i \omega \tau}\big)e^{im\omega\tau} d\tau,
\end{align}
and
\begin{align}\label{eq1777aa}
&s_n=\frac{1}{2\pi}\int_{0}^{2\pi} f_0\big(i\omega A_2 e^{i \omega \tau}-i\omega\bar{A}_2 e^{-i \omega \tau}\big)e^{in\omega\tau} d\tau.
\end{align}
In the Fourier series (\ref{eq1777a}) and (\ref{eq1777aa}), the terms $r_1 e^{i\omega\tau}$ and, $s_1 e^{i\omega\tau}$ respectively, are only the resonant terms. To eliminate secular terms, the resonant terms in Eq. (\ref{ca5}) must be removed. This yields:
\begin{subequations}
\begin{align}
&
2 \zeta_1 \omega A_1 + 2 i \omega A^\prime_1 
+ A_1 - p A_2+ \frac{q}{2}(\bar{A}_1-\bar{A}_2)
+ 3 \kappa_{13} A^2_1\bar{A}_1\nonumber
\\& + 10 \kappa_{15} A^3_1 \bar{A}_1^2 
- \omega^2 A_1+ \sigma_1 r_1 = 0,\label{ca55a} \\
&2i\zeta_2 \omega A_2 + 2 i \omega A^\prime_2 
+ \kappa_{21}A_2 - \mu p A_1+\mu p A_2+ \frac{\mu q}{2}(\bar{A}_2-\bar{A}_1)+ 3 \kappa_{23} A^2_2\bar{A}_2\nonumber\\&+ 10 \kappa_{25} A^3_2 \bar{A}_2^2 - \omega^2 A_2 + \sigma_2 s_1 = 0.\label{ca55b}
\end{align}\label{ca55}
\end{subequations}
Introducing the following polar forms below:
\begin{align}\label{polar}
    A_1=\frac{1}{2} a_1 e^{i \alpha_1}, \quad\quad\quad A_2=\frac{1}{2} a_2 e^{i \alpha_2},
\end{align}
for separating the real and imaginary parts. Eq. (\ref{polar}) gives us $A^\prime_1=\frac{1}{2}\big(a^\prime_1+ia_1{\alpha^\prime_1}\big)e^{i\alpha_1}$ and $A^\prime_2=\frac{1}{2}\big(a^\prime_2+ia_2{\alpha^\prime_2}\big)e^{i\alpha_2}$, where $^\prime$ shows the time derivative. A new form $u_1 = a_1 \cos \psi_1$ and $u_2 = a_2 \cos \psi_2$ are derived by incorporating an additional parameter, specifically $\psi_1 =\omega \tau + \alpha_1$ and $\psi_2 =\omega \tau + \alpha_2$ respectively. In this context, the term "dry friction" is expressed as follows for the scenario $a_1,a_2 > 0$, The dry friction term is expressed as follows:
\begin{align}\label{15A}
 r_1&=\frac{1}{2\pi}\int_0^{2\pi} f_0(-a_1\sin\psi_1)e^{-i\psi_1}e^{i\alpha_1}d\psi_1,\quad \Rightarrow \quad \frac{2i}{\pi}e^{i\alpha_1},
  \end{align}
and
\begin{align}\label{15AA}
 s_1&=\frac{1}{2\pi}\int_0^{2\pi} f_0(-a_2\sin\psi_2)e^{-i\psi_2}e^{i\alpha_2}d\psi_2,\quad \Rightarrow \quad \frac{2i}{\pi}e^{i\alpha_2}.
  \end{align}
For the steady-state case, the time derivative of $A_1$ \& $A_2$ will be equal to zero. So substituting the $A^\prime_1=A^\prime_2=0$ into Eq. (\ref{ca5}) for the steady case, and then using the Eqs. (\ref{polar}), (\ref{15A}), and (\ref{15AA}) into Eq. (\ref{ca5}). Then, after a long but smooth calculation, we will get the following form of equations:
\begin{subequations}
\begin{align}
&\frac{5}{8} \kappa_{15} a_1^5 + \frac{3}{4} \kappa_{13} a_1^3 - p a_2e^{{i}(\alpha_2-\alpha_1)}- \frac{q}{2} a_2 e^{-{i}(\alpha_2+\alpha_1)} + \frac{q}{2}a_1 e^{-2{i} \alpha_1} \nonumber\\&+2 i\zeta_1\omega a_1-\omega^2a_1+a_1 + \frac{4i\sigma_1}{\pi}=0
,\label{ca6a} \\
&\frac{5}{8} \kappa_{25} a_2^5+ \frac{3}{4} \kappa_{23} a_2^3-\mu pa_1e^{i(\alpha_1-\alpha_2)}-\frac{q}{2}\mu a_1e^{-i(\alpha_1+\alpha_2)}+\frac{q}{2}\mu a_2e^{-2{i}\alpha_2}\nonumber\\&+2i\zeta_2\omega a_2-\omega^2 a_2+\mu pa_2+\kappa_{21}a_2+\frac{4i\sigma_2}{\pi}=0.\label{ca6b}
\end{align}\label{ca6}
\end{subequations}
Separating the real and imaginary parts of Eqs. (\ref{ca6a}) \& (\ref{ca6b}), which gives us the following system of equations:
\begin{subequations}
\begin{align}
&\frac{q}{2}a_1\cos(2\alpha_1)-\frac{q}{2}a_2\cos(\alpha_2+\alpha_1)+\frac{3}{4}\kappa_{13}a_1^3+\frac{5}{8}\kappa_{15}a_1^5\label{ca7a}\\&+(1-\omega^2)a_1-pa_2\cos(\alpha_2-\alpha_1)=0,\nonumber\\
&2\omega\zeta_1 a_1-pa_2\sin(\alpha_2-\alpha_1)-\frac{q}{2}a_1\sin(2\alpha_1)+\frac{q}{2}a_2\sin(\alpha_2+\alpha_1)+\frac{4}{\pi}\sigma_1=0,\label{ca7b}\\
&\frac{q}{2}\mu a_2\cos(2\alpha_2)-\frac{q}{2}\mu a_1\cos(\alpha_1+\alpha_2)+\frac{3}{4}\kappa_{23}a_2^3+\frac{5}{8}\kappa_{25}a_2^5\label{ca7c}\\&+(\kappa_{21}+\mu p-\omega^2)a_2-\mu p a_1\cos(\alpha_1-\alpha_2)=0,\nonumber\\
&2\omega\zeta_2 a_2-\mu pa_1\sin(\alpha_1-\alpha_2)-\frac{q}{2}\mu a_2\sin(2\alpha_2)+\frac{q}{2}\mu a_1\sin(\alpha_1+\alpha_2)+\frac{4}{\pi}\sigma_2=0.\label{ca7d}
\end{align}\label{ca7}
\end{subequations}
Note that Eqs. (\ref{ca7a}) \& (\ref{ca7b}) are the real and imaginary parts of Eq. (\ref{ca6a}) and Eqs. (\ref{ca7c}) \& (\ref{ca7d}) are the real and imaginary parts of Eq. (\ref{ca6b}) respectively. The stability of the solutions of Eqs. (\ref{ca7}) is judged from the Jacobian of Eqs. (\ref{ca55}).

\section{Dynamical Analysis}\label{sec4}
In this section, we analyzed the considered system analytically and numerically and also performed a comparison analysis of analytical and numerical solutions. We have divided this section into two parts. Subsection (\ref{different_masses}) describes the analysis for different masses $ m_1\neq m_2$, and Subsection (\ref{same_masses}) describes the behavior of the considered system for the same masses $m_1= m_2$.
\subsection{Bifurcation analysis for different masses (\texorpdfstring{$m_1\neq m_2$}{m₁ \neq m₂})}\label{different_masses}
This subsection explains the results of the bifurcation analysis obtained analytically and numerically for a 2DOF parametric oscillators with different masses. 

\begin{figure}
    \centering
    \begin{subfigure}{0.49\textwidth}
        \includegraphics[width=\linewidth]{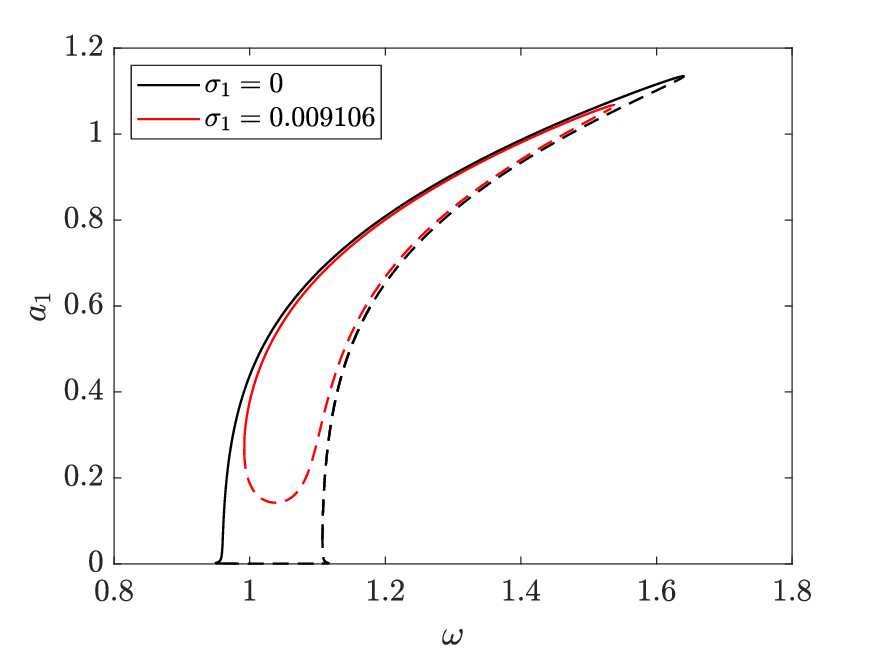}
        \caption{ }
        \label{fig:diffent_mass_analytical}
    \end{subfigure}
    \begin{subfigure}{0.49\textwidth}
        \includegraphics[width=\linewidth]{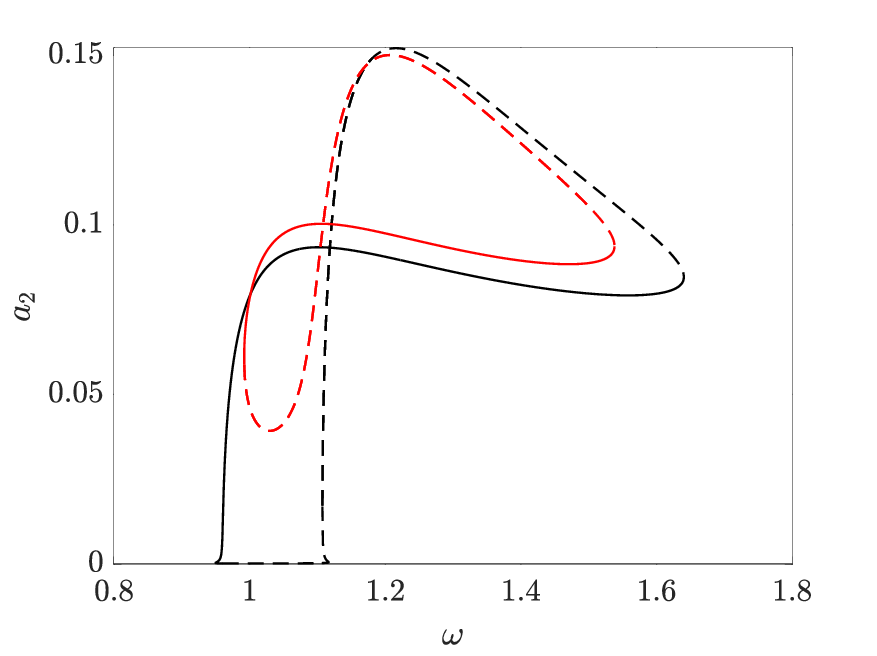}
        \caption{ }
        \label{fig:diff_mass_analytical_2}
    \end{subfigure}
\caption{Comparing the analytical results from CxA for the first mass (a) and the second mass (b) in the case of $m_1\neq m_2$. }
    \label{fig:diff_mass_analytical}
\end{figure}

\begin{figure}
    \centering
    \begin{subfigure}{0.49\textwidth}
        \includegraphics[width=\linewidth]{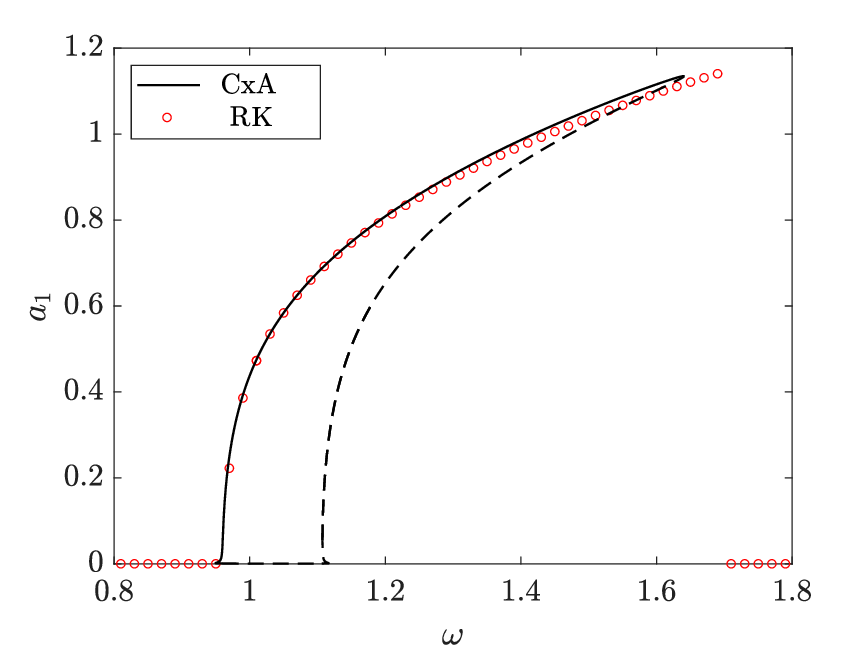}
        \caption{ }
        \label{fig:diff_mass_compare_1_sigma_0}
    \end{subfigure}
    \begin{subfigure}{0.49\textwidth}
        \includegraphics[width=\linewidth]{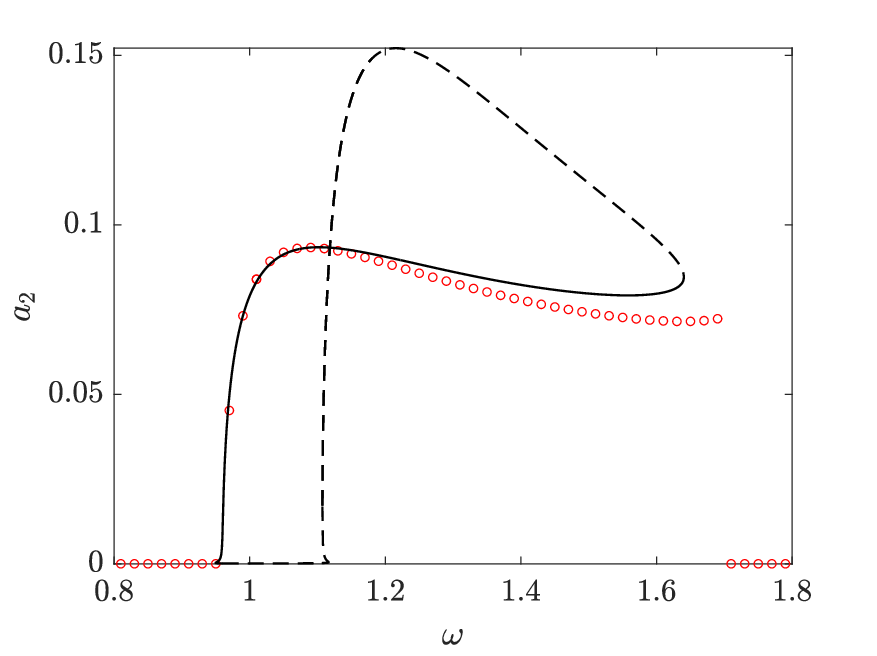}
        \caption{ }
        \label{fig:diff_mass_compare_2_sigma_0}
    \end{subfigure}
\caption{Comparing the analytical results with a Runge-Kutta sweep simulation for the first mass (a) and the second mass (b) where $\sigma_1=0$ and $m_1\neq m_2$. }
    \label{fig:diff_mass_compare_sigma_0}
\end{figure}

\begin{figure}
    \centering
    \begin{subfigure}{0.51\textwidth}
        \includegraphics[width=\linewidth]{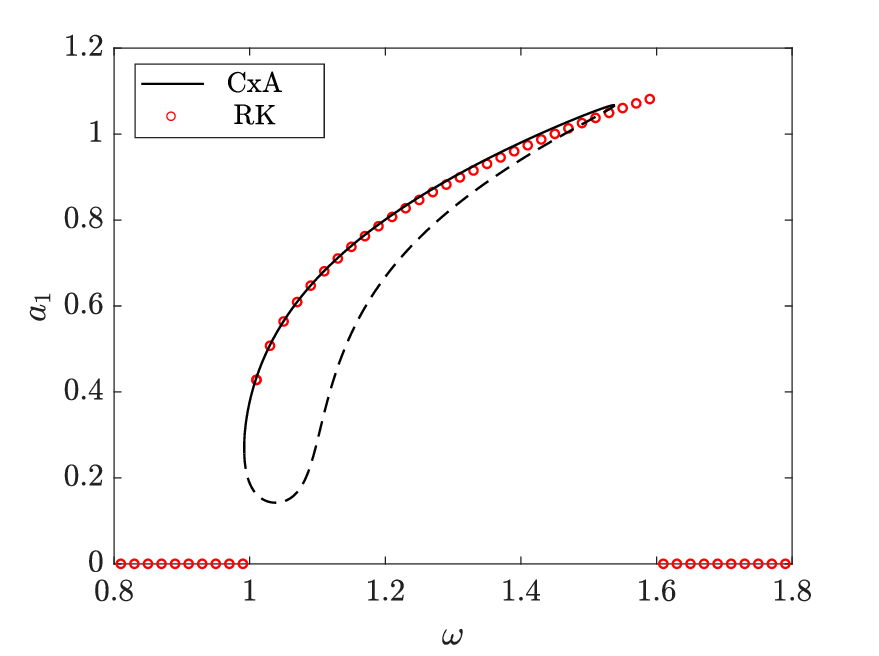}
        \caption{ }
        \label{fig:diff_mass_compare_1_sigma_09a}
    \end{subfigure}
    \begin{subfigure}{0.47\textwidth}
        \includegraphics[width=\linewidth]{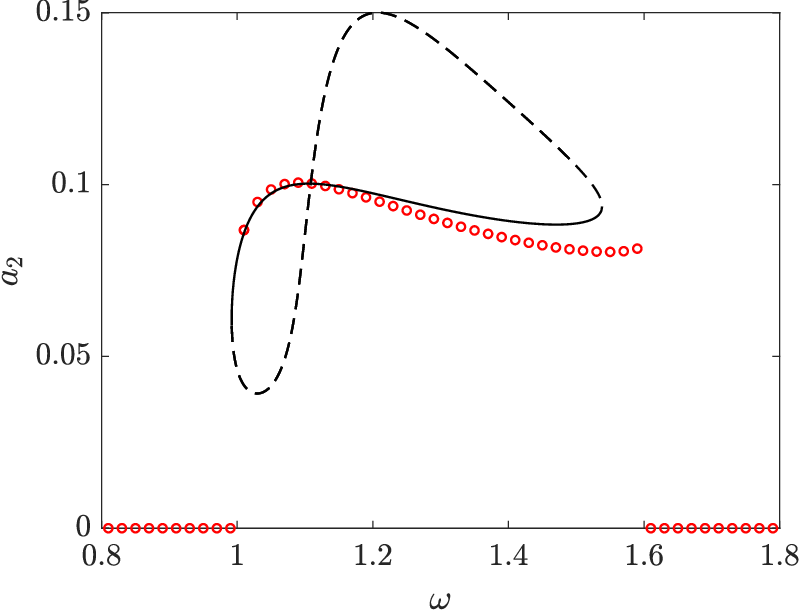}
        \caption{ }
        \label{fig:diff_mass_compare_2_sigma_09b}
    \end{subfigure}
\caption{Comparing the analytical results with a Runge-Kutta sweep simulation for the first mass (a) and the second mass (b) where $\sigma_1=0.009106$  and $m_1\neq m_2$.}
    \label{fig:diff_mass_compare_sigma_09}
\end{figure}

For the case of unequal masses, the parameter set A is held in Table \ref{tab2}. A continuation technique \cite{cont1} is used to solve Eqs. (\ref{ca7a}-\ref{ca7d}) numerically. The results are plotted in Figs. \ref{fig:diff_mass_analytical} in the case of friction and no friction ($\sigma_1=0$). The solid curves represent the stable solutions, while the dashed curves indicate unstable solutions, and the stability is computed from the Jacobian of Eqs. (\ref{ca7a}-\ref{ca7d}). The amplitude curve of the first mass $a_1$ is similar to the 1DOF case investigated in \cite{p5}. The second mass shows a fold where the amplitude saturates just below $a_2=0.1$. To compare the validity, the EOMs  Eqs. (\ref{2a}-\ref{2b}) are simulations with a Runga-Kutta scheme in Figs. \ref{fig:diff_mass_compare_sigma_0} and  Figs. \ref{fig:diff_mass_compare_sigma_09} for the cases without and with friction. The excitation $\omega$ in the RK-scheme is increased stepwise from $0.8$ to $1.8$. The RMS of the displacement multiplied by $\sqrt2$ after steady-state is recorded at each frequency step. It can be seen that the full EOMs closely follow the analytical curve, especially near $\omega=1$. For $\omega$ near the jump, there is a slight deviation.

\begin{figure}
    \centering
    \begin{subfigure}{0.8\textwidth}
        \centering
        \includegraphics[width=\textwidth]{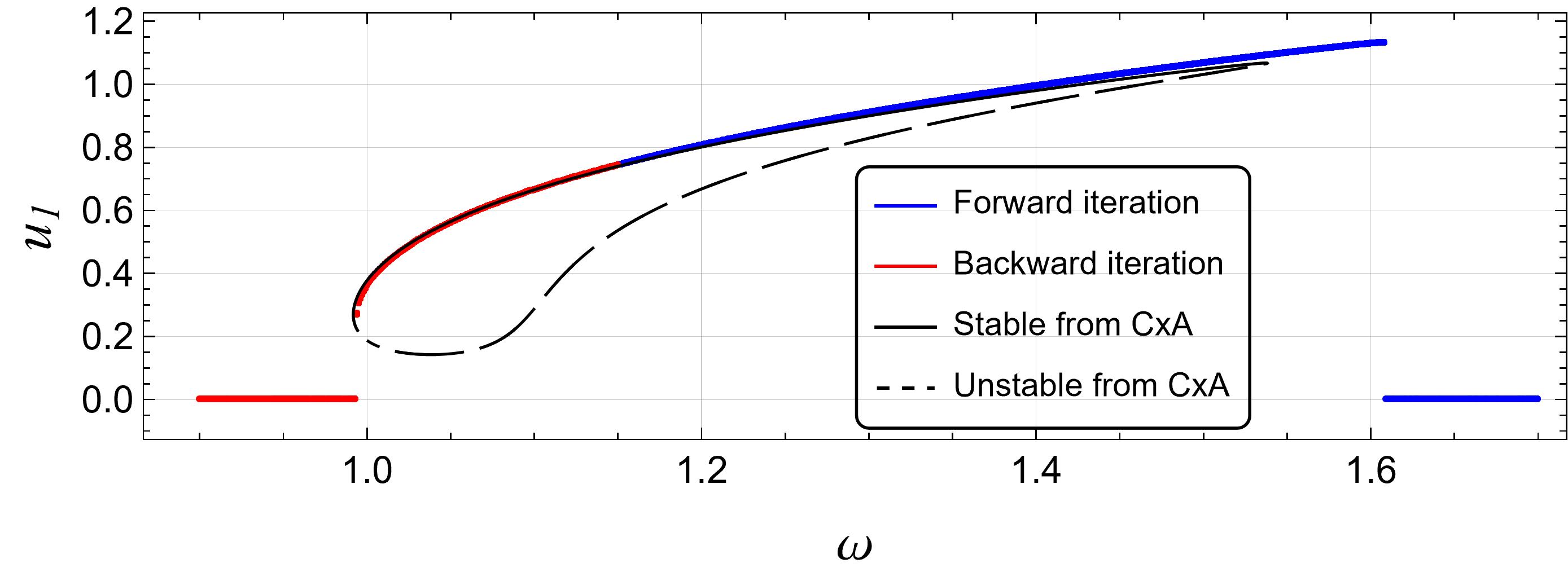}
        \caption{}
        \label{bif4a}
    \end{subfigure}
    \vspace{0.5cm}
    \begin{subfigure}{0.8\textwidth}
        \centering
        \includegraphics[width=\textwidth]{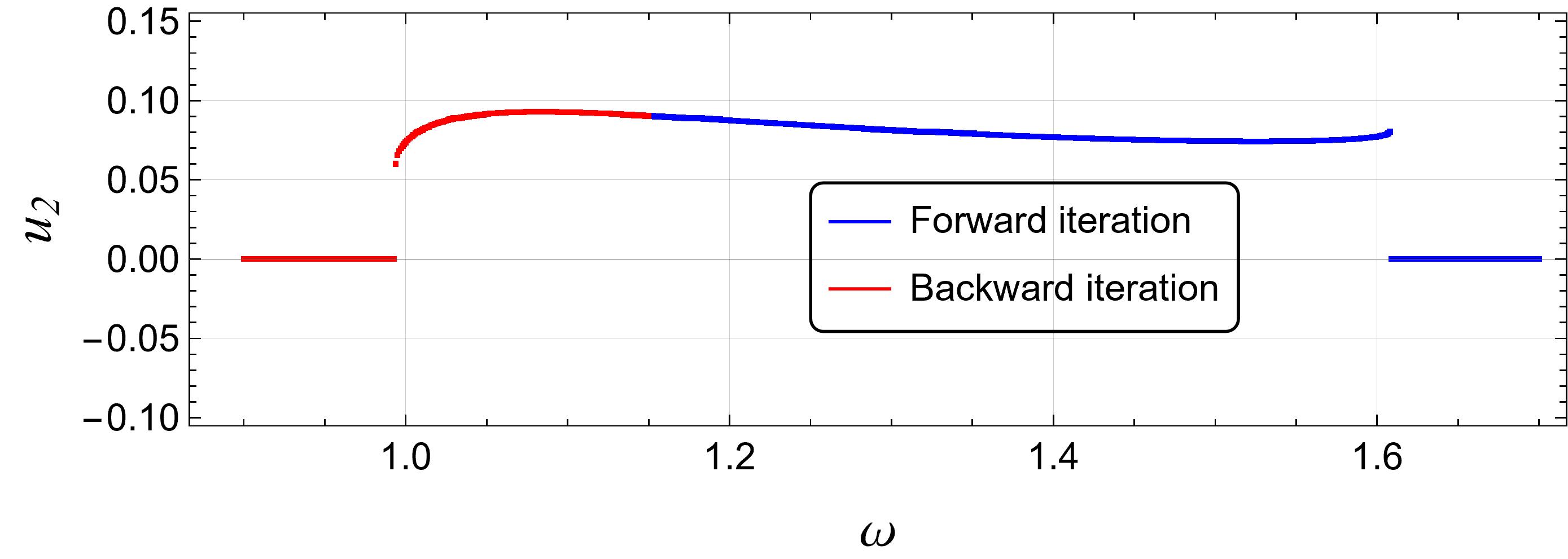}
        \caption{}
        \label{bif4b}
    \end{subfigure}
    \vspace{0.5cm}
    \begin{subfigure}{0.8\textwidth}
        \centering
        \includegraphics[width=\textwidth]{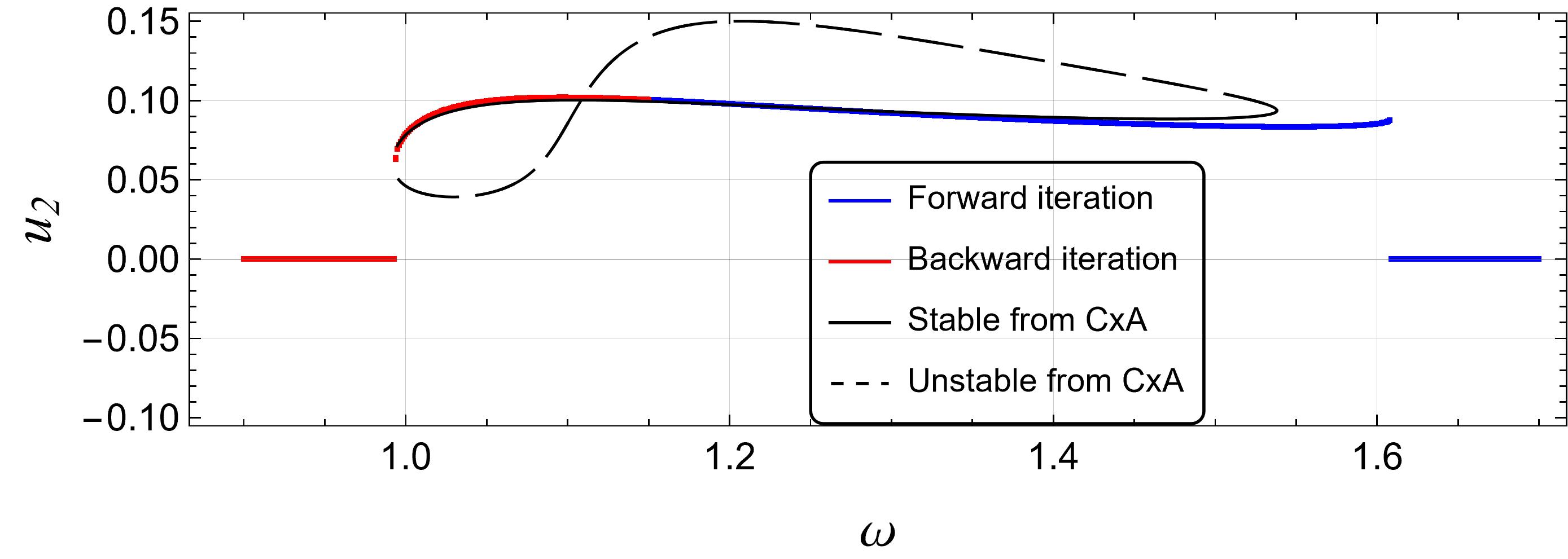}
        \caption{}
        \label{bif4c}
    \end{subfigure}
     \vspace{0.5cm}
    \begin{subfigure}{0.8\textwidth}
        \centering
        \includegraphics[width=\textwidth]{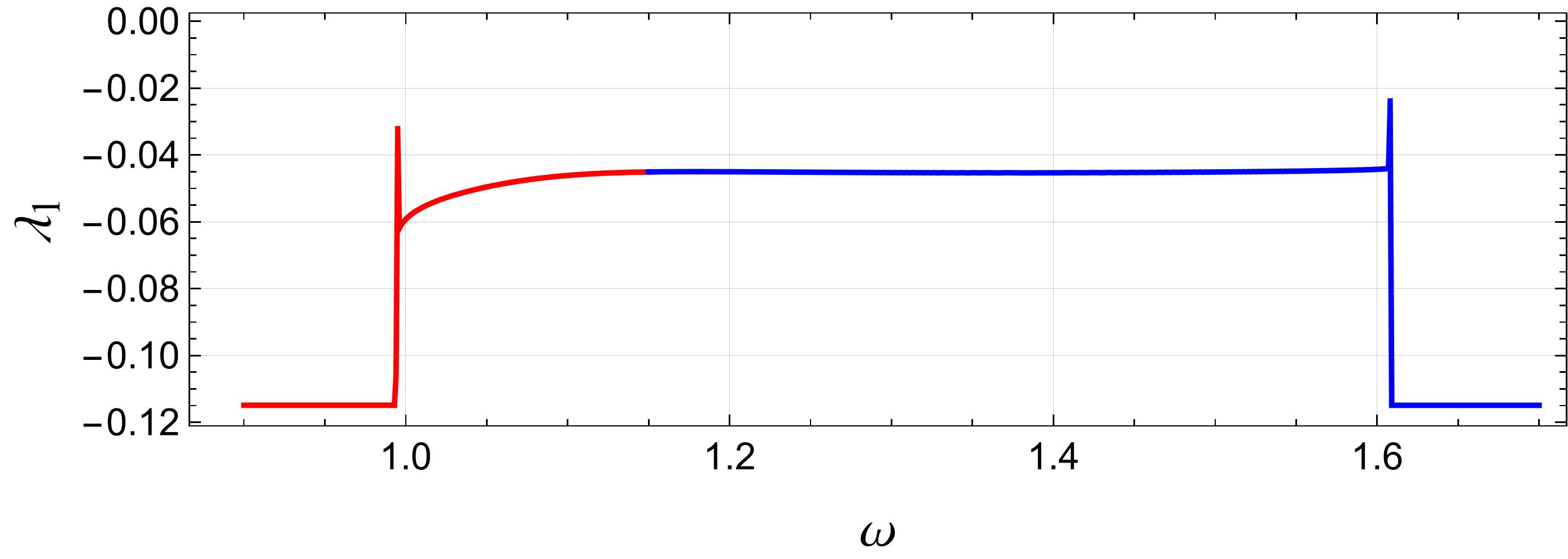}
        \caption{}
        \label{bif4d}
    \end{subfigure}
    \caption{Illustrating the bifurcation plot of the local maxima of $u_1$ for coordinates $u_1$ (a) and $u_2$ (b), the local maxima of $u_2(\tau)$ (c), and the largest Lyapunov exponent $\lambda_1$ (d) as the excitation frequency $\omega$ ranges from $1.15$ to $0.90$ (for backward) and $1.15$ to $1.70$ (for forward) and analytical results obtained from the CxA method: A comparison analysis.
    }
     \label{bif4}
\end{figure}

Fig. \ref{bif4} displays bifurcation diagrams obtained by using a Poincaré map that shows the dynamic behavior of a parametrically excited two-degree-of-freedom (2-DOF) system exhibiting fifth-degree stiffness nonlinearity. The previous three Figs. \ref{fig:diff_mass_analytical}, \ref{fig:diff_mass_compare_sigma_0}, and \ref{fig:diff_mass_compare_sigma_09} be also seen as bifurcation diagrams obtained from other methods, and we have compared them in Fig. \ref{bif4}. The excitation frequency $\omega$ is utilized as the control parameter to examine forward and backward sweeps, with outcomes presented for three observables. We have compared the approximate analytical results obtained by the CxA method with the numerical results in Figs. \ref{bif4a} and \ref{bif4c}, where we observed a good agreement between analytical and numerical results.

Figs. \ref{bif4a} and \ref{bif4b} illustrate the bifurcation diagrams of the local maxima for the primary and secondary coordinates, $u_1$ and $u_2$, respectively, along with the bifurcation diagram of the largest Lyapunov exponent $\lambda_1$ \ref{bif4d} with the bifurcation parameter $\omega$. Lyapunov exponents are used to quantify a system’s sensitivity to initial conditions. They help distinguish between periodic, quasi-periodic, and chaotic behavior. This makes them essential for analyzing the long-term stability of nonlinear systems. The maxima were derived from the solutions by Poincaré sampling at each excitation period, with maxima identified when the corresponding velocities intersect zero with a negative slope. Fig. \ref{bif4c} presents an auxiliary diagram for $u_2$, obtained when the velocity of the second oscillator is zero.

 The numerical continuation was executed utilizing both forward (blue) and backward (red) frequency sweeps. During the forward sweep, $\omega$ was adjusted from 1.15 to 1.70, whereas in the backward sweep, it ranged from 1.15 to 0.90. The initial conditions were established as \(u_1=0.55\), \(u'_1=0.55\), \(u_2=1.1\), and \(u'_2=-1.1\).

 The bifurcation diagrams indicate a stable trivial solution for low values of $\omega$, succeeded by the onset of large-amplitude periodic oscillations with increasing frequency.  The pronounced reliance on the frequency sweep direction underscores the multistable characteristics of the system's response. The seamless forward advancement contrasts with the delayed bifurcation during the backward sweep, indicating the existence of fold bifurcations and the potential separation of basins of attraction.
 
\begin{figure}[H]
    \centering
    \begin{subfigure}{0.49\textwidth}
        \includegraphics[width=\linewidth]{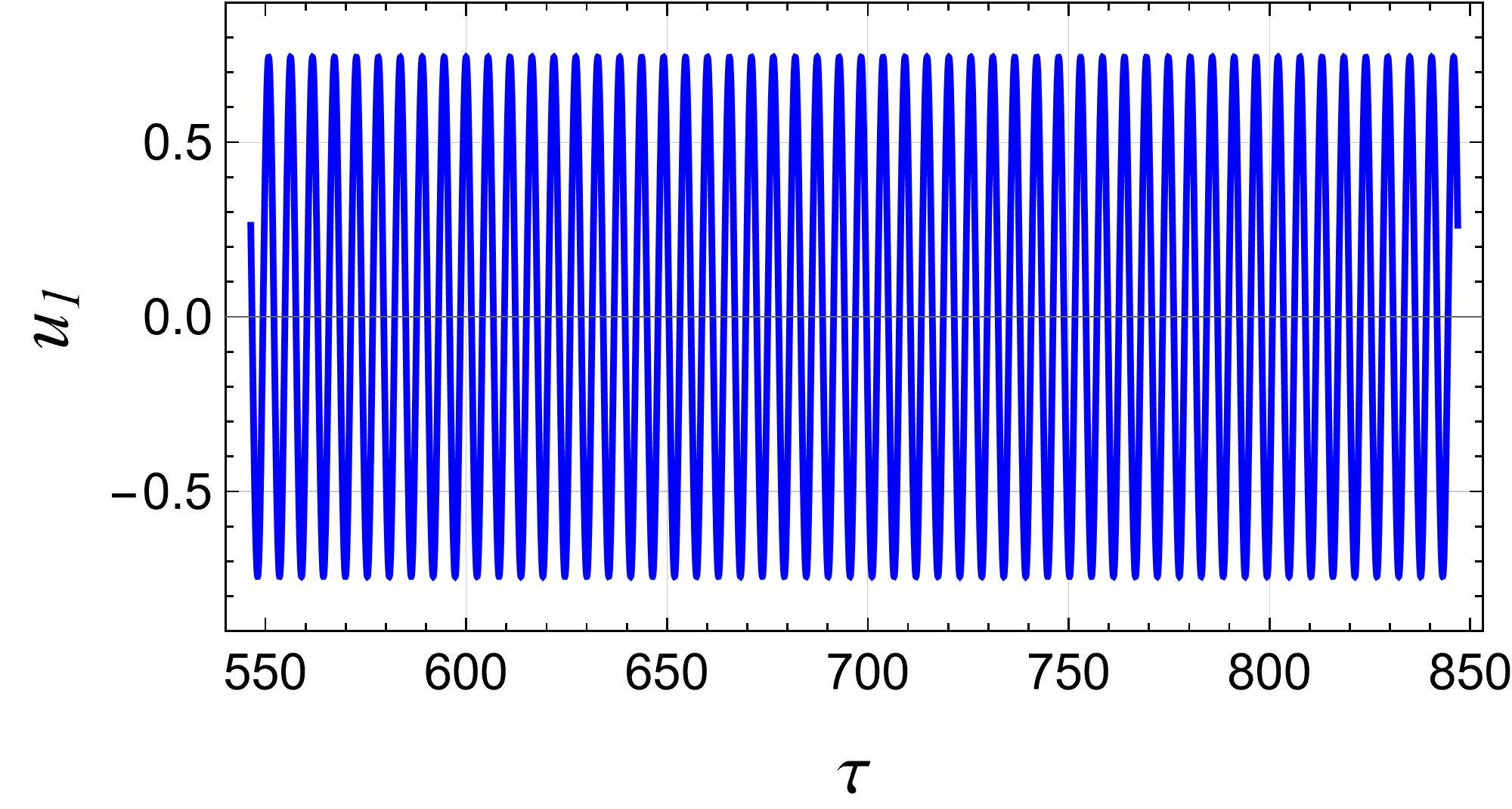}
        \caption{}
        \label{fig12aa}
    \end{subfigure}
    \begin{subfigure}{0.49\textwidth}
        \includegraphics[width=\linewidth]{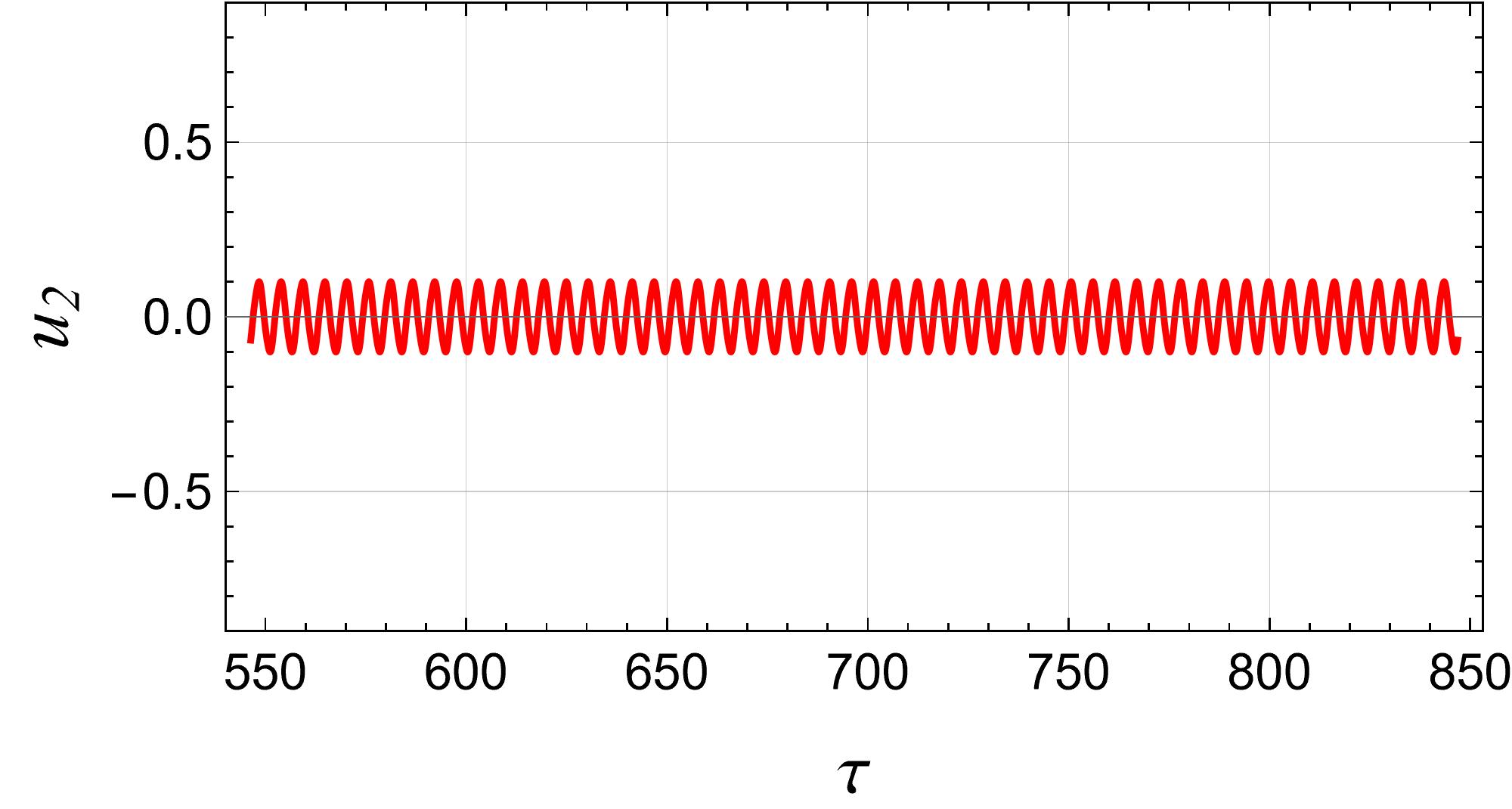}
        \caption{}
        \label{fig12bb}
    \end{subfigure}
    \medskip 
    \begin{subfigure}{0.32\textwidth}
        \includegraphics[width=\linewidth]{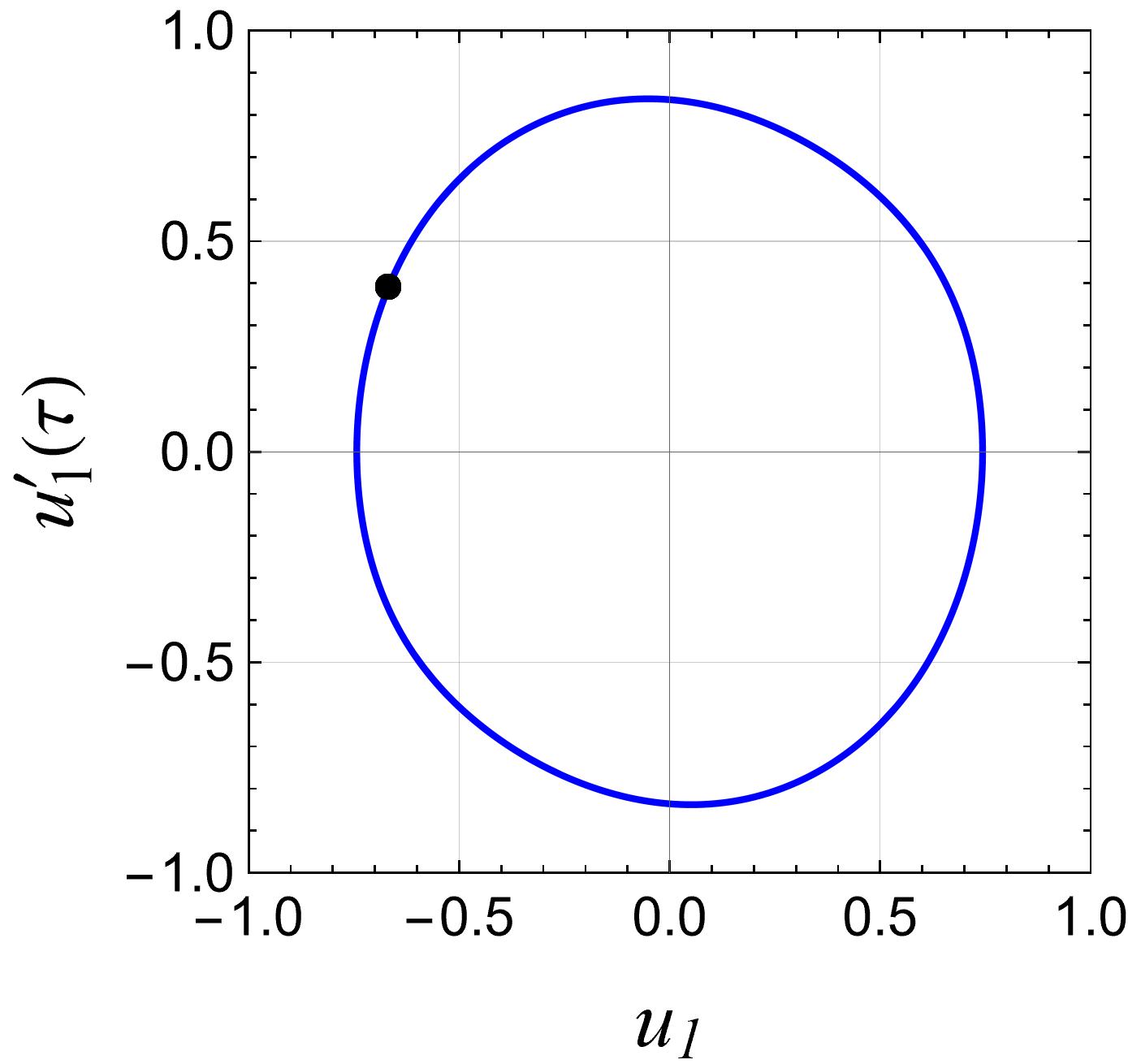}
        \caption{}
        \label{fig12cc}
    \end{subfigure}
    \begin{subfigure}{0.32\textwidth}
        \includegraphics[width=\linewidth]{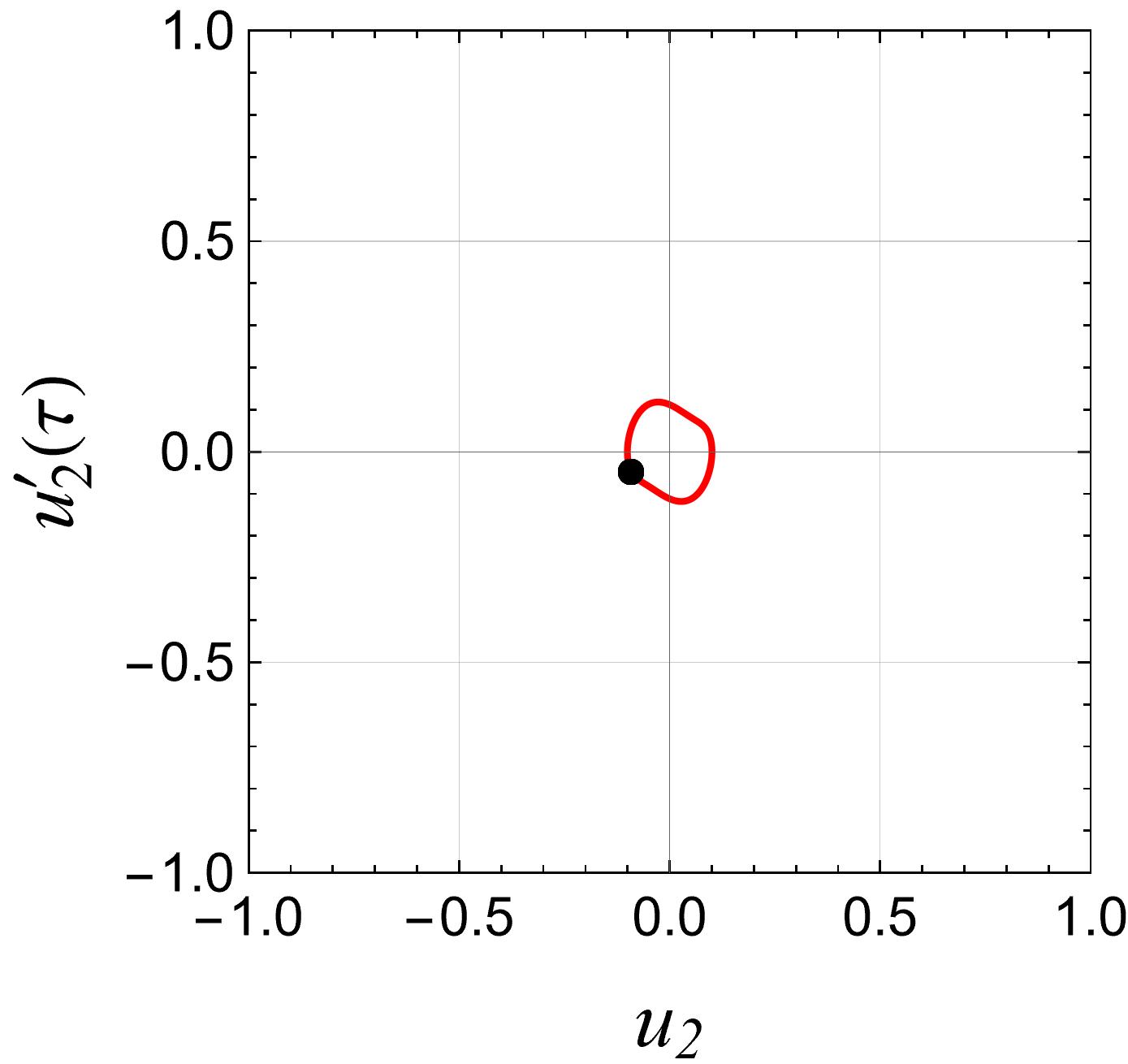}
        \caption{}
        \label{fig12dd}
    \end{subfigure}
     \begin{subfigure}{0.32\textwidth}
        \includegraphics[width=\linewidth]{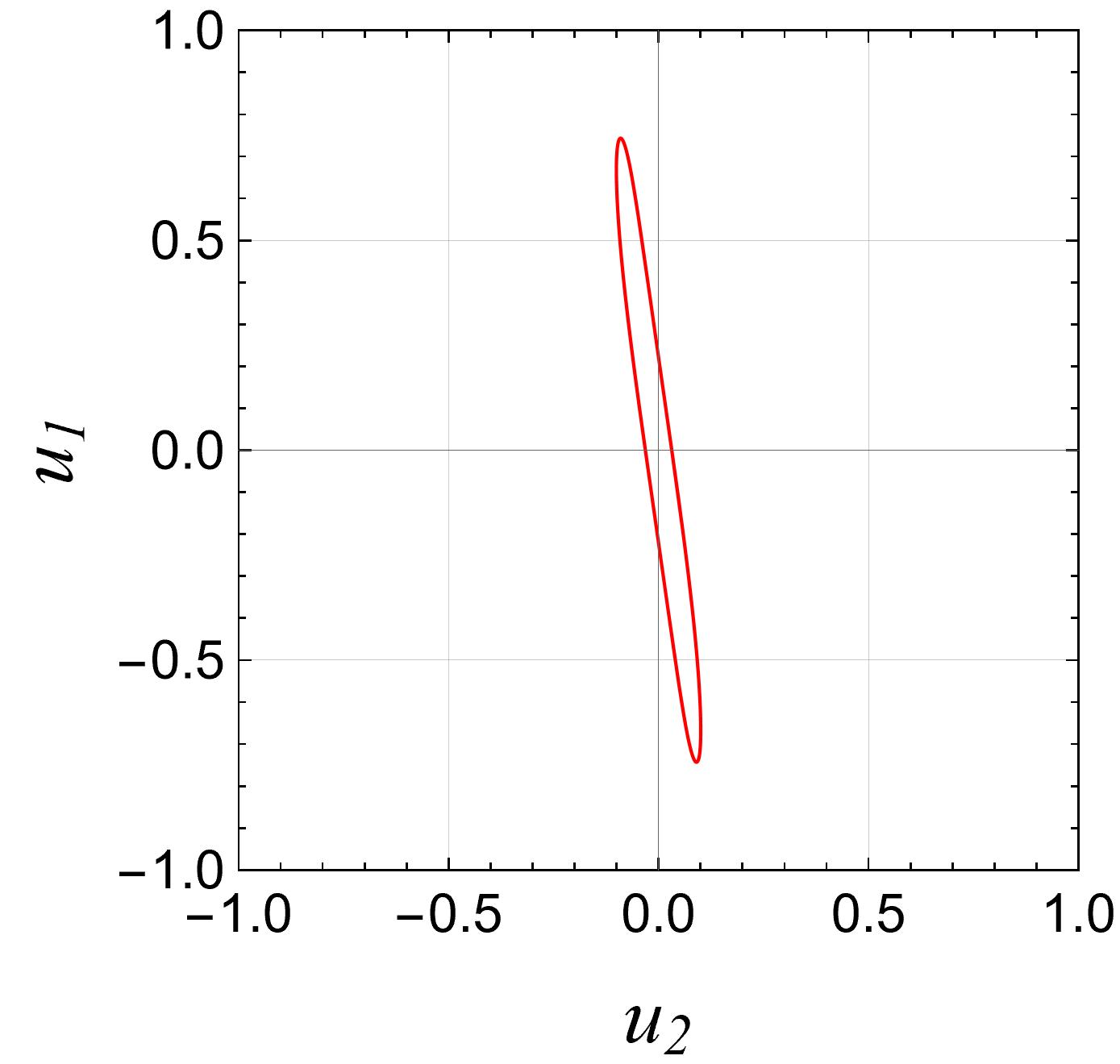}
        \caption{}
        \label{fig12ee}
    \end{subfigure}
\caption{The time-domain plots (a–b) show steady-state oscillations for \( u_1 \) and \( u_2 \) after an initial transient phase of 100 periods, \( \omega = 1.15 \), 
The phase-space plots and single points in Poincaré maps (c) and (d) display closed-loop trajectories, indicating periodic, stable motion in antiphase mode. 
}
    \label{fig121}
\end{figure}

\begin{figure}[H]
    \centering
    \begin{subfigure}{0.8\textwidth}
        \centering
        \includegraphics[width=\textwidth]{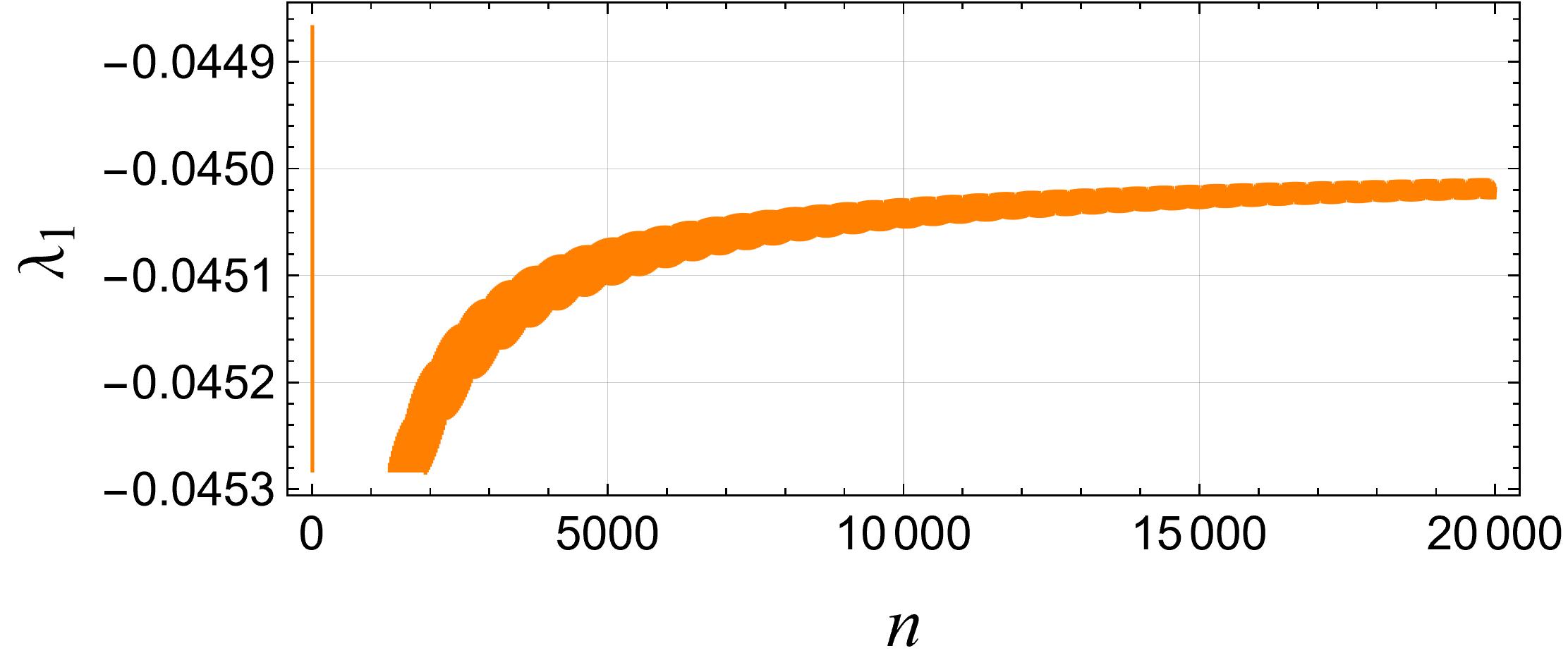}
    \end{subfigure}
    \caption{The maximal Lyapunov exponent $\lambda_1$ as a function of the number of external forcing periods $n$ for \( \omega = 1.15 \).}
     \label{larg1}
\end{figure}

Fig. \ref{fig121} illustrates the steady-state dynamics of the 2DOF system under parametric excitation at a driving frequency of $\omega = 1.15$, with initial conditions are \(u_1=0.55\), \(u'_1=0.22\), \(u_2=1.1\), and \(u'_2=-1.1\).  Following the elimination of transients exceeding 100 excitation cycles, the system dynamics were examined during the next 50 periods.
 Figs. \ref{fig12aa} and \ref{fig12bb} illustrate the time-domain responses of $u_1$ and $u_2$, respectively, demonstrating continuous periodic oscillations characterized by varying amplitudes and frequency components.  The steady-state trajectories indicate that both oscillators are synchronized to the external excitation frequency, albeit with amplitude and phase discrepancies due to modal coupling and stiffness asymmetry.
The associated phase portraits are illustrated in Figs. \ref{fig12cc} and \ref{fig12dd} for the $(u_1, \dot{u}_1)$ and $(u_2, \dot{u}_2)$ planes, respectively. Both trajectories create closed loops, validating the regularity noted in the temporal histories and also illustrating the associated Poincaré maps, sampled with velocity components $\dot{u}_1$ and $\dot{u}_2$ intersecting zero in the negative direction. The maps in Figs. \ref{fig12cc} and \ref{fig12dd}, represented in $(u_1, \dot{u}_1)$ and $(u_2, \dot{u}_2)$ coordinates, exhibit distinct singular points, signifying a phase-locked periodic orbit. Fig. \ref{fig12ee} further depicts the projection in the $(u_2, u_1)$ plane, demonstrating a robust correlation while maintaining a non-identical coupling between the two coordinates. The periodic nature of the solution is validated by a positive greatest Lyapunov exponent $\lambda_1$, with the method of computation illustrated in Fig. \ref{larg1}, and the total number of periods of external forcing is denoted by $n$ (on the x-axis).

\subsection{Bifurcation analysis for same masses (\texorpdfstring{$m_1=m_2$}{m₁= m₂})}\label{same_masses}
This subsection explains the results of the bifurcation analysis obtained analytically and numerically for a 2DOF parametric oscillators with the same masses. 

\begin{figure}[H]
    \centering
    \begin{subfigure}{0.49\textwidth}
        \includegraphics[width=\linewidth]{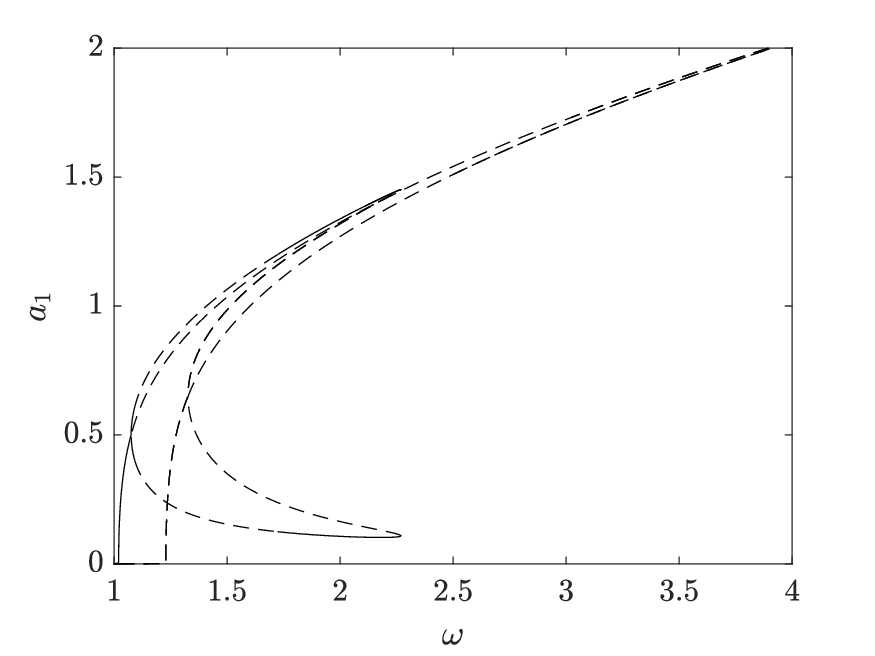}
        \caption{ }
        \label{fig:same_mass_1_analytical}
    \end{subfigure}
    \begin{subfigure}{0.49\textwidth}
        \includegraphics[width=\linewidth]{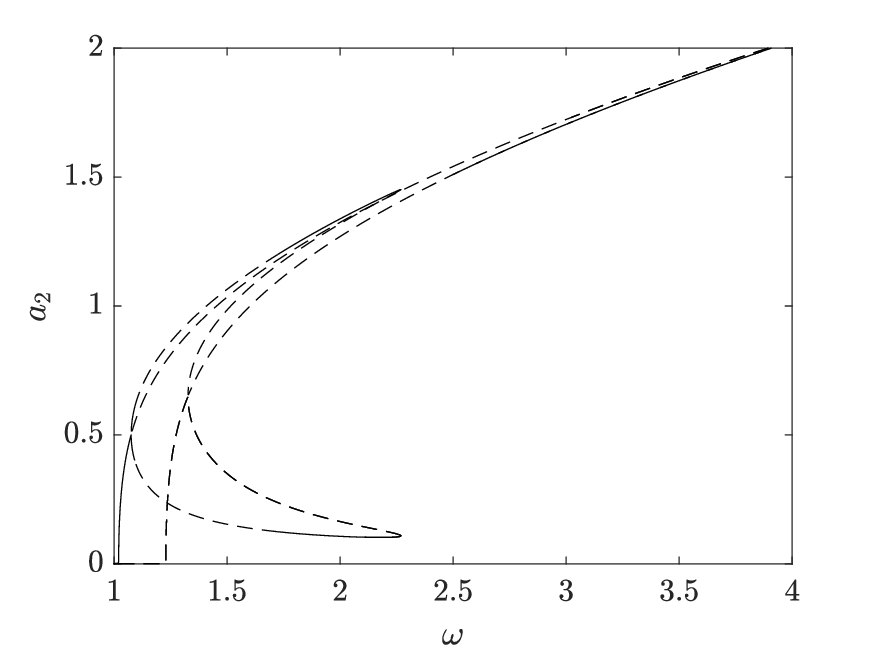}
        \caption{ }
        \label{fig:same_mass_2_analytical}
    \end{subfigure}
        \begin{subfigure}{0.49\textwidth}
        \includegraphics[width=\linewidth]{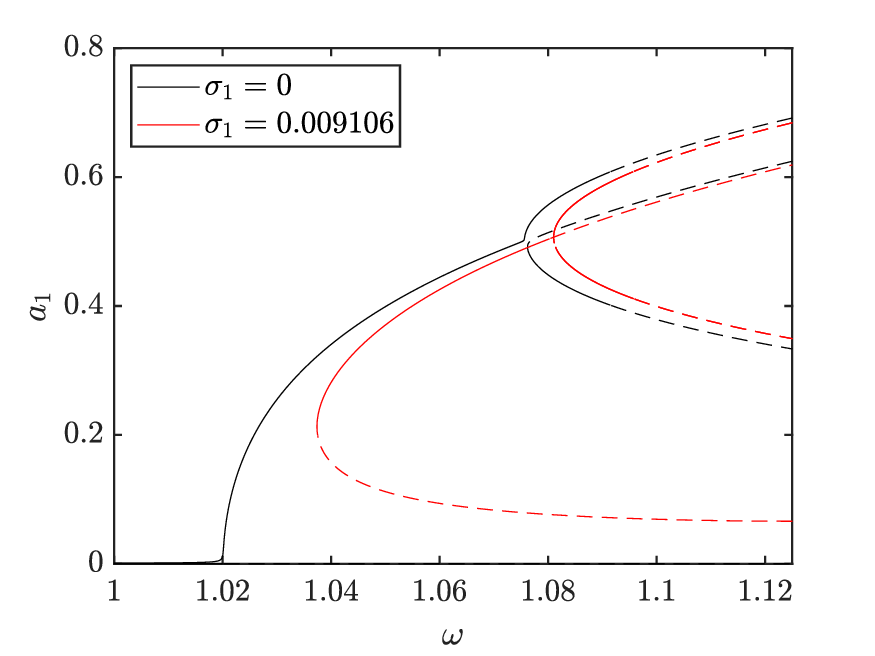}
        \caption{ }
        \label{fig:same_mass_1_zoom_analytical}
    \end{subfigure}
           \begin{subfigure}{0.49\textwidth}
        \includegraphics[width=\linewidth]{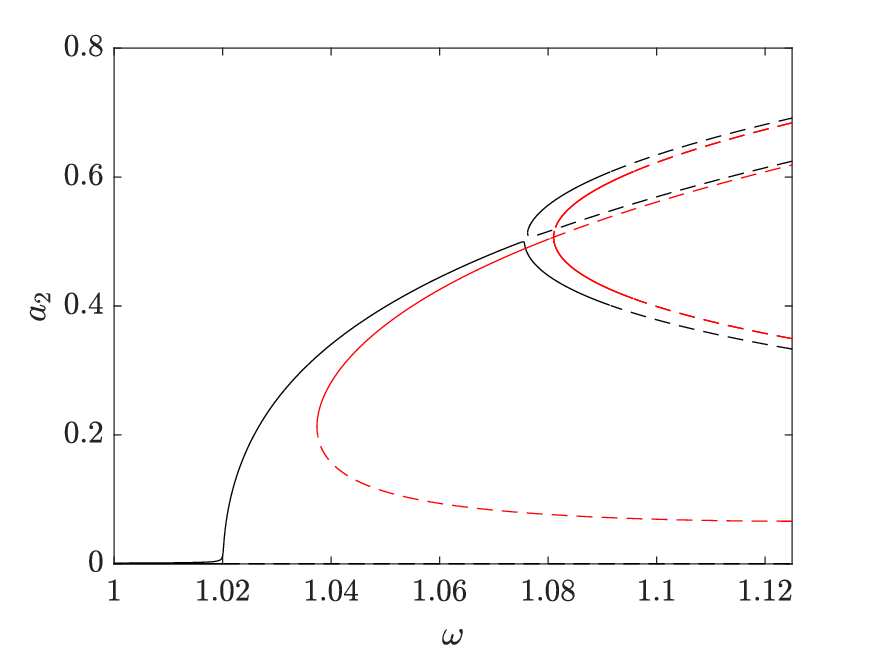}
        \caption{ }
        \label{fig:same_mass_2_zoom_analytical}
    \end{subfigure}
\caption{Comparison of the analytical results (\ref{ca7}) from the CxA method with and without dry friction: (a,b) represents the results without dry friction, and (c,d) represents the behavior of the system with dry friction for $\sigma_1=0.009106$ and without dry friction for $\sigma_1=0$, and $m_1=m_2$. }
    \label{fig:same_mass_analytical}
\end{figure}

\begin{figure}
    \centering
    \begin{subfigure}{0.49\textwidth}
        \includegraphics[width=\linewidth]{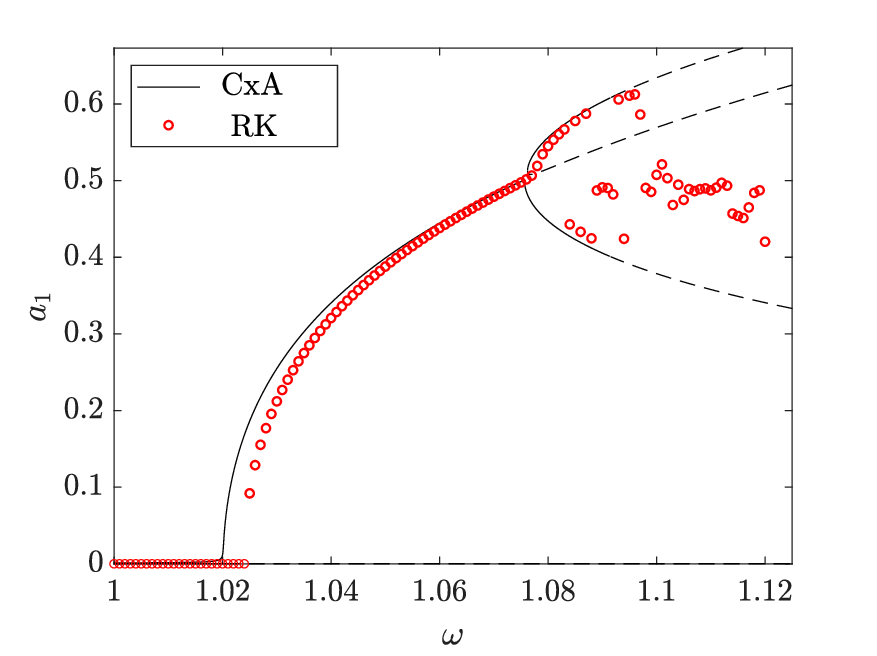}
        \caption{ }
        \label{fig:same_mass_1_compare_sigma_0}
    \end{subfigure}
    \begin{subfigure}{0.49\textwidth}
        \includegraphics[width=\linewidth]{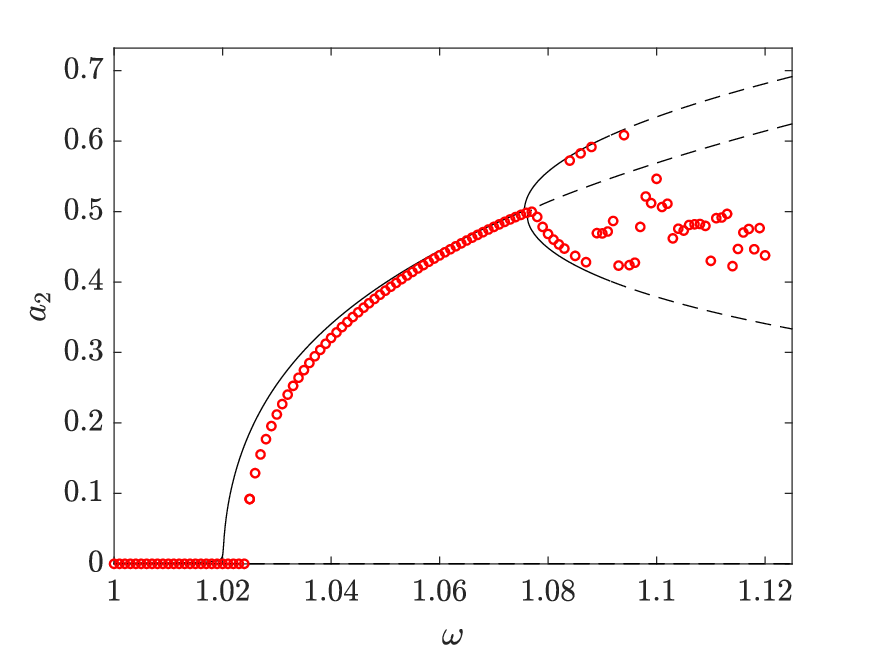}
        \caption{ }
        \label{fig:same_mass_2_compare_sigma_0}
    \end{subfigure}
\caption{ Comparison analysis of analytical results from CxA with the RMS of the RK simulation for $m_1$ (a) and for $m_2$ (b) without dry friction $\sigma_1=\sigma_2=0$, in the case of the same masses, $m_1=m_2$.}
    \label{fig:same_mass_compare_sigma_0}
\end{figure}

\begin{figure}
    \centering
    \begin{subfigure}{0.49\textwidth}
        \includegraphics[width=\linewidth]{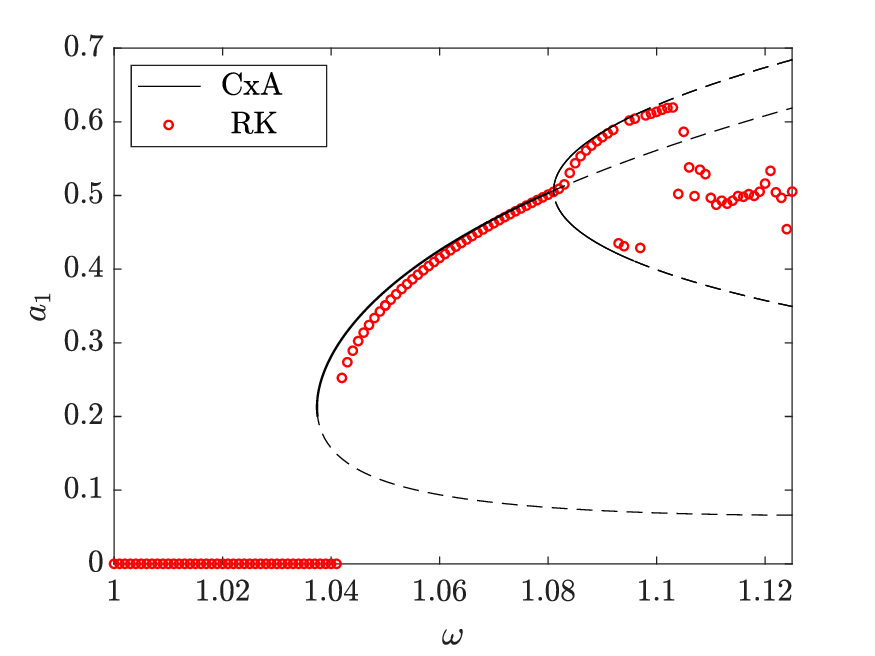}
        \caption{ }
        \label{fig:same_mass_1_compare_sigma_09}
    \end{subfigure}
    \begin{subfigure}{0.49\textwidth}
        \includegraphics[width=\linewidth]{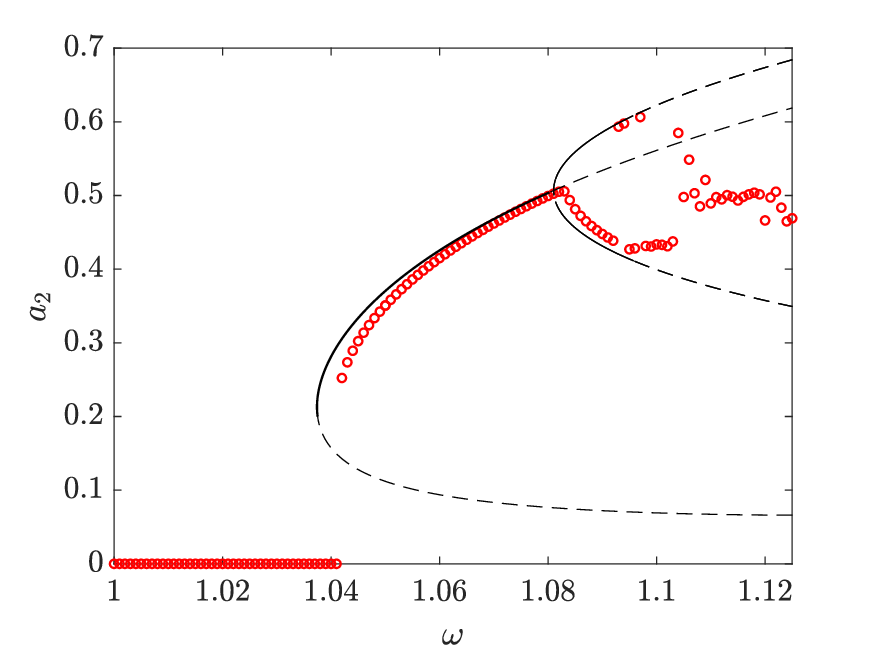}
        \caption{ }
        \label{fig:same_mass_2_compare_sigma_09}
    \end{subfigure}
\caption{Comparison analysis of analytical results from CxA with the RMS of the RK simulation for $m_1$ (a) and for $m_2$ (b) with dry friction $\sigma_1=0.009106$, and $\sigma_2=\mu \sigma_1$, in the case of the same masses, $m_1=m_2$. }
    \label{fig:same_mass_compare_sigma_09}
\end{figure}

\begin{figure}[H]
    \centering
    \begin{subfigure}{0.8\textwidth}
        \centering
        \includegraphics[width=\textwidth]{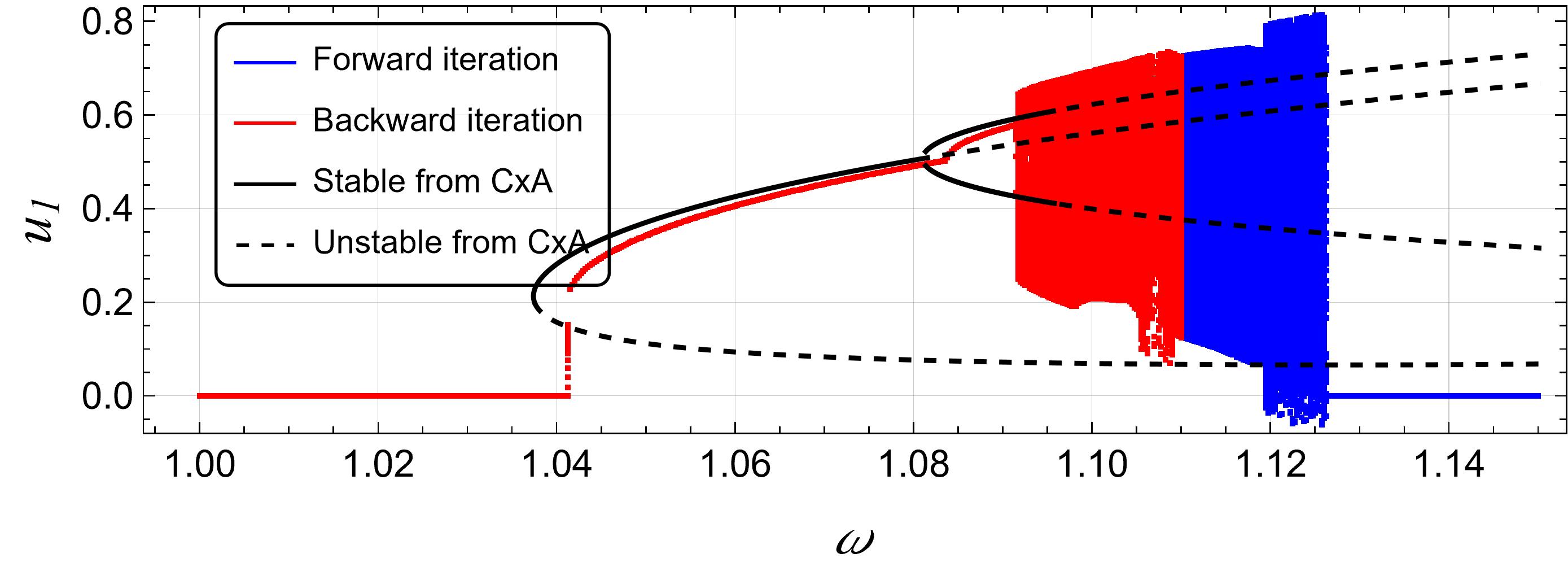}
        \caption{}
        \label{bif2a}
    \end{subfigure}
    \vspace{0.5cm}
    \begin{subfigure}{0.8\textwidth}
        \centering
        \includegraphics[width=\textwidth]{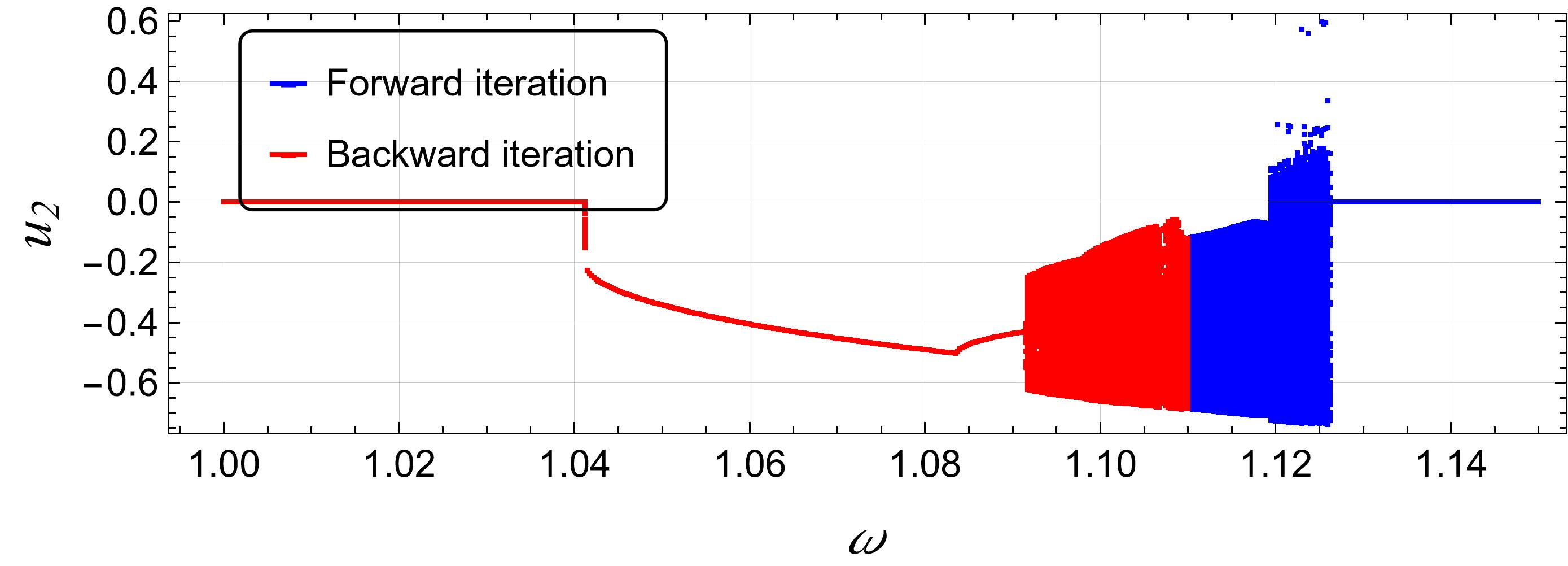}
        \caption{}
        \label{bif2b}
    \end{subfigure}
    \vspace{0.5cm}
    \begin{subfigure}{0.8\textwidth}
        \centering
        \includegraphics[width=\textwidth]{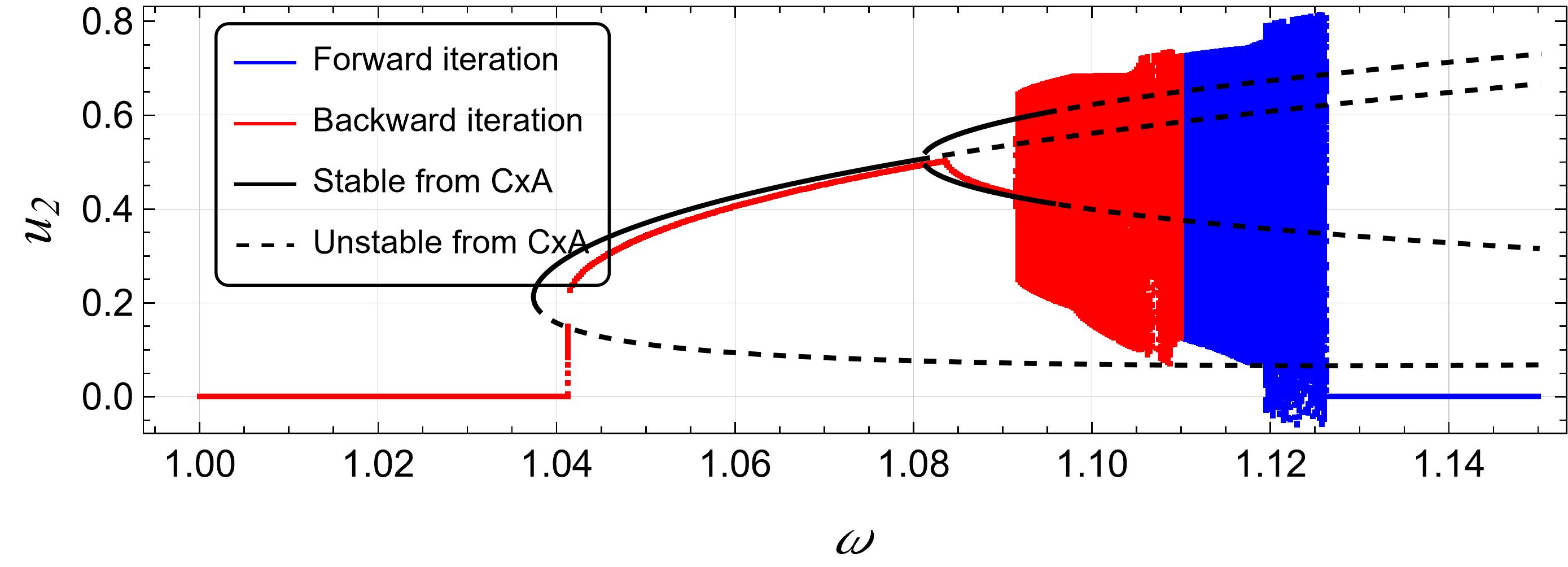}
        \caption{}
        \label{bif2c}
    \end{subfigure}
    \vspace{0.5cm}
    \begin{subfigure}{0.8\textwidth}
        \centering
        \includegraphics[width=\textwidth]{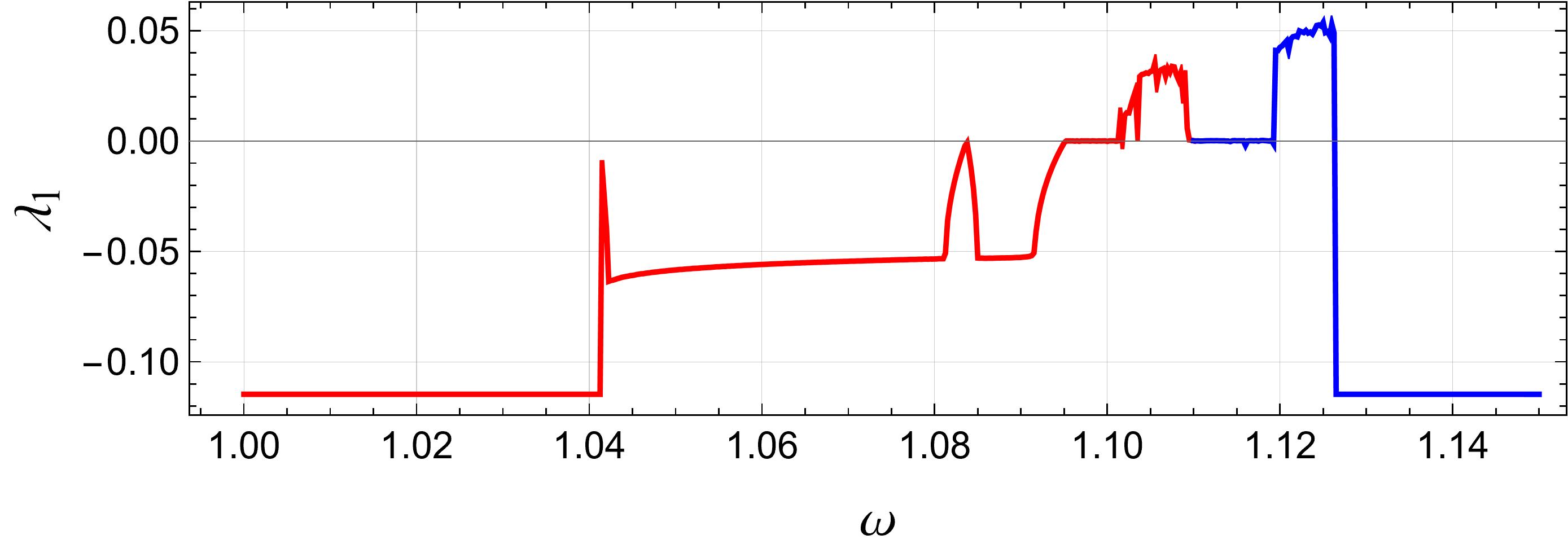}
        \caption{}
        \label{bif2d}
    \end{subfigure}
    \caption{Illustrating the bifurcation plot of the local maxima of $u_1$ for coordinates $u_1$ (a) and $u_2$ (b), the local maxima of $u_2(\tau)$ (c), and the largest Lyapunov exponent $\lambda_1$ (d) as the excitation frequency $\omega$ ranges from $1.11$ to $1.0$ (for backward) and $1.11$ to $1.15$ (for forward) and analytical results obtained from the CxA method: A comparison analysis.
    }
    \label{bif2}
\end{figure}

For the case of equal masses, the parameter set B holds in Table \ref{tab2}. The analytical results are plotted in Figs. \ref{fig:same_mass_analytical} in the case of friction and no friction ($\sigma_1=0$). Figs. \ref{fig:same_mass_1_analytical}  and \ref{fig:same_mass_2_analytical} show the same frequency response as the system is now symmetrical. Only the case without friction is shown in these figures, as they are close to the curves with friction. Compared to the case of unequal mass, there are solutions over a much wider frequency range; however, most of the branches are unstable except closer to $\omega=1$. When zoomed in, in Figs. \ref{fig:same_mass_1_zoom_analytical} and \ref{fig:same_mass_2_zoom_analytical}, it can be seen that $\omega=1.08$, the stable branch has a pitchfork bifurcation before becoming unstable. For the case with friction, the pitchfork shifts to the right. On the two stable branches of the pitchfork, the response becomes unstable starting from $\omega$=1.1. Observing the eigenvalues of the Jacobian, a Hopf bifurcation is observed where complex eigenvalues go from negative to positive real values. This hints at a quasi-periodic response in the time domain, where the amplitude will modulate.
\noindent The RMS of the RK simulation is plotted in Figs. \ref{fig:same_mass_compare_sigma_0} without friction and Fig. \ref{fig:same_mass_compare_sigma_09} with friction. For no friction, it can be seen that after the pitchfork bifurcation, the first mass tends to the upper branch Fig. \ref{fig:same_mass_1_compare_sigma_0} while the other mass tends to the lower branch. The same can be seen in the case of friction. Once both the upper and lower branches become unstable, the RMS does not follow a trend dictated by the branches. A chaotic response appears, which will be studied later.
Fig. \ref{bif2} displays bifurcation diagrams computed for $\omega$ as a control parameter. Diagrams depict the intricate nonlinear response of a parametrically excited 2DOF system featuring stiffness nonlinearity. The diagrams in Figs. \ref{bif2a} and \ref{bif2b} are created using Poincaré sections obtained at the local maxima of the primary coordinate $u_1$, identified when the velocity $\dot{u}_1$ passes zero with a negative slope, and the bifurcation diagram of the largest Lyapunov exponent $\lambda_1$ \ref{bif2d} with the bifurcation parameter $\omega$. Moreover, the diagram in Fig. \ref{bif2c} was computed at the local maxima of $u_2$, while its velocity was zero $(u_1'=0)$. The system's reaction is analyzed by both forward (blue) and backward (red) frequency sweeps throughout the range $\omega\in[1.0, 1.15]$, demonstrating the system's sensitivity to initial conditions. The initial conditions are \(u_1=0.99\), \(u'_1=0.88\), \(u_2=1.1\), \(u'_2=0\), and, emphasizing its multistable characteristics. In the low-frequency range of the control parameter, the system demonstrates a stable trivial solution; however, once it $\omega$ surpasses around $1.04$, a branch of periodic oscillations arises. Close to $\omega\approx1.09$, an abrupt change transpires, resulting in complex dynamics characterized by the presence of dense windows in the bifurcation diagrams. This regime signifies the emergence of quasi-periodicity and chaos.

Rich bifurcation dynamics are observed, encompassing periodic, quasiperiodic, and chaotic behaviors. Moreover, the bifurcation diagrams obtained from the Poincaré map align closely with the behavior indicated by the largest Lyapunov exponent. In particular, a positive exponent corresponds to chaotic motion, a zero value reflects quasiperiodic dynamics, and a negative exponent indicates the presence of stable periodic orbits.

Based on bifurcation diagrams from Fig. \ref{bif2}, selected cases of system motion were analyzed using time series, phase plots, and Poincaré maps.
\begin{figure}[H]
    \centering
    \begin{subfigure}{0.49\textwidth}
        \includegraphics[width=\linewidth]{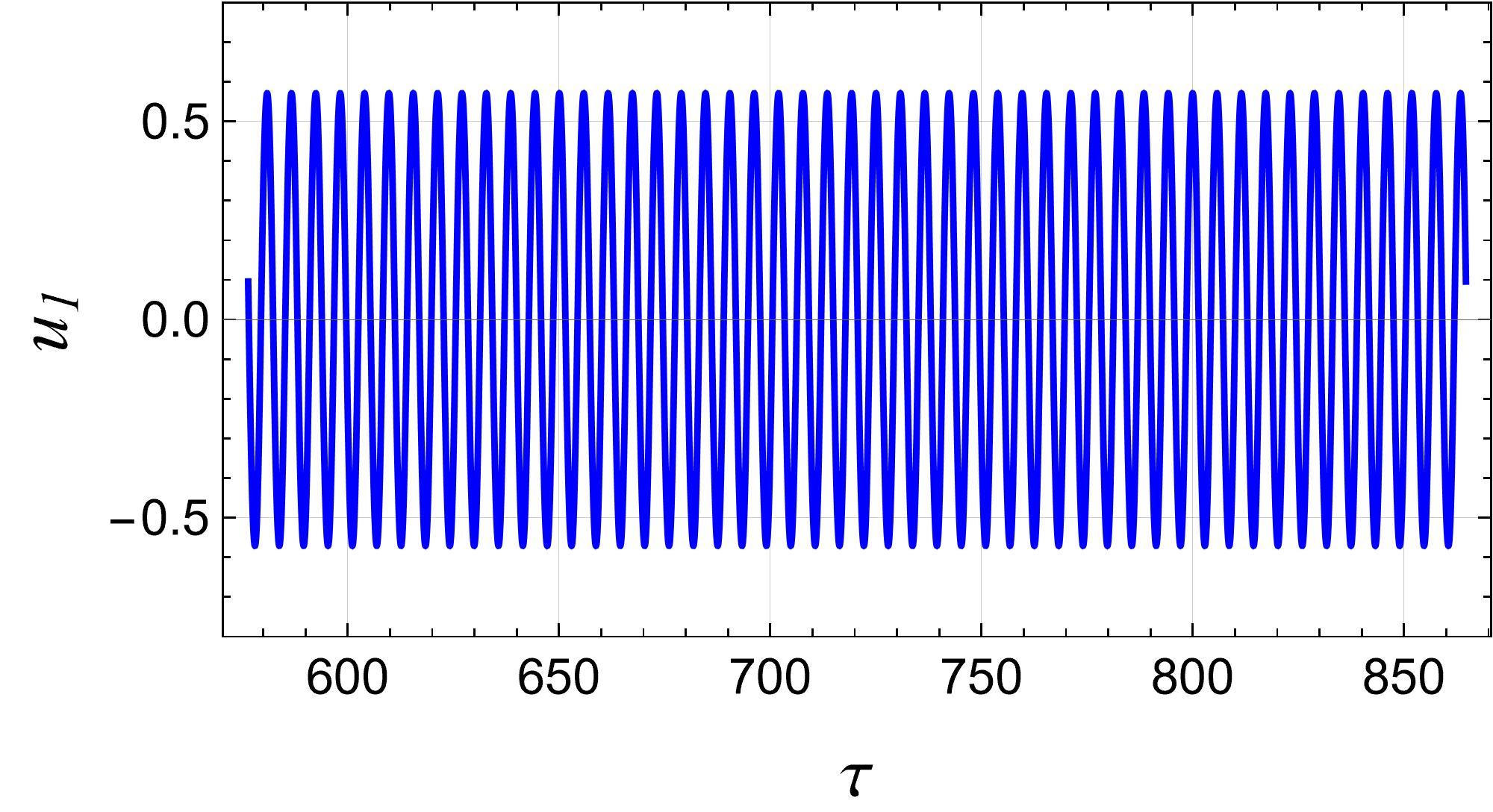}
        \caption{}
        \label{fig6a}
    \end{subfigure}
    \begin{subfigure}{0.49\textwidth}
        \includegraphics[width=\linewidth]{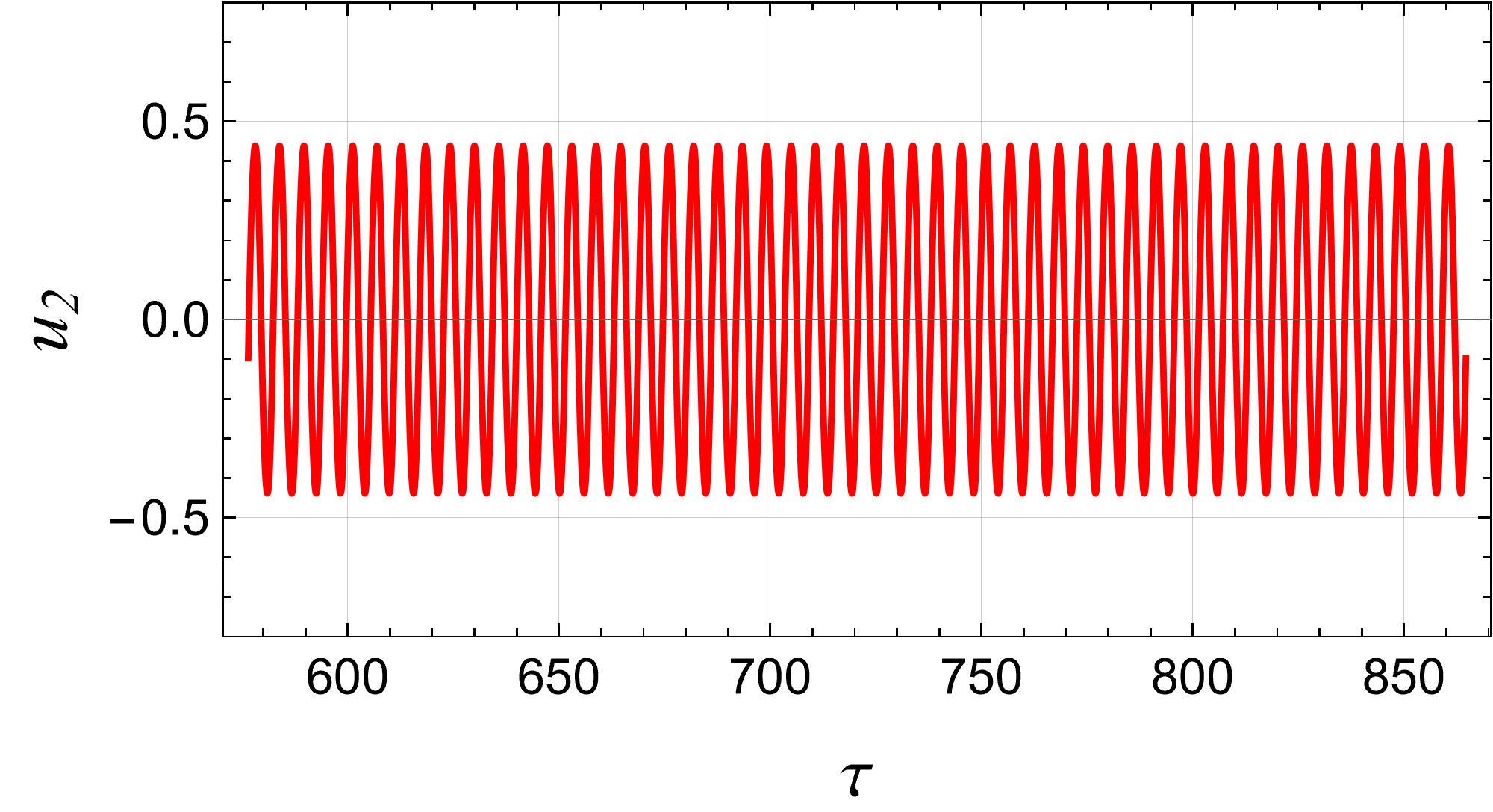}
        \caption{}
        \label{fig6b}
    \end{subfigure}
    \medskip 
    \begin{subfigure}{0.32\textwidth}
        \includegraphics[width=\linewidth]{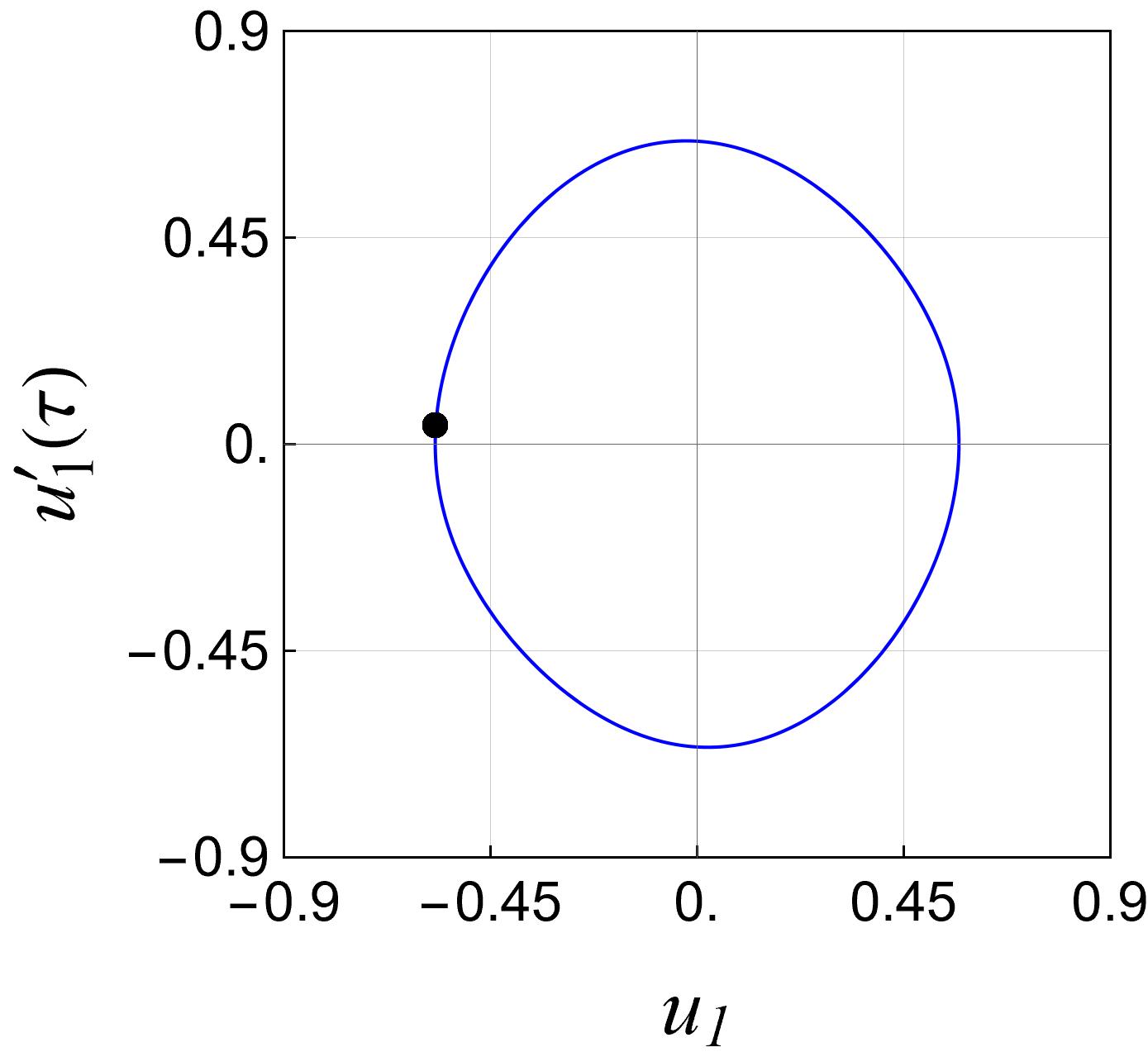}
        \caption{}
        \label{fig6c}
    \end{subfigure}
    \begin{subfigure}{0.32\textwidth}
        \includegraphics[width=\linewidth]{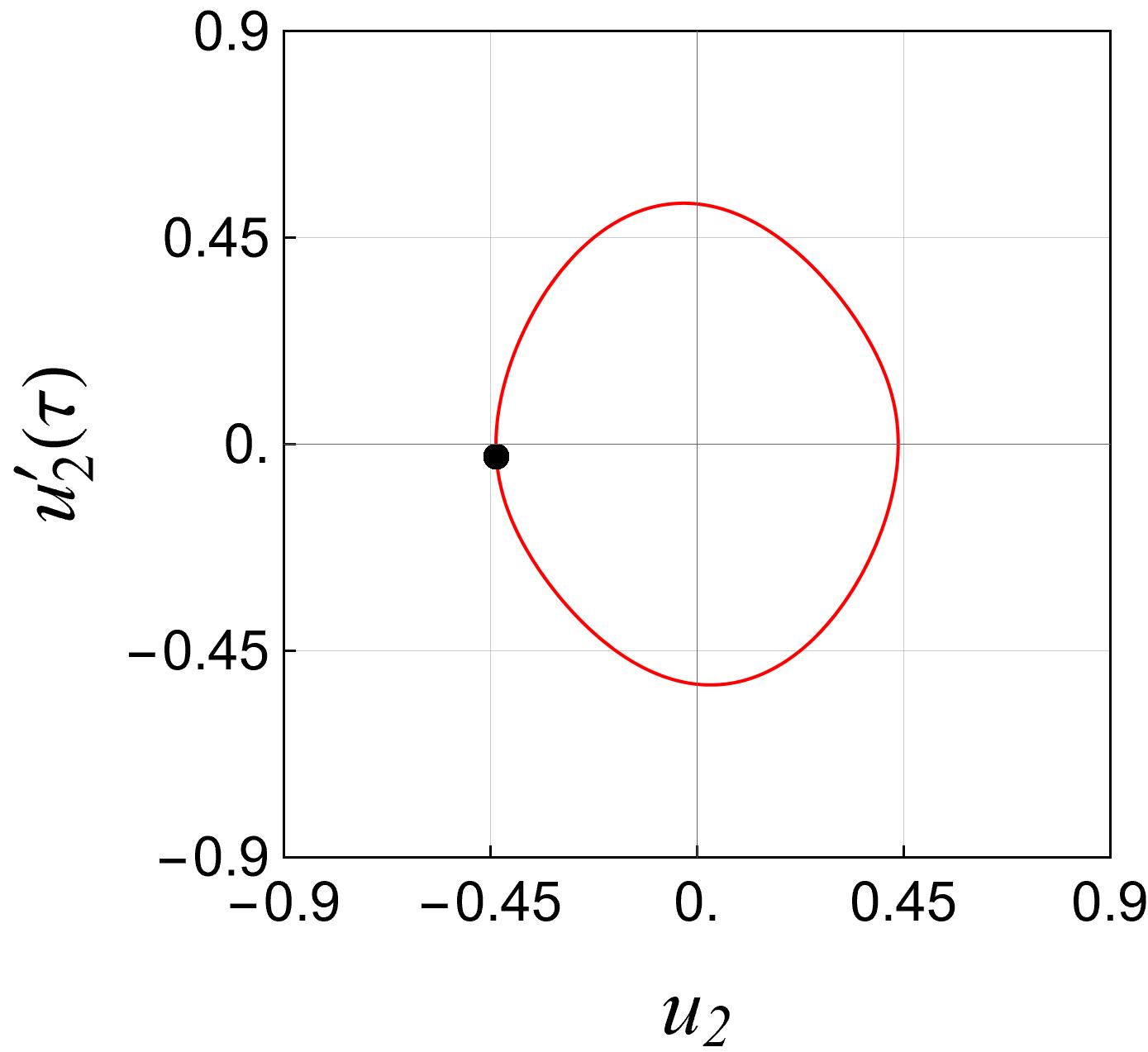}
        \caption{}
        \label{fig6d}
    \end{subfigure}
     \begin{subfigure}{0.32\textwidth}
        \includegraphics[width=\linewidth]{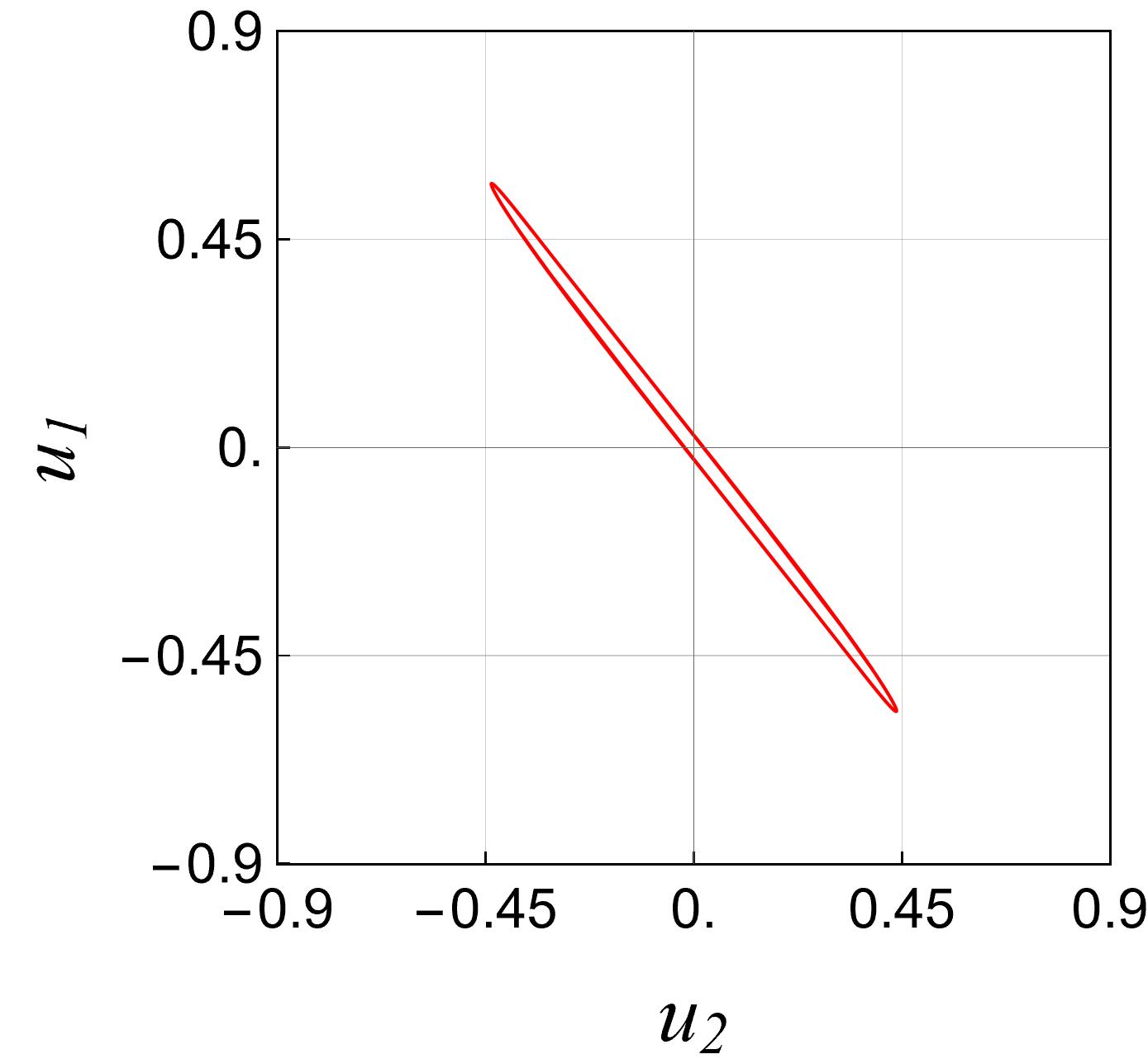}
        \caption{}
        \label{fig6e}
    \end{subfigure}
\caption{The time-domain plots (a–b) show steady-state oscillations for \( u_1\) and \( u_2\) after an initial transient phase of 100 periods, \( \omega = 1.09 \).
The phase-space plots with Poincaré maps (c-d) display closed-loop trajectories, indicating periodic, stable motion in antiphase mode (e). Single Poincaré points also state for regular oscillations.}
    \label{fig6}
\end{figure}

\begin{figure}[H]
    \centering
    \begin{subfigure}{0.8\textwidth}
        \centering
        \includegraphics[width=\textwidth]{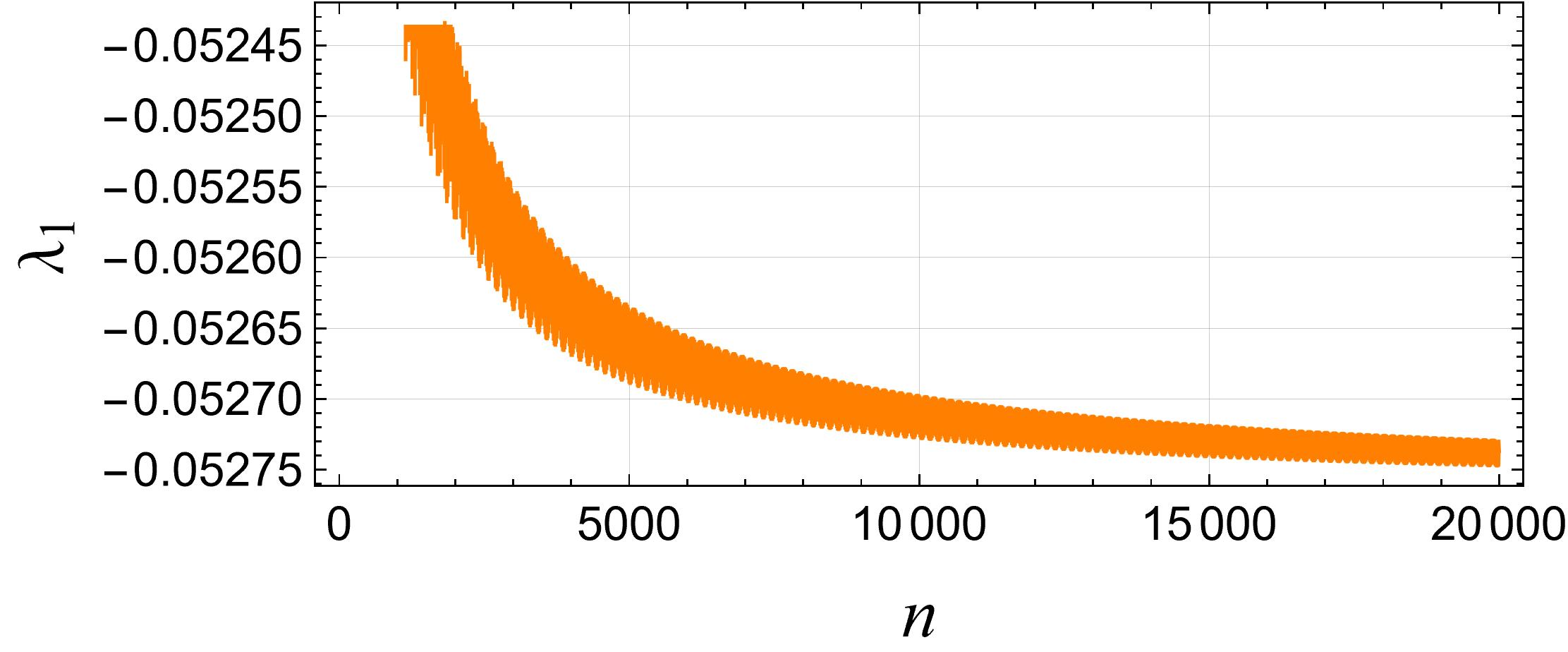}
    \end{subfigure}
    \caption{The maximal Lyapunov exponent $\lambda_1$ as a function of the number of external forcing periods $n$ for \( \omega = 1.09 \)}
     \label{larg2}
\end{figure}

\noindent Time series displayed in Figs.~\ref{fig6a} and \ref{fig6b} were calculated for a driving frequency \( \omega = 1.09 \) for the initial conditions are \(u_1=0.20\), \(u'_1=0.98\), \(u_2=0.56\), and \(u'_2=0.045\), and the system experiences periodic oscillations. The phase plots in Figs. \ref{fig6c} and \ref{fig6d} show closed orbits typical for periodic motion. Furthermore, from  Fig. \ref{fig6e}, it can be seen that the oscillations are close to the antiphase mode due to the negative slope of the $u_1-u_2$ curve. The Poincaré maps are presented as black dots in  Figs. \ref{fig6c} and \ref{fig6d}. Maps in Fig. \ref{fig6c} were obtained when $u_2$ archives its local maximum, while those in Fig. \ref{fig6d} were obtained when $u_1$ archives its local maximum (corresponding to Fig. \ref{bif2b}).
Single points observed in all Poincaré maps indicate regular motion in the system. The periodic behavior of the outcome is validated by a positive greatest Lyapunov exponent $\lambda_1$ for $\omega=1.09$, and taking the values of the parameters from Table \ref{tab2}, with the method of computation illustrated in Fig. \ref{larg2}, where n is the total number of periods of external forcing.

\begin{figure}[H]
    \centering
    \begin{subfigure}{0.49\textwidth}
        \includegraphics[width=\linewidth]{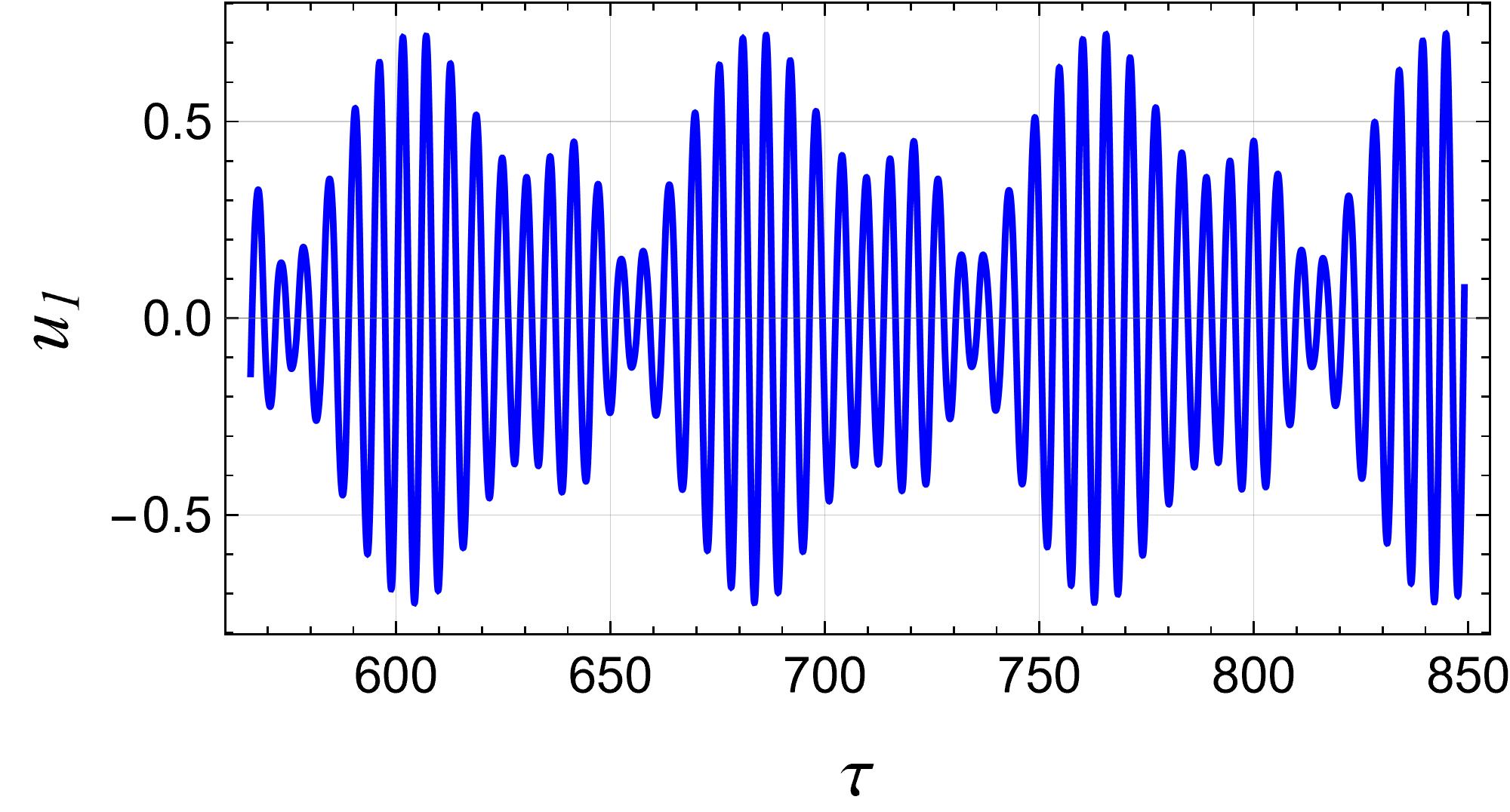}
        \caption{}
        \label{fig7a}
    \end{subfigure}
    \begin{subfigure}{0.49\textwidth}
        \includegraphics[width=\linewidth]{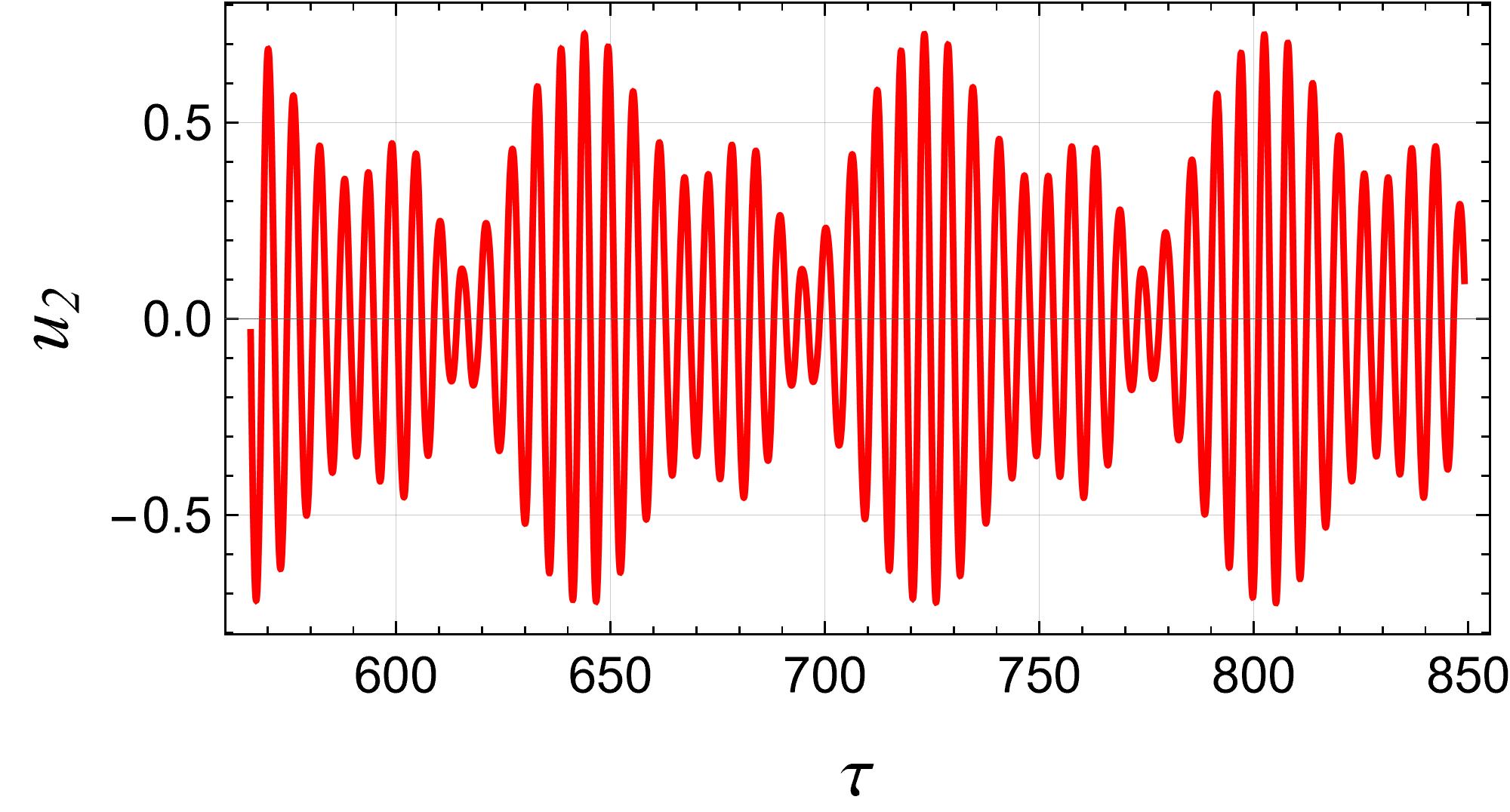}
        \caption{}
        \label{fig7b}
    \end{subfigure}
    \medskip 
    \begin{subfigure}{0.32\textwidth}
        \includegraphics[width=\linewidth]{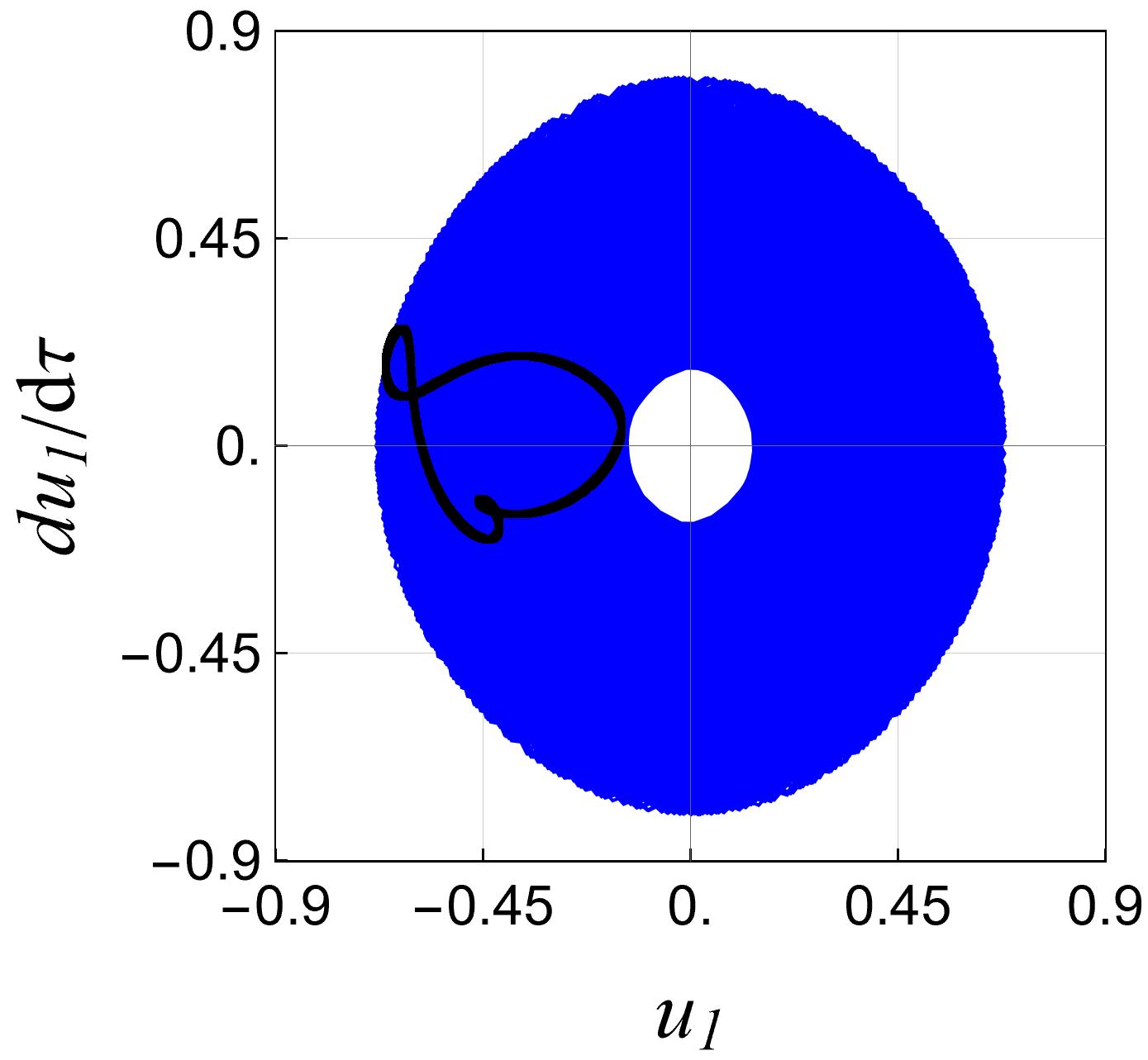}
        \caption{}
        \label{fig7d}
    \end{subfigure}
     \begin{subfigure}{0.32\textwidth}
        \includegraphics[width=\linewidth]{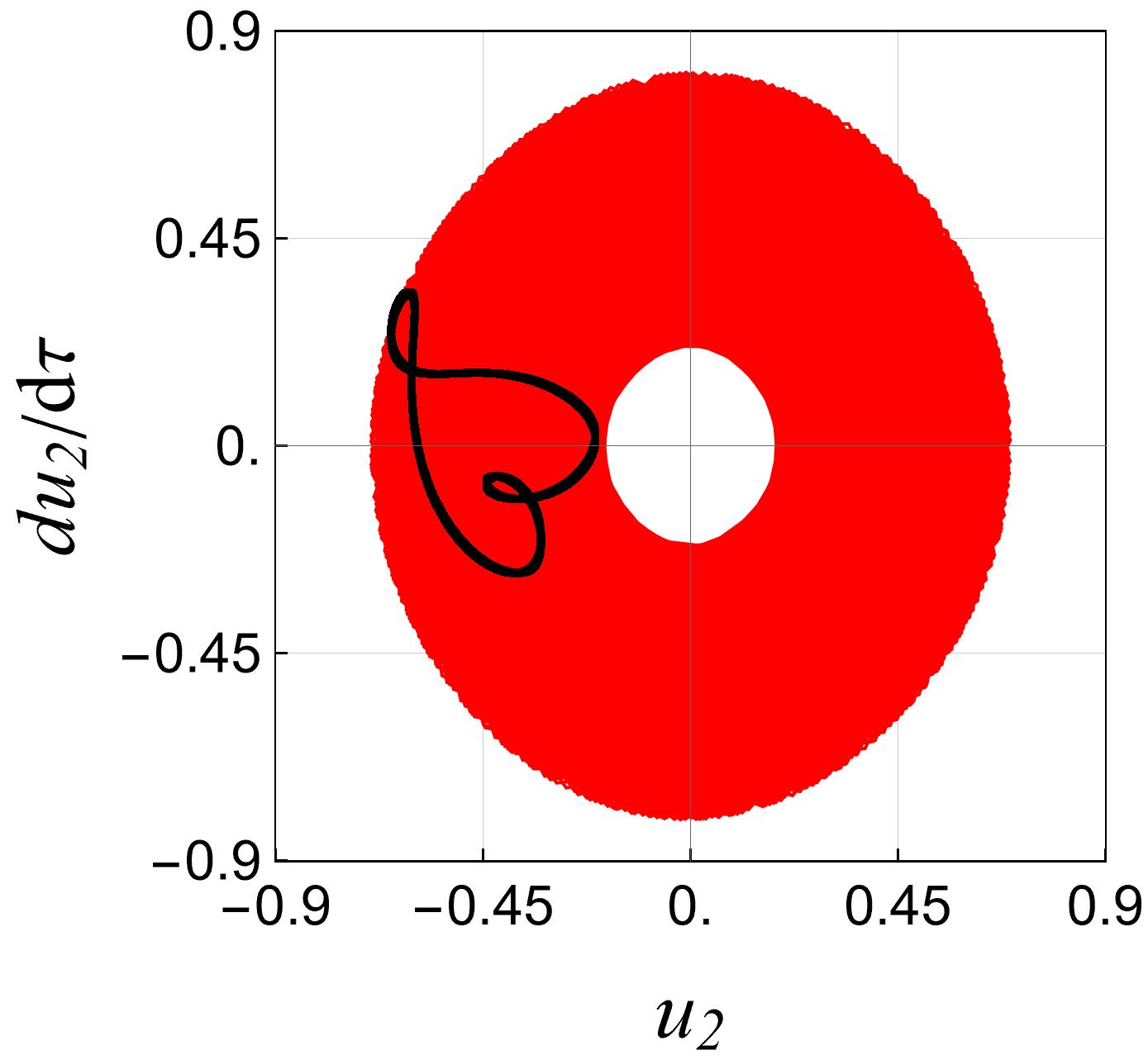}
        \caption{}
        \label{fig7e}
    \end{subfigure}
     \medskip 
    \begin{subfigure}{0.32\textwidth}
        \includegraphics[width=\linewidth]{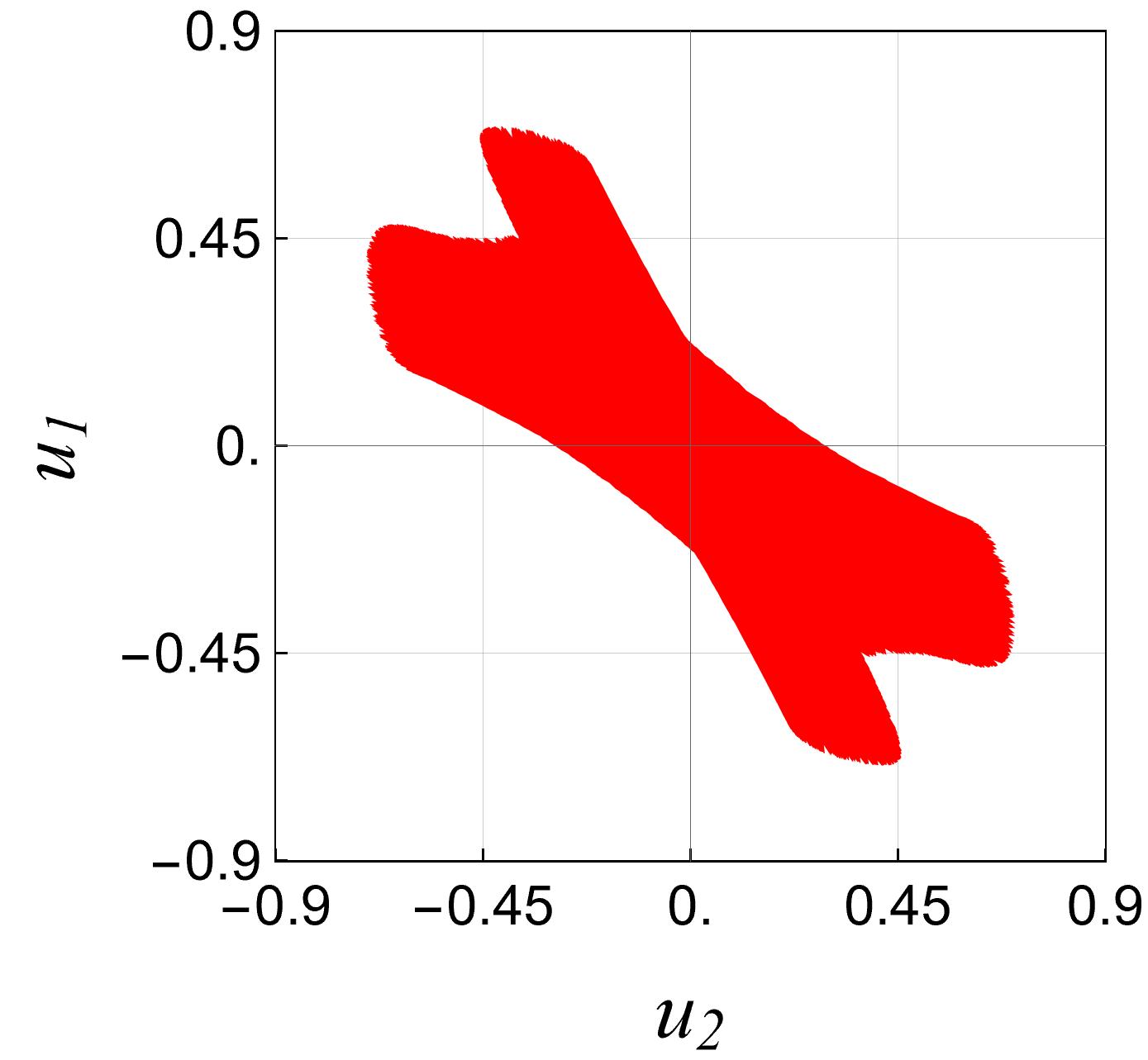}
        \caption{}
        \label{fig7f}
    \end{subfigure}
    \caption{The time-domain plots (top) show near steady-state oscillations for \( u_1 \) and \( u_2 \) after an initial transient phase of 100 periods, \( \omega = 1.10 \).
    The phase-space plots (bottom) display multiple closed-loop trajectories, and the Poincaré section indicates a quasi-periodic motion.}
    \label{fig7}
\end{figure}

\begin{figure}[H]
    \centering
    \begin{subfigure}{0.8\textwidth}
        \centering
        \includegraphics[width=\textwidth]{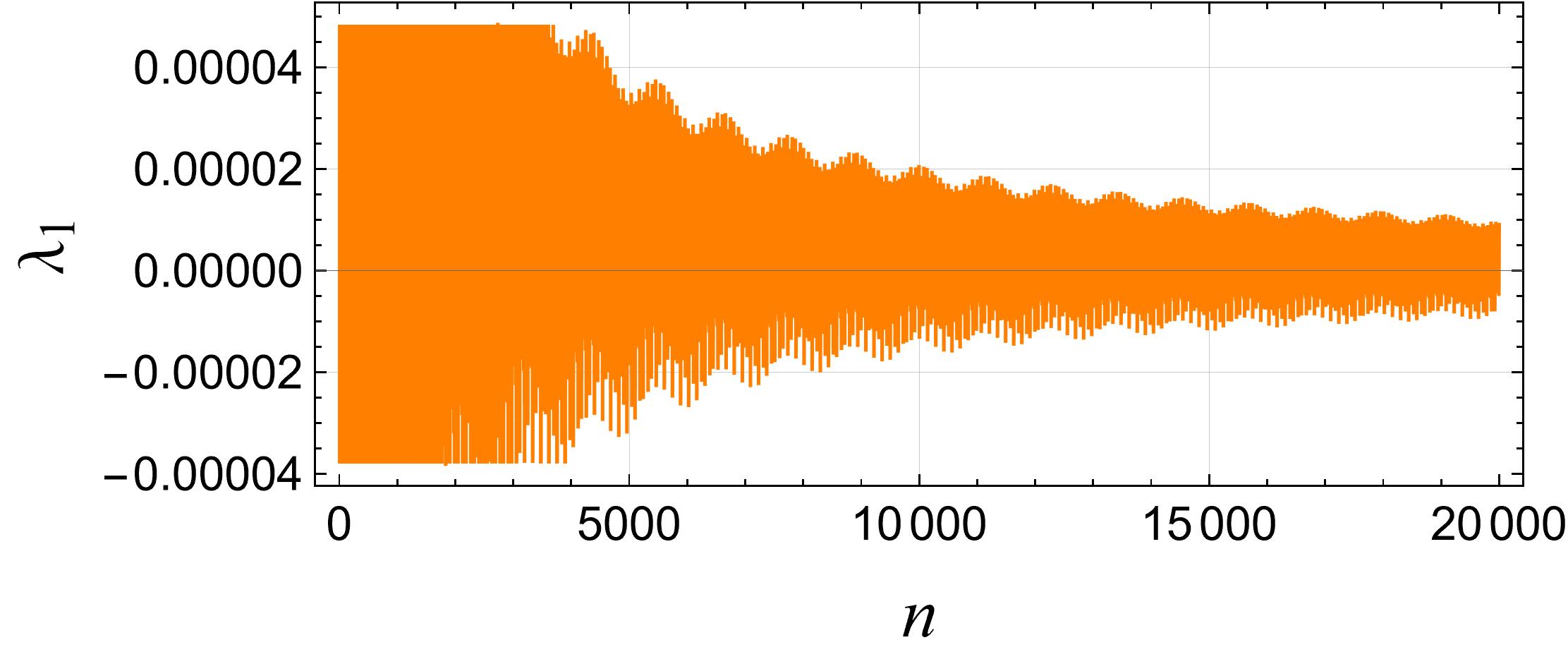}
    \end{subfigure}
    \caption{The maximal Lyapunov exponent $\lambda_1$ as a function of the number of external forcing periods $n$ for \( \omega = 1.10 \)}
     \label{larg3}
\end{figure}

A motion is presented in Figs. \ref{fig7a} and \ref{fig7b} show oscillations with fluctuating amplitude and is obtained for frequency $\omega=1.10$, and the initial conditions are \(u_1=0.75\), \(u'_1=0.25\), \(u_2=0.11\), and \(u'_2=0.79\). 
The phase trajectories with Poincaré maps are displayed in Figs. \mbox{\ref{fig7d}-\ref{fig7e}} and are composed of many closely spaced and overlapping orbits, which is untypical for periodic motion. The fluctuating amplitude indicates quasi-periodic motion, and it is confirmed by Poincaré maps. Poincaré points form closed curves, confirming the presence of quasi-periodicity resulting from frequency components that are mutually incommensurable, meaning their ratios are irrational and the trajectory never exactly repeats, although it remains bounded and orderly. Oscillations exhibit near antiphase character in Fig. \ref{fig7f}. Fig. \ref{larg3} presents the quasi-periodic motion, which is validated by a positive greatest Lyapunov exponent $\lambda_1$ for the $\omega=1.10$, as a function of the number of periods of forcing $n$.

\begin{figure}[H]
    \centering
    \begin{subfigure}{0.49\textwidth}
        \includegraphics[width=\linewidth]{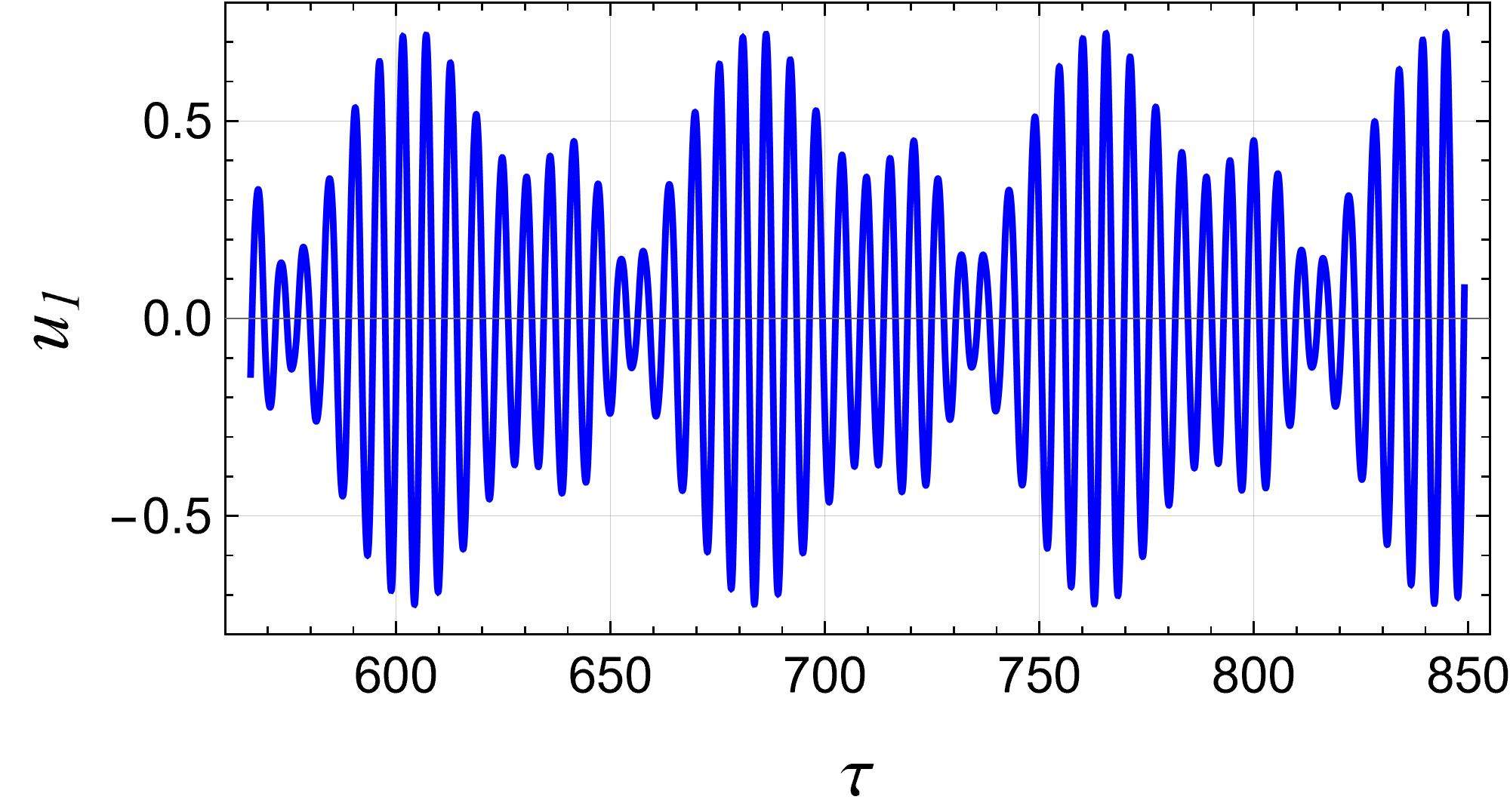}
        \caption{}
        \label{fig8a}
    \end{subfigure}
    \begin{subfigure}{0.49\textwidth}
        \includegraphics[width=\linewidth]{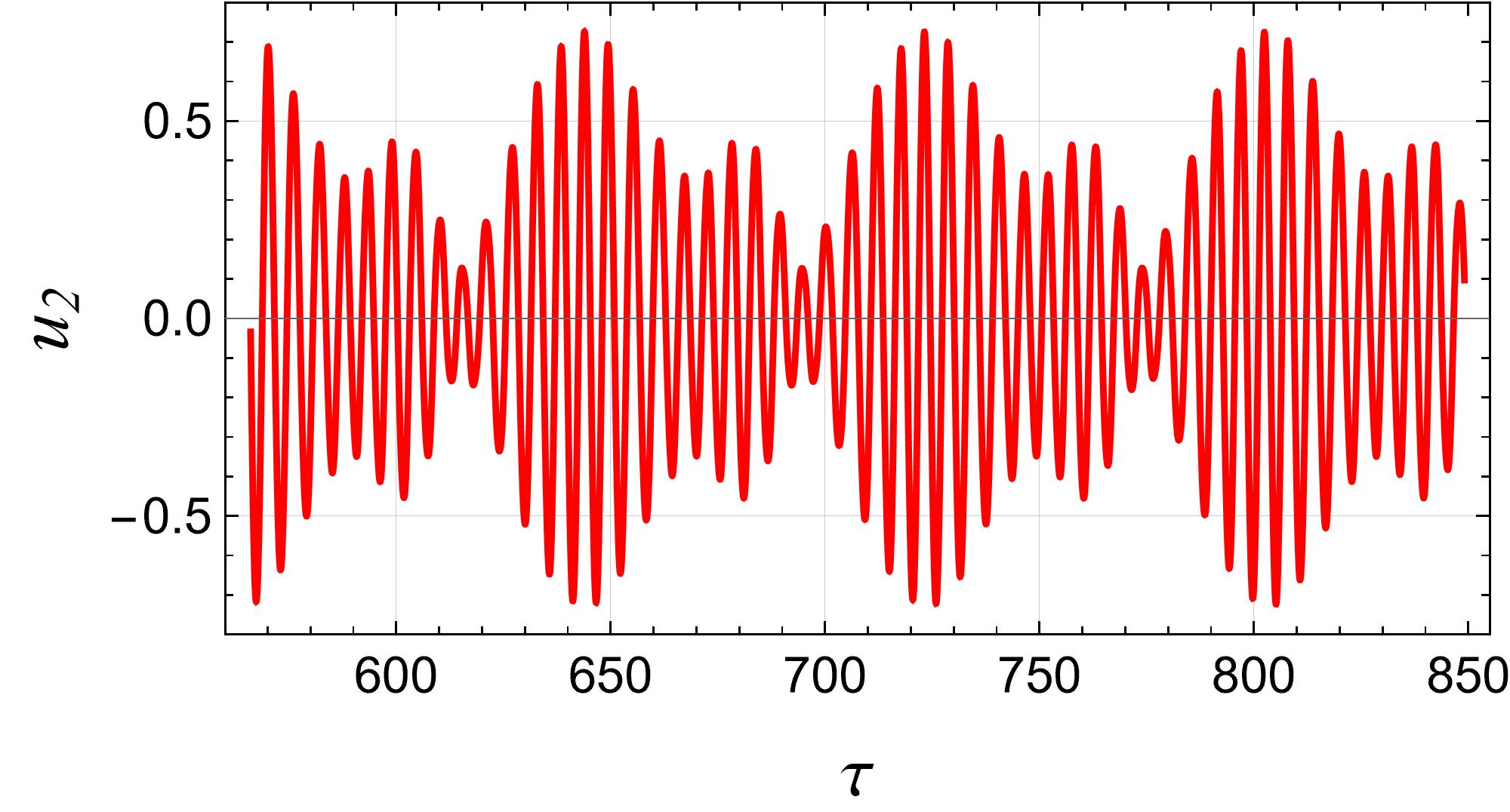}
        \caption{}
        \label{fig8b}
    \end{subfigure}
    \medskip 
    \begin{subfigure}{0.32\textwidth}
        \includegraphics[width=\linewidth]{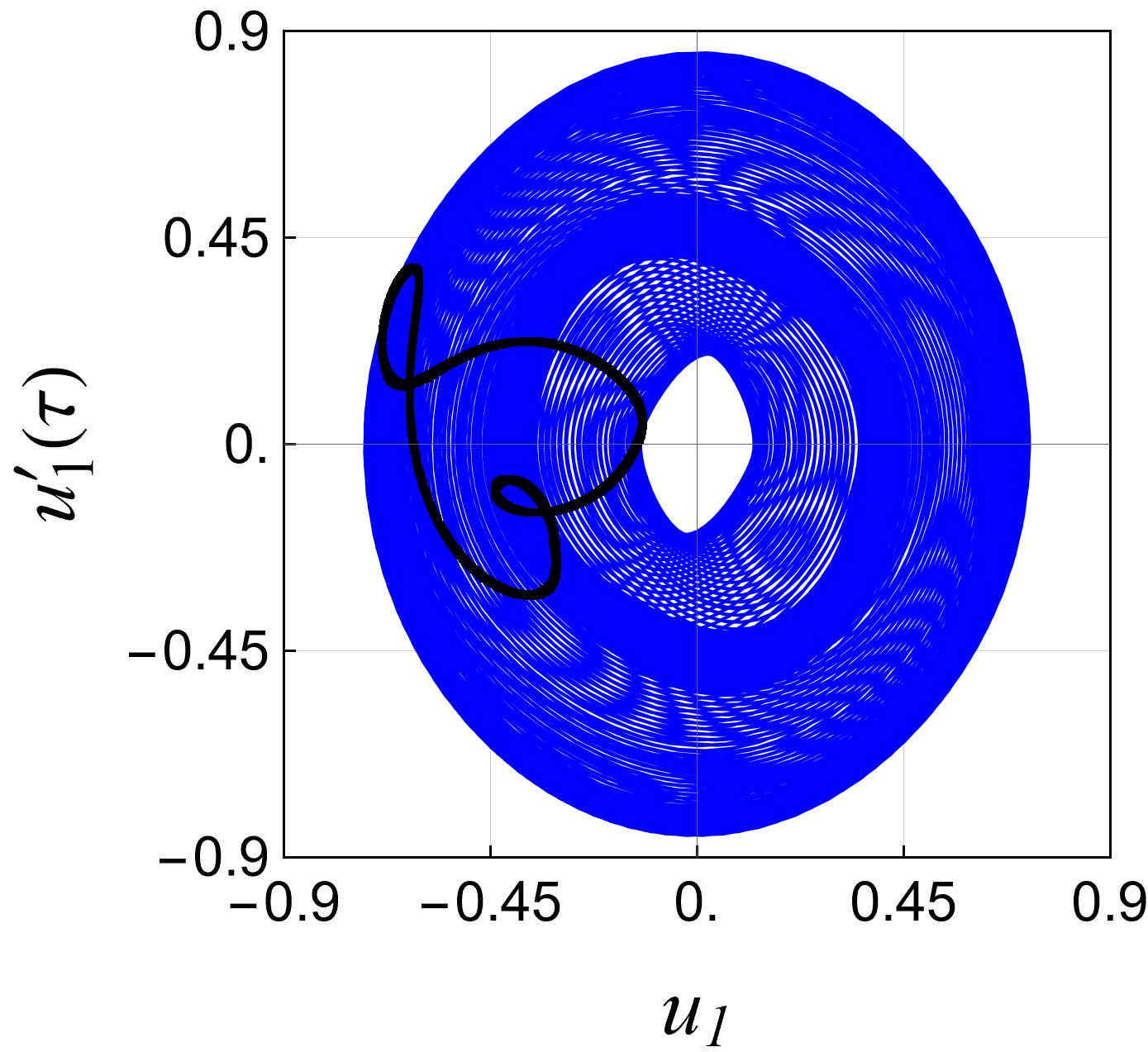}
        \caption{}
        \label{fig8c}
    \end{subfigure}
    \begin{subfigure}{0.32\textwidth}
        \includegraphics[width=\linewidth]{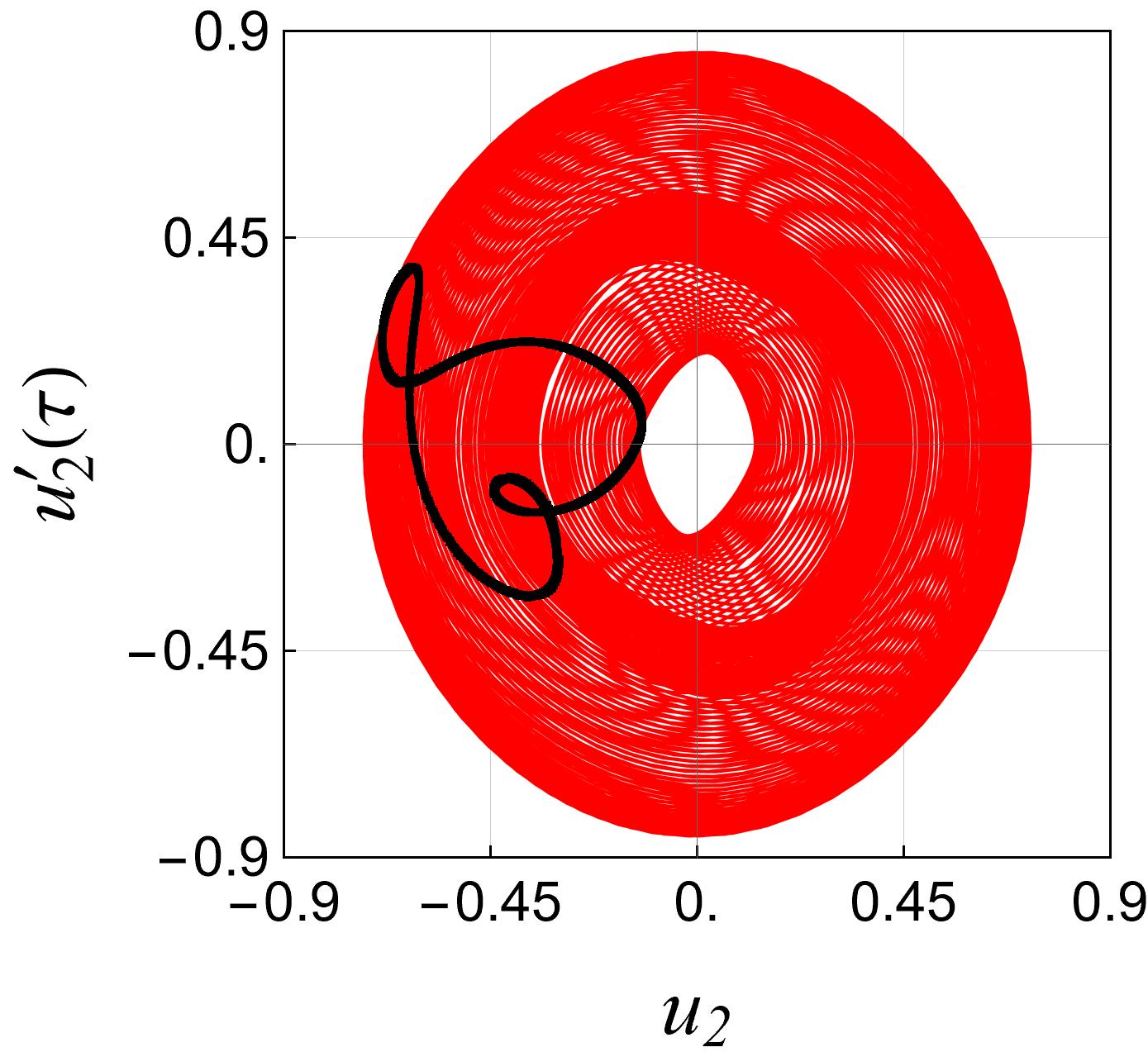}
        \caption{}
        \label{fig8d}
     \end{subfigure}
     \begin{subfigure}{0.32\textwidth}
        \includegraphics[width=\linewidth]{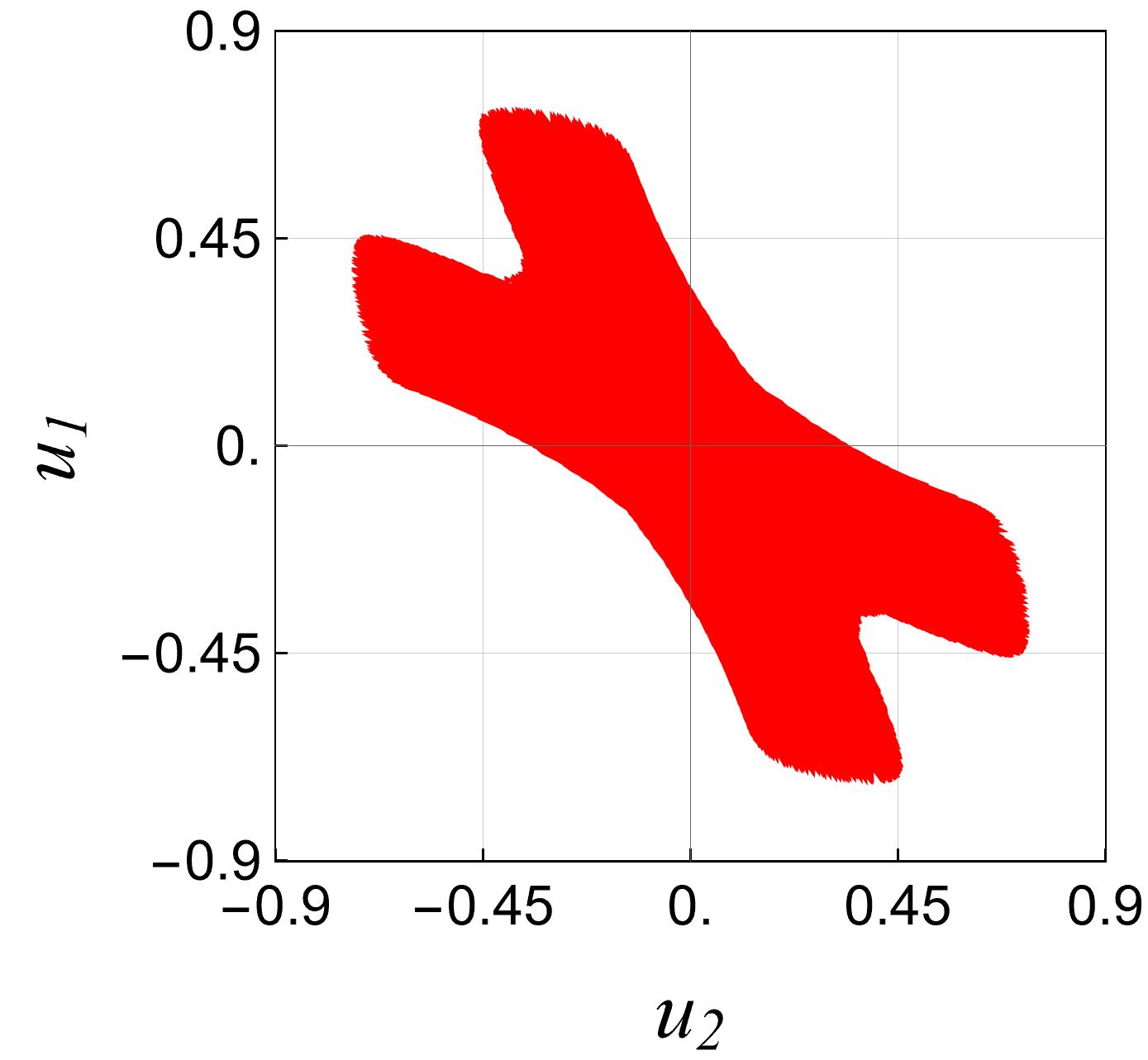}
        \caption{}
        \label{fig8e}
    \end{subfigure}
\caption{The time-domain plots (top) show near steady-state oscillations for \( u_1 \) and \( u_2\) after an initial transient phase of 100 periods, \( \omega = 1.11 \).
The phase-space plots (bottom) display an uncertain trajectory, while the Poincaré maps indicate quasi-periodicity.}
    \label{fig8}
\end{figure}

\begin{figure}
    \centering
    \begin{subfigure}{0.8\textwidth}
        \centering
        \includegraphics[width=\textwidth]{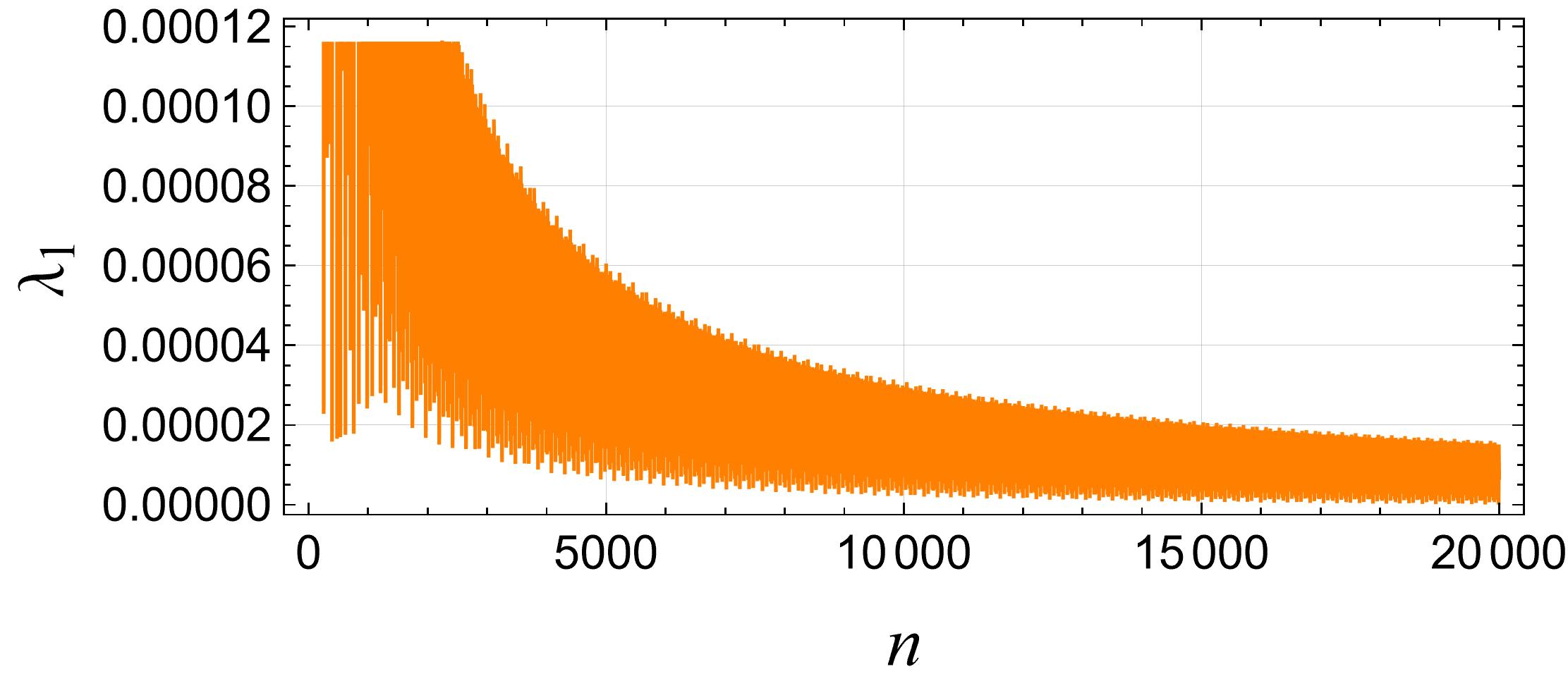}
    \end{subfigure}
    \caption{The maximal Lyapunov exponent $\lambda_1$ as a function of the number of external forcing periods $n$ for \( \omega = 1.11 \)}
     \label{larg4}
\end{figure}

\noindent In Fig. \ref{fig8}, the time series plots might at first glance suggest the chaotic motion of the oscillators. However, the phase portraits and Poincaré maps clearly indicate that the system operates in a quasi-periodic pattern for the examined frequency $\omega=1.11$, and the initial conditions are \(u_1=0.99\), \(u'_1=0.88\), \(u_2=1.1\), and \(u'_2=0\). The quasi-periodic behavior of the results is confirmed by a positive greatest Lyapunov exponent $\lambda_1$ for $\omega=1.11$, as a function of the number of periods of forcing $n$, and assuming the values of involved parameters from Table \ref{tab2}, with the method of computation presented in Fig. \ref{larg4}.

\begin{figure}
    \centering
    \begin{subfigure}{0.49\textwidth}
        \includegraphics[width=\linewidth]{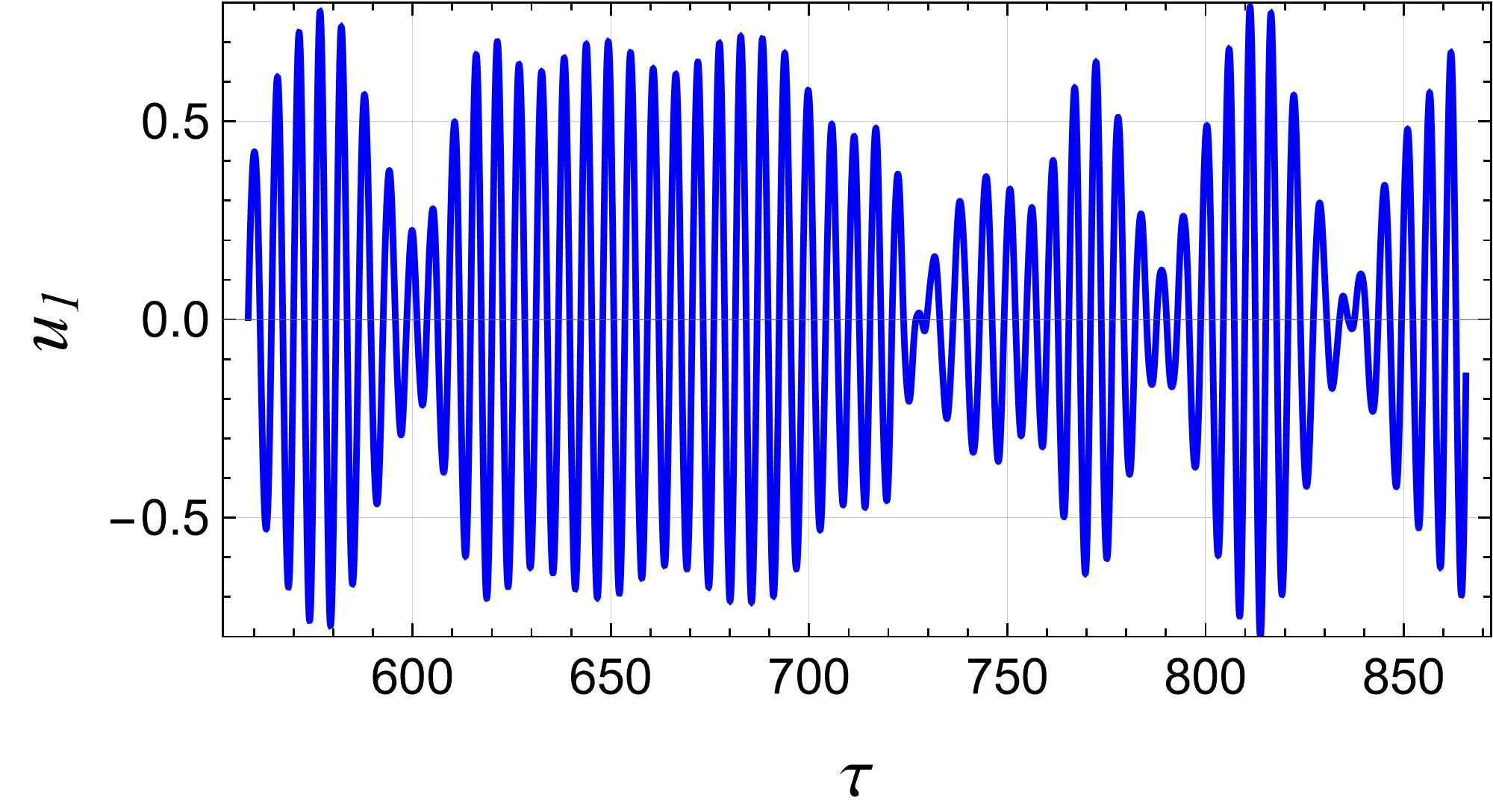}
        \caption{}
        \label{fig12a}
    \end{subfigure}
    \begin{subfigure}{0.49\textwidth}
        \includegraphics[width=\linewidth]{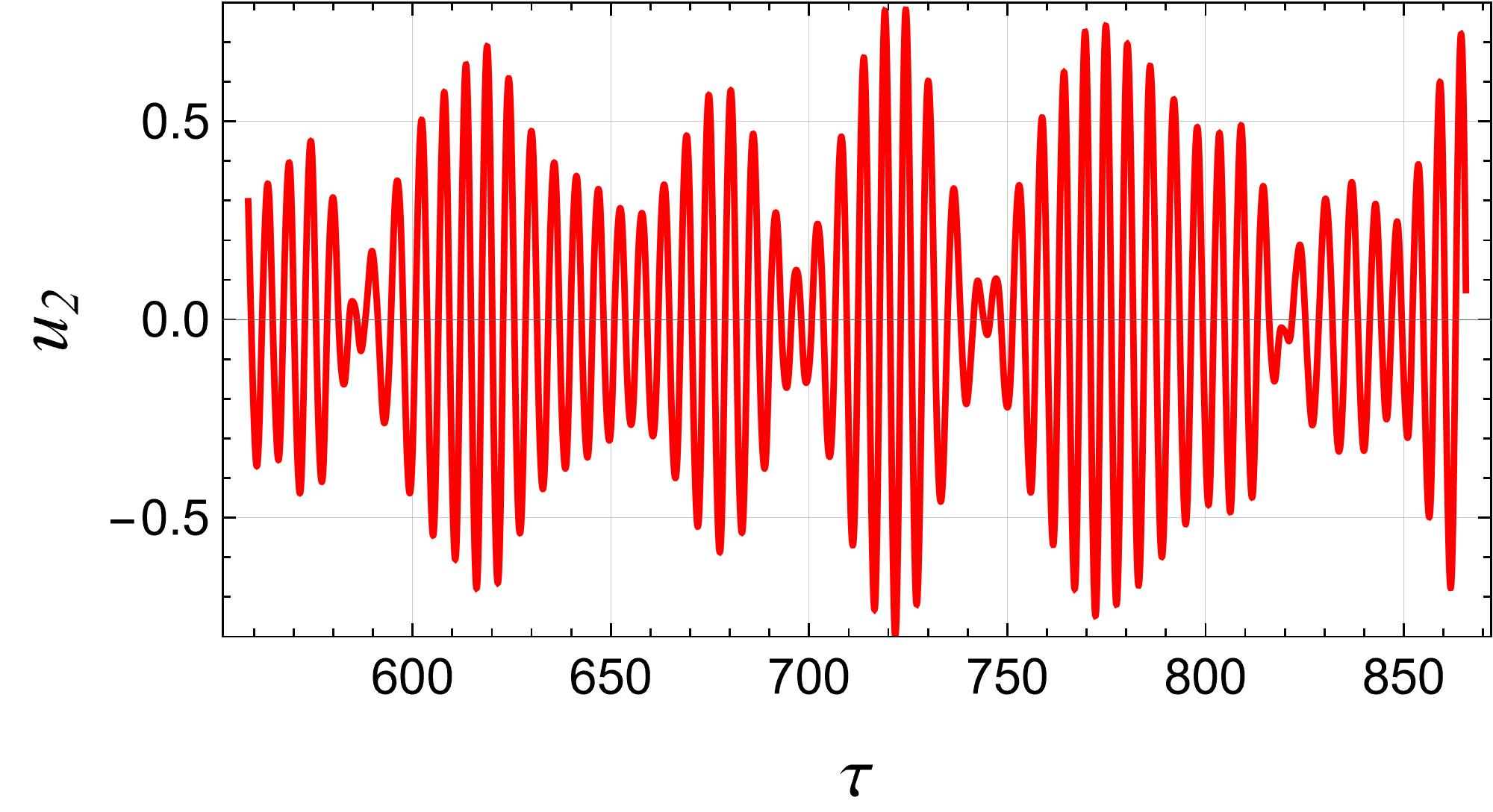}
        \caption{}
        \label{fig12b}
    \end{subfigure}
    \medskip 
    \begin{subfigure}{0.32\textwidth}
        \includegraphics[width=\linewidth]{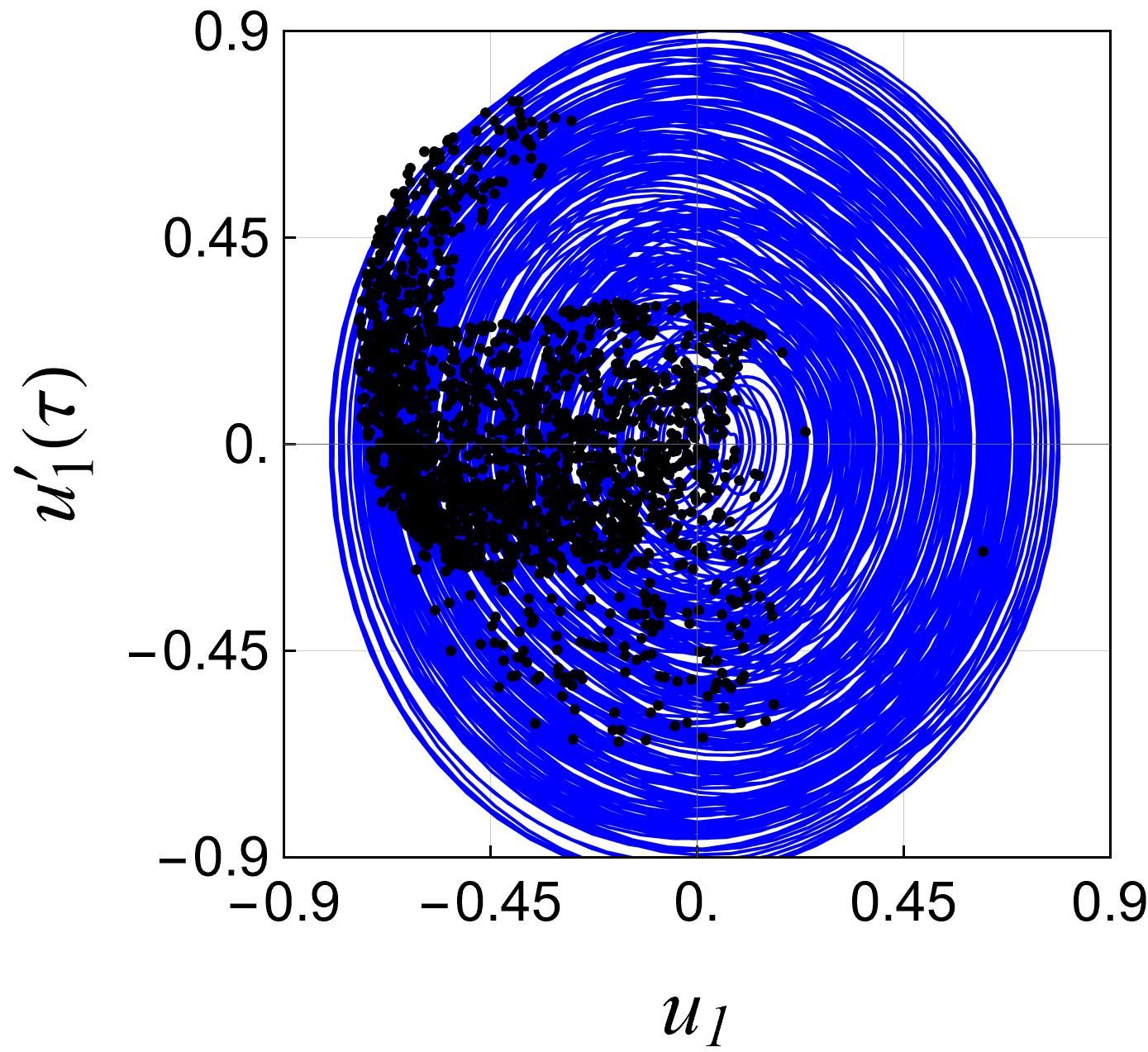}
        \caption{}
        \label{fig12c}
    \end{subfigure}
    \begin{subfigure}{0.32\textwidth}
        \includegraphics[width=\linewidth]{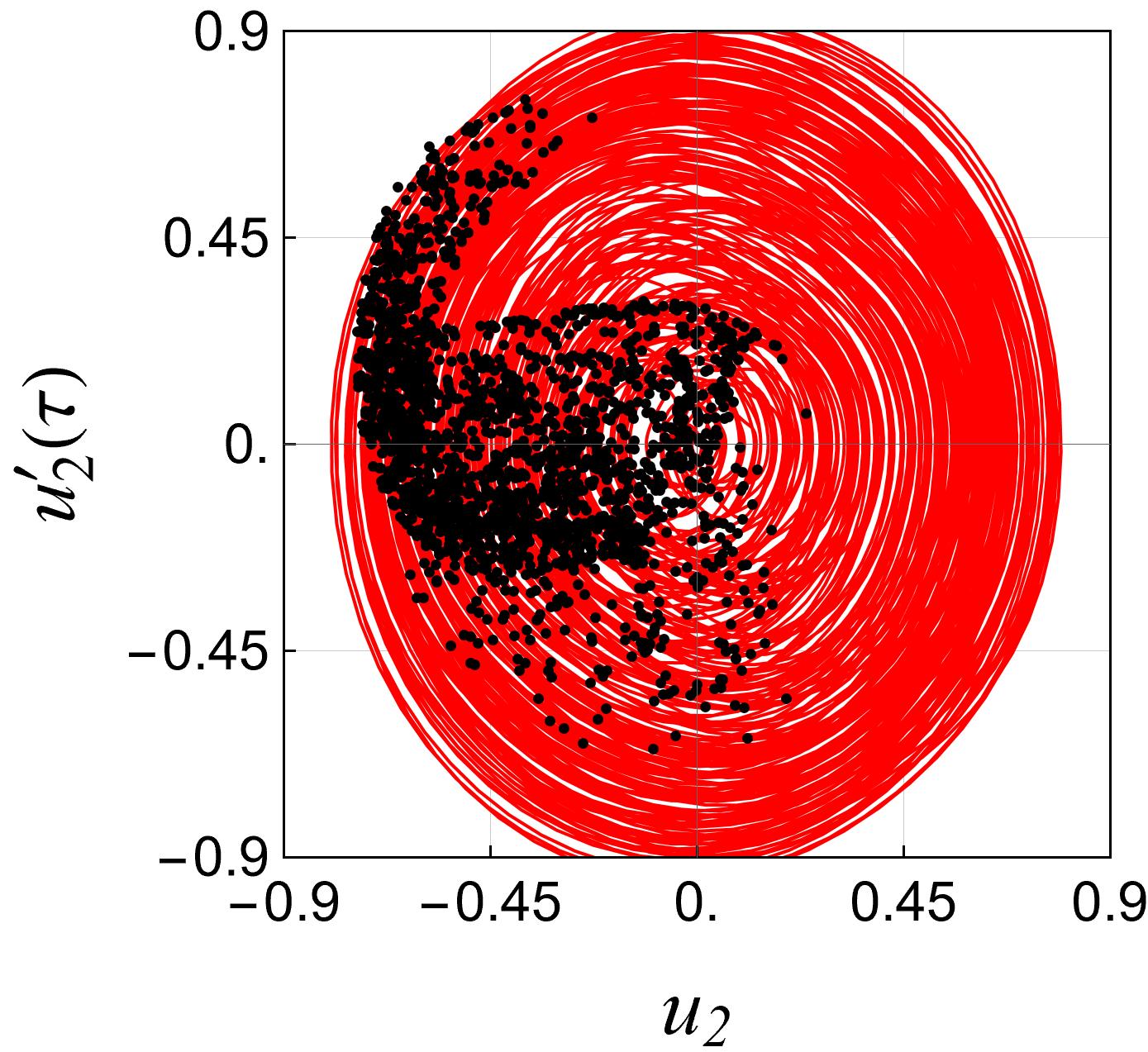}
        \caption{}
        \label{fig12d}
     \end{subfigure}
     \begin{subfigure}{0.32\textwidth}
        \includegraphics[width=\linewidth]{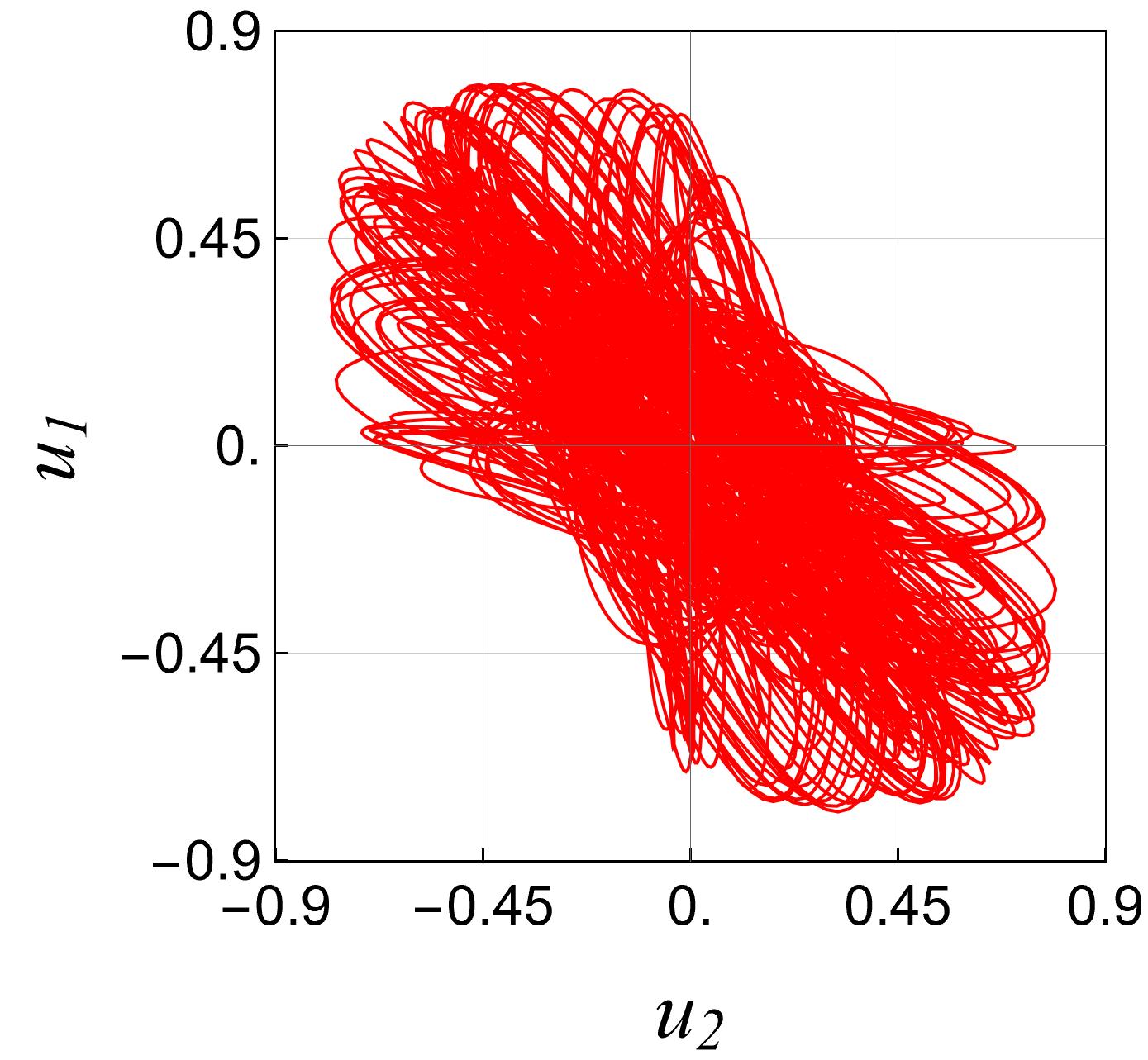}
        \caption{}
        \label{fig12e}
    \end{subfigure}
    \caption{The time-domain plots (top) show near steady-state oscillations for \( u_1 \) and \( u_2 \) after an initial transient phase of 100 periods, \( \omega = 1.125 \).
    The phase-space plots (bottom) display an uncertain trajectory, and the Poincaré section indicates a chaotic regime.}
    \label{fig12}
\end{figure}

\noindent It is evident that the studied case represents a further evolution of the quasi-periodic motion analyzed in Fig. \ref{fig7}. The oscillation amplitudes increase with the increase of the control parameter $\omega$, resulting, finally, in non-periodic motion presented in Fig. \ref{fig12}.
 
\begin{figure}
    \centering
    \begin{subfigure}{0.8\textwidth}
        \centering
        \includegraphics[width=\textwidth]{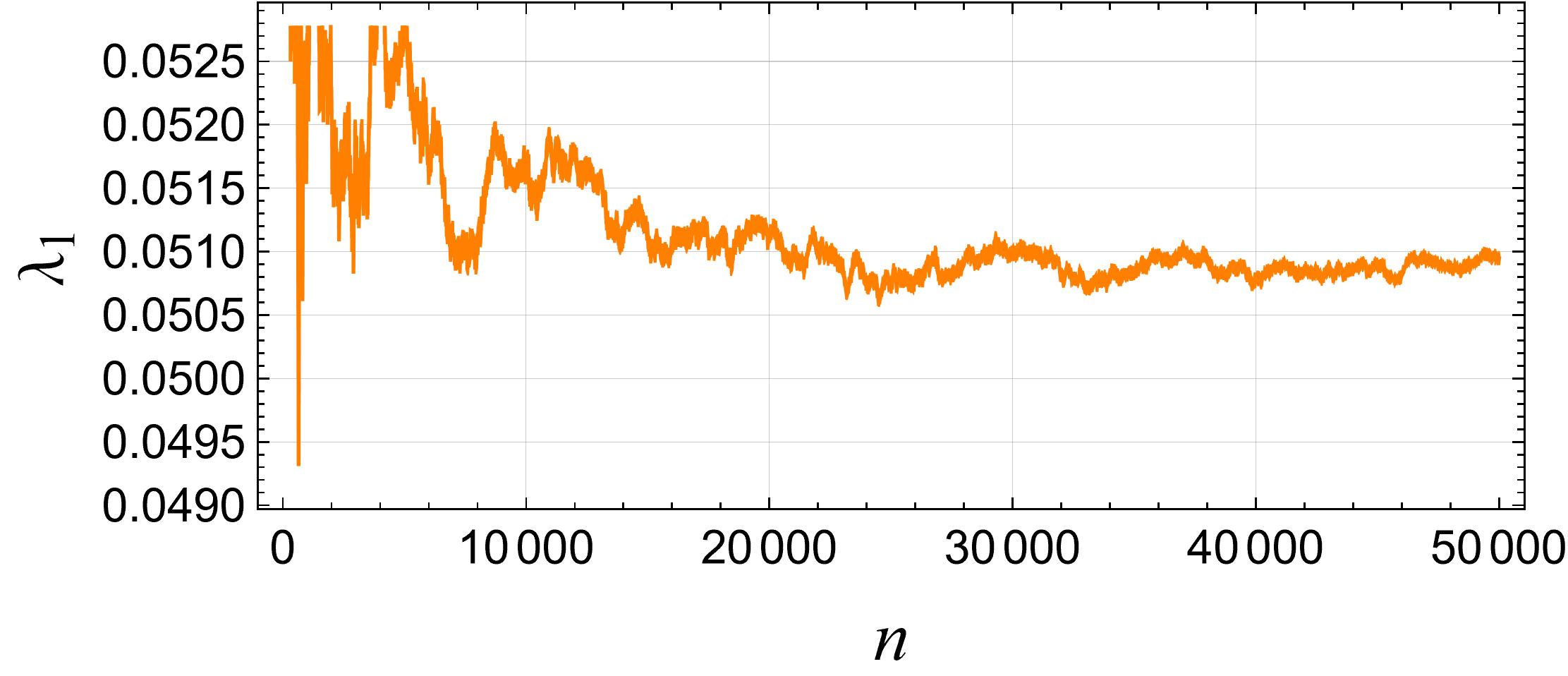}
    \end{subfigure}
    \caption{The maximal Lyapunov exponent $\lambda_1$ as a function of the number of external forcing periods $n$ for \( \omega = 1.125 \)}
     \label{larg5}
\end{figure}

The case of system motion considered in Fig. \ref{fig12} meets the characteristics of a chaotic regime for \(\omega=1.125\) and the initial conditions are \(u_1=0.99\), \(u'_1=0.88\), \(u_2=1.1\), and \(u'_2=0\). The time series and phase trajectories show a lack of repeatability, and the Poincaré maps form a chaotic attractor. It is worth noting that quasi-periodic and chaotic windows look similar in bifurcation diagrams in Fig. \ref{bif2}, and only further analysis via Poincaré maps and Lyapunov exponents can shed light on the exact type of system oscillations. The system’s chaotic behavior, as observed in the results, is validated by the variation of the largest Lyapunov exponent $\lambda_1$ with respect to the number of forcing periods $n$, as illustrated in Fig. \ref{larg5}.
    
\section{Conclusion}\label{sec5}
This paper presents a thorough analytical and numerical investigation of the detailed complex dynamic behavior of a nonlinear two-degree-of-freedom parametric oscillator that includes dry friction, magnetic stiffness, and time-periodic coupling from a rectangular cross-section rotating beam. The extension of earlier published work \cite{p4,p5} 1DOF models to a more representative coupled system effectively captured the interaction of nonlinear restoring forces and nonsmooth friction dynamics. Analytical solutions derived from the Complex Averaging method have been validated by numerical solutions, demonstrating a variety of dynamical regimes, including periodic, quasi-periodic, and chaotic responses. We have plotted the Lyapunov exponents to explore the behavior of the system's motion. Lyapunov exponents provide a measure of how small differences in initial conditions evolve over time. They help identify whether the system settles into regular or chaotic motion. This makes them a valuable tool for understanding the stability of complex dynamic systems. The system behaves similarly to the 1DOF case for the case of different masses, while it displays complex bifurcation patterns and multistability, which are significantly affected by the order of nonlinear stiffness, friction intensity, and mass arrangement. These findings enhance the theoretical comprehension of coupled nonlinear oscillators with nonsmooth properties and establish a solid basis for practical applications. The insights acquired are pertinent to the advancement of efficient vibration control systems, nonlinear energy harvesters, and complicated mechanical structures that utilize multistable and resonance-enhanced dynamics.
\newline
Future work—the complexity and richness of the observed dynamics motivate further analytical, numerical, and experimental efforts. Building on the present CxA–numerical framework, our next planned research will focus on the application of additional semi-analytical techniques, such as the Method of Multiple Scales (MMS) for capturing slow–fast dynamics and the Harmonic Balance Method (HBM) for accurately predicting periodic and quasi-periodic responses. We have already successfully applied MMS and HBM to our earlier published work, a 1DOF parametric oscillator with dry friction \cite{p5}. On the numerical side, advanced continuation algorithms and basin-of-attraction mapping will be employed to explore global bifurcations and hidden attractors. Furthermore, experimental validation using an extended version of the 1DOF setup to a full 2DOF configuration is a key objective, enabling direct comparison of theoretical predictions with real-world behavior.

\subsection*{Author contribution statement}
{\textbf{Muhammad Junaid-U-Rehman:} Conceptualization, Investigation, Methodology, Validation, Software, Visualization, Writing—Original draft \textbf{Grzegorz Kudra:} Conceptualization, Methodology, Validation, Software, Visualization, Writing—Reviewing and Editing, Supervision, Project Administration; \textbf{Krystian Polczyński:} Investigation, Data Curation, Validation, Software; \textbf{Kevin Dekemele:} Investigation, Methodology, Software, Formal analysis; \textbf{Jan Awrejcewicz:} Supervision, Formal Analysis.}
\subsection*{Acknowledgements}
This research has received funding from the National Science Center, Poland, under the grant PRELUDIUM 23 No. DEC-2024/53/N/ST8/00400. This article was completed while the first author, Muhammad Junaid-U-Rehman, is a PhD student at the Interdisciplinary Doctoral School of Lodz University of Technology, Poland.
For the purpose of Open Access, the authors have applied a CC-BY public copyright
license to any Author Accepted Manuscript (AAM) version arising from this submission.
\subsection*{Declaration of competing interest}
The authors affirm that no personal relationships or recognized competing financial interests could have influenced the research provided in this study.
\subsection*{Data availability}
{Data will be provided upon request.}

\end{document}